\documentclass[12pt,a4paper,twoside,prd,showpacs,nofootinbib,preprintnumbers]{revtex4-1}

\newcommand{\beq}{\begin{equation}}
\newcommand{\eeq}{\end{equation}}
\usepackage{graphicx}
\usepackage{xcolor}
\usepackage{amssymb,amsmath,bm,natbib}
\newcommand{\nexp}{PIONEER}

\newcommand{\eps}{\varepsilon}

\usepackage{amsmath}
\newcommand{\Lagr}{\mathcal{L}}

\begin{document}

\title{Testing Lepton Flavor Universality  with Pion, Kaon, Tau, and Beta Decays}

\author{Douglas Bryman}
\email{doug@triumf.ca}
\affiliation{Department of Physics and Astronomy, University of British Columbia, Vancouver, BC, Canada V6T 1Z1}
\affiliation{TRIUMF, Vancouver, BC, Canada V6T 2A3}
\author{Vincenzo Cirigliano}
\email{cirigliano@lanl.gov}
\affiliation{Theoretical Division, Los Alamos National Laboratory, Los Alamos, NM 87545}
\author{Andreas Crivellin}
\email{ andreas.crivellin@cern.ch}
\affiliation{Physik-Institut, Universit\"at Z\"urich,
	Winterthurerstrasse 190, CH-8057 Z\"urich, Switzerland}
\affiliation{Paul Scherrer Institut, CH--5232 Villigen PSI, Switzerland}
\affiliation{CERN Theory Division, CH--1211 Geneva 23, Switzerland}
\author{Gianluca Inguglia}
\email{Gianluca.Inguglia@oeaw.ac.at}
\affiliation{Institute of High Energy Physics, Austrian Academy of Sciences, Vienna, Austria, 1050}

\preprint{CERN-TH-2021-184, LA-UR-21-30608, PSI-PR-21-25, ZU-TH 54/21}

\begin{abstract}
We present an overview of searches for violation of lepton flavor universality  with focus on low energy precision probes using  pions, kaons, tau leptons, and nuclear beta decays. The current experimental results are reviewed,  the theoretical status within the context of the Standard Model is summarized, and future prospects (both experimental and theoretical) are discussed. We review the implications of these measurements for physics beyond the Standard Model by performing a global model-independent fit to modified $W$ couplings to leptons and four-fermion operators. We also discuss new physics in the context of simplified models and review Standard Model extensions with focus on those that can explain a possible deviation from unitarity of the Cabibbo-Kobayashi-Maskawa quark mixing  matrix.
\end{abstract}

\maketitle

\newpage
\tableofcontents
\newpage
\section{Introduction}

The Standard Model (SM) of particle physics describes the known constituents of matter: the three generations (or flavors) of quarks and leptons, as well as their interactions (excluding gravity). Its final missing ingredient, the Higgs boson, was discovered at the Large Hadron Collider (LHC) at CERN in 2012 ~\cite{ATLAS:2012yve,CMS:2012qbp}. However, it is clear that the SM cannot be the ultimate fundamental theory of nature. In addition to many theoretical arguments for the existence of new physics, the SM e.g.~cannot account the existence of Dark Matter (DM) or Dark Energy established at cosmological scales nor for neutrino masses or the existence of exactly three generations of fermions. 

Therefore, the search physics beyond the SM is a prime subject of current research. There are, in general, two ways to search for new particles and interactions: direct searches at high energy colliders (such as the LHC) and indirect searches for quantum effects in precision observables (see Particle Data Group~\cite{ParticleDataGroup:2020ssz}). Concerning the latter, one specially promising avenue is to search for the violation of (approximate) symmetries of the SM. In this way, one is very sensitive to new physics (NP) that does not necessarily respect these symmetries and thus leads to sizable effects even if the mass scale is quite high. Furthermore, the symmetries of the SM can be exploited to obtain more precise predictions as in most cases theoretical and parametric uncertainties are reduced. 

In the SM, the gauge interactions are the same for all flavors, i.e. they respect lepton flavor universality (LFU), which is in fact only broken by the Higgs Yukawa couplings. As these couplings are very small (at most, of the order of one percent for the tau lepton), LFU is an approximate accidental symmetry of the SM (at the Lagrangian level). However, the impact of the lepton masses, originating from the Higgs Yukawa couplings after electroweak (EW) symmetry breaking on the lifetimes of charged leptons is enormous due to kinematic effects. Therefore,  LFUV implies interactions with different couplings to electrons, muons and tau leptons (disregarding phase space effects) that directly distinguish among the charged leptons at the Lagrangian level. 

In recent years experimental tests of LFU have accumulated intriguing hints for  physics effects not included  in the SM (see Ref.~\cite{Crivellin:2021sff} for a short review).\footnote{Note that interesting hints for new scalar particles have emerged recently at the LHC~\cite{Buddenbrock:2019tua,Crivellin:2021ubm,Fischer:2021sqw}. } In particular, the measurements of the ratios of branching ratios (Br)  $R(D^{(*)})$=Br[$ B\to D^{(*)}\tau\nu_\tau$]/Br[$B\to D^{(*)}\ell\nu_\ell$]~\cite{Lees:2012xj,Aaij:2017deq,Abdesselam:2019dgh} , where $\ell=\mu, e$, and  $R(K^{(*)})$=Br[$B\to K^{(*)} \mu^+ \mu^-$]$/$Br[$B\to K^{(*)} e^+ e^-$]~\cite{Aaij:2017vbb,Aaij:2019wad,LHCb:2021trn} deviate from the SM expectation by more than $3\sigma$~\cite{Amhis:2019ckw,Murgui:2019czp,Shi:2019gxi,Blanke:2019qrx,Kumbhakar:2019avh} and $4\sigma$~\cite{Alguero:2019ptt,Aebischer:2019mlg,Ciuchini:2019usw,Arbey:2019duh}, respectively.\footnote{Even though the ratios $R(D)$ and $R(D^*)$ point towards violation of $\mu-\tau$ LFU motivating NP effects, in the observables discussed in this review we do not include them. The NP effects required are so large that these ratios cannot be explained by modified $W\ell\nu$ couplings which are more stringently constrained by $\tau$ decays. Therefore, effects in 2-quark-2-lepton operators are required for explaining $R(D)$ and $R(D^*)$ which in general have no direct correlations with the tests of LFUV discussed in this review, unless a flavor symmetry is assumed. However, some of these  scenarios give rise to large radiative corrections to $\tau$ decays, such that they are excluded by low energy probes of LFUV~\cite{Feruglio:2016gvd}.} In addition, anomalous magnetic moments $(g-2)_\ell$ ($\ell=e,\mu,\tau$) of charged leptons are intrinsically related to LFU violation as they are chirality flipping quantities. Here, there is the long-standing discrepancy in $(g-2)_\mu$ of $4.2\sigma$~\cite{Bennett:2006fi,Muong-2:2021ojo,Aoyama:2020ynm} which can be considered as a hint of LFU violation (LFUV), since, if compared to $(g-2)_e$, the bound from the latter on flavor blind new physics (NP) is much more stringent.  In addition, there is a hint for LFUV in the difference of the forward-backward asymmetries ($\Delta A_{\rm FB}$) in $B\to D^*\mu\nu$ vs $B\to D^*e\nu$~\cite{Bobeth:2021lya,Carvunis:2021dss}. As another possible indication of LFUV, CMS observed an excess in non-resonant di-electron pairs with respect to di-muons~\cite{Sirunyan:2021khd}. Furthermore, the possible deficit in first-row unitarity of the Cabibbo-Kobayashi-Maskawa (CKM) matrix, known as the Cabibbo angle anomaly (CAA) can also be viewed as a sign of LFUV~\cite{Coutinho:2019aiy,Crivellin:2020lzu}. 

The connection of the CAA and LFUV can be seen as follows. The determination of $V_{ud}$ from $\beta$ decays, which is most relevant for a possible explanation of the CAA, is affected by a modified $W \mu\nu$ coupling~\cite{Crivellin:2020lzu}. Importantly, a modification of the $W\mu\nu$ couplings, if not compensated by an effect in $W e\nu$, would also affect, for example,  the ratios of decay rates $R^{\pi}_{e/\mu} = \frac{\Gamma(\pi\to e \nu)}{\Gamma(\pi\to \mu\nu)}$ or $R^{\tau}_{e/\mu} =\frac{\Gamma(\tau\to e \nu\overline\nu)}{\Gamma(\tau\to \mu\nu\overline\nu)}$ which provide the best tests of LFU. In fact, recent global fits to electroweak (EW) observables and tests of LFU show a preference for a value of $R^{\pi}_{e/\mu}$ smaller than its SM expectation~\cite{Coutinho:2019aiy,Crivellin:2020ebi}. Furthermore, $R(K^{*})$ can be correlated to $R^{\pi}_{e/\mu}$~\cite{Capdevila:2020rrl} and a combined explanation of the deficit in the first-row CKM unitarity and the CMS excess in di-electrons even predicts that $R^{\pi}_{e/\mu}$ should be smaller than its SM value~\cite{Crivellin:2021rbf}.

These considerations provide additional motivation for us to review, summarize and re-examine the different searches for LFUV in the charged current with focus on pion, kaon, tau, and beta decays. In the next section we will discuss the experimental and theoretical status of these processes. We will then consider the impact on NP searches, first in a model independent way by a global analysis of modified $W\ell\nu$ coupling and four-fermion operators, and then by considering different NP models with focus on the ones that can explain the CAA in Sec.~\ref{NPsection}.\footnote{Here we will focus on models with heavy NP components such that the effective Lagrangian only contains SM fields. However,  light but massive right-handed neutrinos~\cite{Bryman:2019bjg,deGouvea:2015euy,NA62:2020mcv,NA62:2021bji}, Majorons~\cite{Lessa:2007up}, as well as light Dark Matter candidates~\cite{Altmannshofer:2019yji,Elahi:2021jia} can also  have an impact on the tests of LFUV studied in this review; these effects are searched for in  pion, kaon, and $\tau$ decay experiments~\cite{Aguilar-Arevalo:2017vlf,Aguilar-Arevalo:2019owf,Bryman:2021ilc}.} We then give an outlook for future experimental and theoretical prospects in Sec.~\ref{sec:future} before we conclude in Sec.~\ref{sec:conclusions}.

\section{Standard Model Theory and Observables}
\label{SMT}
\subsection{Light Meson decays}

The  ratios of the decay rates 
\begin{equation}
R^{P}_{e/\mu}  = \Gamma[ P \to e \bar{\nu}_e (\gamma) ]/  \Gamma[ P \to \mu \bar{\nu}_\mu (\gamma) ]\,,  
\end{equation}
with  $P=\pi,K$, provide some of the most stringent tests of LFU of the SM gauge interactions. In the SM, the decay rates $\Gamma[ P \to e \bar{\nu}_e (\gamma)] $  are  helicity-suppressed due to the $V-A$ structure of the charged current. Moreover, their ratios can be calculated with an extraordinary precision at the $10^{-4}$ level~\cite{Marciano:1993sh,Finkemeier:1995gi,Cirigliano:2007xi,Cirigliano:2007ga} because, to a first approximation, the strong interaction dynamics cancel out in the ratio $R^{P}_{e/\mu}$ and the hadronic structure dependence  appears only through EW corrections. Due to these  features and the precise experimental measurements, the ratios $R^{P}_{e/\mu}$ are very sensitive probes of all SM extensions that induce non-universal corrections to $W$-lepton couplings as well as $\bar e\nu \bar ud$, $\bar e\nu \bar us$ operators, in particular, if they generate a pseudo-scalar current or induced scalar current~\cite{Campbell:2003ir}. 

The most recent theoretical calculations of $R^{P}_{e/\mu}$~\cite{Cirigliano:2007xi,Cirigliano:2007ga} are based on Chiral Perturbation 
Theory (ChPT), the low-energy effective field theory (EFT) of QCD~\cite{Weinberg:1978kz,Gasser:1983yg,Gasser:1984gg}, generalized to include virtual photons and light charged leptons~\cite{Knecht:1999ag}. This framework provides a controlled expansion of the decay rates in terms of a power-counting scheme characterized by the dimensionless ratio $Q \sim m_{\pi,K, \mu}/\Lambda_\chi$, with $\Lambda_\chi$ $ \sim 4 \pi F_\pi$ $\sim 1.2 $\ ${\rm GeV}$ (where $F_\pi \simeq 92.4$~MeV is the pion decay constant), and the electromagnetic  coupling $e$. In this setup one can write
\begin{eqnarray}
R^{P}_{e/\mu} &= & \bar R^{P}_{e/\mu}
\, \Bigg[   1 + 
\Delta_{e^2 Q^0}^{P} +   \Delta_{e^2 Q^2}^{P}  + \Delta_{e^2 Q^4}^{P}   +  ...  + \Delta^{P}_{e^4 Q^0} {+ ...}
\Bigg] \,,
\end{eqnarray}
with
\begin{eqnarray}
\bar R^{P}_{e/\mu} & =& \frac{m_e^2}{m_\mu^2}  \left(  \frac{m_P^2 - m_e^2}{m_P^2 - m_\mu^2} 
\right)^2 \,.
\label{eq:R0}
\end{eqnarray}
Here we have kept all the terms needed to reach an uncertainty of $\sim 10^{-4}$ for the ratio. The leading electromagnetic corrections  $\Delta_{e^2 Q^0}^{P}$  corresponds to the point-like approximation for pions and kaons, and their expressions are well known~\cite{Kinoshita:1959ha}.   
The hadronic structure dependence first appears through the correction  $\Delta_{e^2 Q^2}^{P} \sim (\alpha/\pi)  (m_P/\Lambda_\chi)^2$, which features both 
the calculable  double-chiral logarithms and an {\it{a priori}} unknown low-energy coupling  constant (LEC), which was estimated in large-$N_C$ QCD ($N_C$ being the number of colors)~\cite{Cirigliano:2007xi,Cirigliano:2007ga} and found to contribute negligibly to the error budget.

\subsubsection{Pion Decays}
\label{piondecays}
In the pion case ($P= \pi^\pm$), one usually defines the ratio to be fully photon-inclusive such that it is infrared safe. As a consequence, one has to include in $R^{P}_{e/\mu}$ terms arising from the structure dependent contribution to  $\pi  \to \ell \bar{\nu}_\ell\gamma$~\cite{Bijnens:1992en}, 
that are formally of $O(e^2 Q^4)$, but are not helicity suppressed 
and behave as $\Delta^{P}_{e^2 Q^4} \sim (\alpha/\pi)\,  (m_P/\Lambda_\chi)^4 \,(m_P/m_e)^2$.  Finally, at the level of uncertainty considered, one needs to include higher order corrections in $\alpha$, namely  $\Delta^{P}_{e^4 Q^0} $. The leading logarithmic correction  $\Delta^{P}_{e^4 Q^0,LL}  = (7/2)   (\alpha/\pi \log m_\mu/m_e)^2$ has been calculated in Ref.~\cite{Marciano:1993sh} and the effect of sub-leading contributions was estimated in Ref.~\cite{Cirigliano:2007xi} 
as $ (\alpha/\pi)^2  \log m_\mu/m_e \sim 0.003\%$.
Numerically one finds $\Delta_{e^2 Q^0}^{\pi} = -3.929 \%$,  $\Delta_{e^2 Q^2}^{\pi} = 0.053 (11) \% $,  $\Delta_{e^2 Q^4}^{\pi} = 0.073 (3) \% $, and $\Delta_{e^4 Q^0}^{(\pi)} = 0.055 (3) \% $, 
which leads to the SM expectation~\footnote{Due to a larger uncertainty estimate in  $\Delta_{e^4 Q^0}^{\pi}$ namely   $\Delta_{e^4 Q^0}^{\pi} = 0.055 (10) \% $, Ref.~\cite{Bryman:2011zz} quotes a final result  of  $  R{(\rm SM)}^{\pi}_{e/\mu}  = (1.2352\pm 0.0002)\times10^{-4}$.} 
\begin{equation}
\label{Remu_SM}
R{(\rm SM)}^{\pi}_{e/\mu}    = 1.23524(15)\times10^{-4}\,.
\end{equation}
We reiterate here that (i) this prediction includes  structure-dependent hard-bremsstrahlung corrections to $\Gamma(\pi^+ \rightarrow e^+ \nu (\gamma))$ which are not helicity suppressed; and (ii)  the dominant uncertainty of the SM prediction  arises from a  low-energy constant in chiral perturbation  theory, followed by the non-leading logarithmic corrections of $O(\alpha^2)$. 

The most accurate measurement of $R^{\pi}_{e/ \mu}$ was reported by the TRIUMF PIENU collaboration~\cite{Aguilar-Arevalo:2015cdf},
\begin{equation}
\label{Remu_exp}
     R({\rm Exp})^{\pi}_{e/\mu} = (1.2344 \pm 0.0023 (\text{stat}) \pm 0.0019 (\text{syst})) \times 10^{-4},
\end{equation}
 at the 0.24\%  precision level. The PDG~\cite{Zyla:2020zbs} average including  previous experiments done at TRIUMF~\cite{Aguilar-Arevalo:2015cdf,Bryman:1985bv,Britton:1993cj} and PSI~\cite{Czapek:1993kc} is
\begin{equation}
\label{Remu_exp_pdg}
     R({\rm\overline{Exp}})^{\pi}_{e/\mu}= (1.2327 \pm 0.0023) \times 10^{-4}.
\end{equation}

The comparison between theory and experiment given in Eq.~(\ref{Remu_SM}) and Eq.~(\ref{Remu_exp_pdg}) provides a stringent test of the $e$--$\mu$ universality of the weak interaction. We chose to express the results in terms of the effective couplings $A_\ell$ multiplying the low-energy charged current  contact interaction 
\begin{equation}
   L_{\rm CC}= A_\ell \bar u \gamma^\mu P_L d \bar \nu_\ell \gamma_\mu P_L \ell\,,  
\end{equation}
where $P_L \equiv (1-\gamma_5)/2$. In the SM at tree level the couplings are given by  $A_\ell = - 2 \sqrt{2} G_F V_{ud}$ and thus satisfy LFU, i.e. $A_\ell / A_{\ell^\prime} = 1$.  
The measurement of $R^\pi_{e/\mu}$ results in 
 \begin{equation}
     \left( \frac{A_\mu}{A_e}  \right)_{ R^{\pi}_{e/ \mu} }= 1.0010 \pm 0.0009  \,,\label{gpi}
 \end{equation}
which is in excellent agreement with the SM expectation. 
A deviation from $A_\ell / A_{\ell^\prime} = 1$ can originate from various mechanisms. In the literature it is common to interpret deviations from $A_\ell/A_{\ell^\prime} = 1$ in terms of flavor-dependent couplings $g_\ell$ of the $W$-boson to 
the leptonic current, in which case $A_\ell \propto g_\ell$. We discuss this scenario in detail in Sec.~\ref{sec:vertex}. We note that in the context of modified $W$-boson couplings LFU tested with $R^{\pi}_{e/\mu} $  probes the couplings of a longitudinally polarized $W$-boson whereas tests using purely leptonic reactions like $\tau\to \ell \nu_\tau \nu_{\ell}$ ($\ell=e,\mu$) test the couplings of transversely polarized $W$-boson and are thus complementary.

\subsubsection{Kaon Decays}
\label{kaondecays}
LFU can also be tested using the ratios 
\begin{eqnarray}  R^{K}_{e/\mu}  &= & \frac{\Gamma\left[ K^+\rightarrow e^+ \nu (\gamma) \right]}{\Gamma\left[K^+ \rightarrow \mu^+ \nu (\gamma)\right]} \,, \textrm{and} \\ 
R^{K\to\pi}_{e/\mu} &=& \frac{\Gamma\left[ K\rightarrow \pi e \nu (\gamma) \right]}{\Gamma\left[K \rightarrow \pi \mu \nu (\gamma)\right]}\,,
\end{eqnarray}
where for $R^{K\to\pi}_{e/\mu} $ both neutral and charged kaons decays, e.g. $K_L\to \pi^\pm\ell^\mp \nu$ and $K^\pm \to \pi^0 \ell^\pm \nu$, are used. 

The calculation of $R^{K}_{e/\mu}$ is  similar to the one of  $R^{\pi}_{e/\mu}$ described in the previous section. 
An important difference concerns the  definition of the infrared-safe decay rate, which requires including part of the radiative decay mode. The radiative  amplitude is the sum of the inner bremsstrahlung ($T_{IB}$) component of $O(e Q)$ and a structure dependent ($T_{\rm SD}$) component of $O(e Q^3)$~\cite{Bijnens:1992en}. While the experimental definition of $R_{e/\mu}^{(\pi)}$ is fully inclusive, the one for  $R^{K}_{e/\mu}$ includes  the effect of $T_{\rm IB}$ in $\Delta_{e^2 Q^0}^{(K)}$ (dominated by soft photons) and excludes the effect of $T_{\rm SD}$. With this definition one finds $\Delta_{e^2 Q^0}^{K} = -3.786 \%$,  $\Delta_{e^2 Q^2}^{K} = 0.135 (11) \% $,  $\Delta_{e^2 Q^4}^{K} =0 $, and $\Delta_{e^4 Q^0}^{K} = 0.055 (3) \% $,  and the SM expectation is~\cite{Cirigliano:2007xi,Cirigliano:2007ga} 
\begin{equation}
\label{RKemu_SM}
   R\text{(SM)}^{K}_{e/\mu}  = (2.477\pm 0.001)\times10^{-5}\,,
\end{equation} 
where  the  final uncertainty accounts for higher order chiral corrections  of expected  size $\Delta_{e^2 Q^2} \times m_K^2/(4 \pi F_\pi)^2$. 

The PDG~\cite{Zyla:2020zbs} average of previous measurements done by the NA62~\cite{Lazzeroni:2012cx} and KLOE~\cite{Ambrosino:2009aa} experiments is
\begin{equation}
\label{RKemu_exp_pdg}
R(\overline{\rm Exp})^{K}_{e/\mu}= (2.488 \pm 0.009) \times 10^{-5}.
\end{equation}
 The comparison of theory and experiment given in Eqns.~(\ref{RKemu_SM}) and (\ref{RKemu_exp_pdg}) corresponds to a test of $e$--$\mu$ universality
 \begin{equation}
  \left( \frac{A_\mu}{A_e} \right)_{R^{K}_{e/\mu} }= 0.9978 \pm 0.0018.  \label{gK}
 \end{equation}

The analogous LFU test based on  the ratios 
 $R^{K\to\pi}_{e/\mu}$ 
 has been discussed by  the Flavianet collaboration~\cite{Antonelli:2009ws}. 
For a given neutral or charged initial state kaon, 
the Fermi constant, $V_{us}$, short-distance radiative corrections,  and the hadronic form factor at zero momentum transfer cancel out 
when taking the ratio  $R^{K\to\pi}_{e/\mu} $. 
Therefore, in the SM this ratio is entirely determined by phase space factors and long-distance radiative corrections~\cite{Cirigliano:2001mk,Cirigliano:2004pv,Cirigliano:2008wn,Seng:2021boy}. The ratios for $K_L$ and $K^\pm$ were found to be consistent 
leading to the following values for $A_\mu/A_e$ \cite{Antonelli:2009ws,Moulson:2017ive,Seng:2021nar}:
 \begin{align}
 \begin{aligned}
    \left(\dfrac{A_\mu}{A_e}\right)_{R^{K_L\to\pi}_{e/\mu}}&=1.0022\pm 0.0024  \,,\textrm{and}\\
    \left(\dfrac{A_\mu}{A_e}\right)_{R^{K^\pm\to\pi^\pm}_{e/\mu}}&=0.9995\pm 0.0026 \,,
    \end{aligned}
 \end{align}
 with the average for $K_{\ell 3}$ decays
 \begin{equation}
    \left(\frac{A_\mu}{A_e}\right)_{R^{K\to \pi}_{e/\mu}}=1.0009 \pm 0.0018\,.
 \end{equation}
The  numbers given above correspond to the recent  analysis of Ref.~\cite{Seng:2021nar}, which uses  experimental input from Ref.~\cite{Antonelli:2009ws} (updated in  Ref.~\cite{Moulson:2017ive} with reduced errors in the charged modes)  and theoretical input on $K_{\ell 3}$ radiative corrections from Refs.~\cite{Cirigliano:2008wn,Seng:2021boy}, which incorporates  a new analysis of $K_{e3}$ modes with reduced uncertainties~\cite{Seng:2021boy}.

Note that $\mu$-$e$ universality can also be determined from $B$ decays such as ${\rm Br}[B\to D^{*}\mu\nu]/{\rm Br}[B\to D^{*}e\nu]$. Even though the relative precision at the \% level~\cite{Glattauer:2015teq,Abdesselam:2017kjf,Waheed:2018djm} is not competitive with the ones obtained from kaon and pion decays, these measures of LFUV are interesting in light of the anomalies in $R(D^{(*)})$ and $\Delta A_{FB}$~\cite{Abdesselam:2017kjf,Bobeth:2021lya,Carvunis:2021dss} as they test different 4-fermion operators.

\subsection{Beta Decays and CKM Unitarity}
\label{betadecays}

The observables testing LFUV discussed so far involve {\it ratios} of purely leptonic or semileptonic meson decays with an electron or muon in the final state. While considering ratios of (semi) leptonic decay rates offers theoretical advantages (e.g. the elements $V_{ud}$ and $V_{us}$ of the Cabibbo-Kobayashi-Maskawa (CKM)~\cite{Cabibbo:1963yz,Kobayashi:1973fv}
 matrix, part of the radiative corrections, and hadronic matrix elements cancel), the high-precision study of absolute semileptonic decay rates can also uncover LFUV effects. 
For example, in the context of corrections to the $W \to \ell \nu_\ell$ vertex, the semileptonic transition  $d(s) \to u e \bar \nu_e$  ($d(s) \to u \mu \bar \nu_\mu$) is sensitive to corrections to the muon (electron) coupling~\cite{Buchmuller:1985jz,Cirigliano:2009wk,Crivellin:2020lzu}{(see Ref.\cite{Bauman:2012fx} for a discussion within supersymmetric models)}. 
In absolute decay rates, these BSM LFUV corrections contaminate the extraction of the CKM elements $V_{ud}$ and $V_{us}$ from measured decay rates. This means that beta decays and the study of CKM unitarity are intertwined with the study of LFUV~\cite{Crivellin:2020lzu}.  
In light of this connection,  we briefly summarize the status of first-row CKM unitarity tests. We will discuss the implications for LFUV BSM interactions in  Sec.~\ref{NPsection}. 

Unitarity of the CKM  matrix~\cite{Cabibbo:1963yz,Kobayashi:1973fv}  implies  $\Delta_{\rm CKM} \equiv |V_{ud}|^2+|V_{us}|^2+|V_{ub}|^2 - 1 = 0$,   where $V_{ud}$, $V_{us}$, $V_{ub}$  represent  the 
 mixing  of   {\it up}  with    {\it down}, {\it strange}, and {\it beauty} quarks, respectively. In practice $|V_{ub} |^2 < 10^{-5}$ can be neglected and CKM unitarity  reduces to the original Cabibbo universality, with the identifications $V_{ud} = \cos \theta_C$ and $V_{us}=  \sin \theta_C$, 
where  $\theta_C$ is the Cabibbo angle~\cite{Cabibbo:1963yz}.
The determination of $V_{uD}$ ($D=d,s$) from various  hadronic weak decays $h_i \to h_f \ell \nu_\ell$ ($\ell = e, \mu$) 
relies on the following schematic formula for the decay rate $\Gamma$,  
\begin{equation}
\Gamma = G_F^2 \times |V_{uD}|^2 \times | M_{\rm had}|^2 \times  (1 + \delta_{\rm ISB} +  \delta_{\rm RC}) \times F_{\rm kin}\,,
\label{eq:master}
\end{equation}
where  $G_F$ is the Fermi constant extracted from muon decay and 
$F_{\rm kin}$ is a phase space factor. Theoretical input comes in the form of  (i) the hadronic matrix elements of the weak current, $M_{\rm had}$,  usually calculated in the isospin limit of QCD (in which {\it up}  and {\it down} quark masses are equal and electromagnetic interactions are turned off);   and (ii)  small \%-level  corrections, $\delta_{\rm IB,RC}$, due to strong isospin-breaking (ISB)   and electromagnetic radiative corrections  (RCs) induced by the exchange of virtual photons and the emission of real photons, characterized by the small expansion parameters $\epsilon_{\rm ISB} \sim (m_u - m_d)/\Lambda_{\rm QCD}$
and $\epsilon_{\rm EM} \sim \alpha / \pi$, respectively, where $\alpha \sim 1/137$ is the electromagnetic fine structure constant. 

\begin{figure*}[t!]
	\begin{center}
		\includegraphics[width=0.9\linewidth]{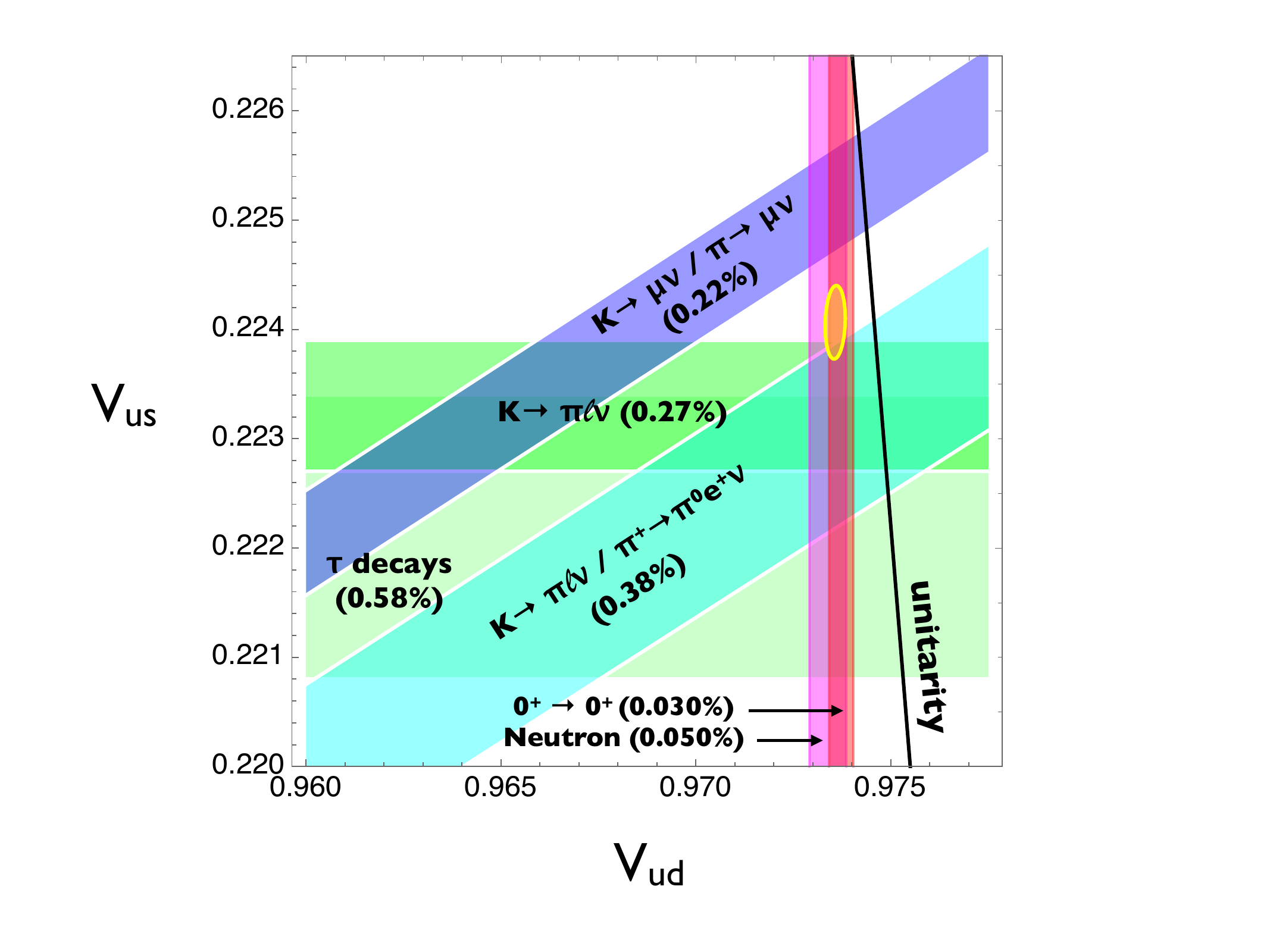}
		\caption{Summary of constraints on $V_{ud}$ and $V_{us}$ (assuming the Standard Model hypothesis) from nuclear, nucleon, meson, and $\tau$  lepton decays.  For each constraint, the one-sigma uncertainty on $V_{us}$ or $V_{ud}$ is given in parenthesis (see text for details).  The one-sigma ellipse from a global fit (with $\chi^2/{\rm d.o.f.} = 2.8$),  depicted in yellow, 
corresponds to  $V_{ud}= 0.97357(27)$ and  $V_{us} = 0.22406(34)$,  implying $\Delta_{\rm CKM} = |V_{ud}|^2 + |V_{us}|^2 - 1 =   (-19.5 \pm 5.3) \times 10^{-4}$. 
		}
		\label{fig:CKM}
	\end{center}
\end{figure*}

Currently, as shown in Fig.~\ref{fig:CKM}, the knowlegde of $V_{ud}$ is dominated by nuclear $0^+ \to 0^+$ decays where the most recent survey~\cite{Hardy:2020qwl} of experimental and theoretical  input leads to  $V_{ud} = 0.97373(31)$.  This incorporates a reduction in the uncertainty on the so-called inner radiative corrections~\cite{Seng:2018yzq,Czarnecki:2019mwq} 
and an increase in uncertainty due to nuclear-structure dependent effects with input from Refs.~\cite{Towner:1992xm,Seng:2018qru,Gorchtein:2018fxl}.
\footnote{Note that the $V_{ud}$ value quoted in the PDG~\cite{Zyla:2020zbs}  does not yet reflect the increased error in the nuclear-structure dependent radiative corrections and therefore has an uncertainty $\delta V_{ud} = 0.00014$.} 

{Thanks to higher precision measurements of the  lifetime~\cite{UCNt:2021pcg} 
and beta asymmetry~\cite{Markisch:2018ndu} (see Ref.~\cite{Dubbers:2021wqv} for a recent review)}, neutron decay is  becoming competitive with super-allowed beta decays concerning the precision with which $V_{ud}$ can be extracted. 
Using the PDG average for the  neutron lifetime (including a scale factor $S=1.6$ to account for tensions among experimental measurements)\footnote{The Particle Data Group excludes the beam lifetime measurements from the current `PDG average’, quoting $\tau_n = 879.4(6) s$. This would change to
$\tau_n= 879.6(8) s$ if one included the beam lifetime measurement result.} and the  post-2002~PDG average\footnote{We follow here the analysis described in the PDG review 
``$V_{ud}$, $V_{us}$, the Cabibbo Angle, and CKM unitarity"{\cite{Zyla:2020zbs}}. A recent comprehensive analysis of beta decays can be found in Ref.~\cite{Falkowski:2020pma}. Adopting input from Ref.~\cite{Falkowski:2020pma} would lead to negligible changes in our global fit presented below.} 
determinations of the axial coupling $g_A = 1.2762(5)$~\cite{Zyla:2020zbs} 
leads to $V_{ud} =  0.97338(33)_{\tau} (32)_{g_A} (10)_{RC} = 0.97338(47)$, with errors originating from the lifetime $\tau_n$, $g_A$, and radiative corrections (RC)~\cite{Zyla:2020zbs}, respectively.
Ongoing and planned neutron experiments aim to reduce the uncertainty in $\tau_n$ and $g_A$ by a factor of few (see \cite{Cirigliano:2019wao} and references therein), which will put the extraction of $V_{ud}$ from neutron decay at the same precision level as superallowed nuclear beta decays. 
Future prospects of improving the extraction of $V_{ud}$ from pion beta decay are discussed in Sec.\ref{sec:future}. 

The most precise value of $V_{us}$ is extracted from  $\Gamma (K \to \pi \ell \nu)$ while $R_A \equiv \Gamma (K \to \mu \nu) / \Gamma (\pi \to \mu \nu)$ provides currently the most  precise  determination of $V_{us}/V_{ud}$~\cite{Marciano:2004uf}. A comprehensive discussion of the experimental and theoretical input up to 2010 can be found in Ref.~\cite{Antonelli:2010yf} and references therein.
Since then, experimental input on the $K^\pm$ branching ratios and form-factor parameters has been updated, as reviewed  in Ref.~\cite{Moulson:2017ive}, while  the most recent theoretical input on the hadronic matrix elements can  be found in Ref.~\cite{Aoki:2019cca} and references therein. Radiative corrections are included according to Refs,~\cite{Cirigliano:2008wn,Seng:2021boy}.  With this input one obtains~\cite{Antonelli:2010yf,Seng:2021nar} $V_{us} = 0.2231(6)$ from $K_{\ell 3}$ decays and  $V_{us}/V_{ud} = 0.2313(5)$ from $R_A$. 
Recently it has been pointed out~\cite{Czarnecki:2019iwz} that $V_{us}/V_{ud}$ can also be obtained through the ratio of vector channel decays $R_V \equiv \Gamma (K \to \pi \ell \nu (\gamma)) / \Gamma (\pi^+ \to \pi^0 e^+ \nu (\gamma))$, leading to $V_{us}/V_{ud} =  0.22908(87)$, with uncertainty dominated by the pion beta decay width, but nonetheless within a factor of two compared to the $R_A$ determination. Future prospects on this front are discussed in Sect.~\ref{sec:future}. Finally, $V_{us}$ can also be extracted from inclusive and exclusive semileptonic decays of the $\tau$ lepton, with final state hadrons carrying strangeness quantum numbers. This leads to a somewhat lower and less precise value $V_{us} = 0.2221(13)$ (see Ref.\cite{Zyla:2020zbs} and references therein).  

Fig.~\ref{fig:CKM} summarizes graphically  the results on $V_{ud}$ and $V_{us}$ discussed so far and reveals that while nuclear and neutron decay lead to a  consistent picture for $V_{ud}$, tensions exist among current determinations of $V_{us}$ ($K_{\ell 3}$ vs.  $K_{\ell 2}$ and $K$ vs. $\tau$ lepton). Moreover, an overall tension with CKM unitarity is also apparent. A global fit leads to $V_{ud}= 0.97357(27)$ and  $V_{us} = 0.22406(34)$, with the  one-sigma ellipse represented in Fig.~\ref{fig:CKM} in yellow, 
implying 
\begin{equation}
\Delta_{\rm CKM} =  (-19.5 \pm 5.3) \times 10^{-4}\,,
\end{equation}
a 3.7-$\sigma$ effect. 
Due to the tension in the input data, the $\chi^2$ per degree of freedom is  2.8, corresponding to a scale factor of $S=1.67$ 
under the assumption that there is no new physics effect.
In Sec.~\ref{NPsection} we will discuss the implications of the current tensions for LFUV interactions. 

\subsection{Tau Decays}
\label{taudecays}
Tests of LFU can also be obtained comparing different $\tau$ decay rates to those of the muon or the pion (kaon). For the $\tau$ we have semileptonic as well as purely leptonic decays at our disposal. While the former can only test $\tau-\mu$ universality efficiently, the latter allow us to assess  $\mu-e$, $\tau-e$ and $\tau-\mu$ universality. If one defines 
\begin{eqnarray}
R^{\tau}_{\mu/e} &=&\frac{\rm Br[\tau^- \to \mu^-\bar{\nu_\mu}\nu_\tau]}{\rm Br[\tau^- \to e^-\bar{\nu_e}\nu_\tau]}\,,\\
R^{\tau \pi (K)}_{\tau/\mu}&=&\frac{\rm Br[\tau \to \pi (K)\nu_\tau]}{\rm Br[\pi (K) \to \mu\nu_\mu]}\,, \\  R^{\tau}_{\tau/\mu}&=& \frac{\rm Br[\tau^- \to e^-\bar{\nu_e}\nu_\tau]}{\rm Br[\mu^- \to e^-\bar{\nu_e}\nu_\mu]}\,,\textrm{and}\\
R^{\tau}_{\tau/e}&=&\frac{\rm Br[\tau^- \to \mu^-\bar{\nu_\mu}\nu_\tau]}{\rm Br[\mu^- \to e^-\bar{\nu_e}\nu_\mu]}\,,
\end{eqnarray}
the LFU ratios can then be expressed in terms of experimentally measured rates and theoretical input. For $\mu-e$ universality one has 
\begin{eqnarray}\label{gmue}
\left( \frac{A_\mu}{A_e}\right)_\tau =\sqrt{ R^{\tau}_{\mu/e} \frac{f(m_e^2/m_\tau^2)}{f(m_\mu^2/m_\tau^2)}}\,,
\end{eqnarray}
with $f(x)=-8x+8x^3-x^4-12x^2 \log x$.
The above expression receives radiative corrections of $O(\alpha/\pi) \times (m_\mu/m_\tau)^2$~\cite{Pich:2013lsa}, which are hence suppressed. 
For $\tau-\mu$ universality one has~\footnote{ In the case of purely leptonic decays, we write the  LFU test in terms of the ratios 
$A_{\ell} / A_{\ell^\prime} \equiv
A_{\ell \ell^{''}} /A_{\ell^{'} \ell^{''}}$ with
$L=A_{\ell^\prime \ell}  \bar \ell ^\prime \gamma^\mu P_L \nu_{\ell^\prime} \bar \nu_\ell \gamma_\mu P_L \ell$.}
\begin{eqnarray}\label{gtaumu}
\left( \frac{A_\tau}{A_\mu}\right)_h = \frac{1-m_\mu^2/m_h^2}{1-m_h^2/m_\tau^2}\sqrt{R^{\tau h} _{\tau/\mu}  \frac{2m_h m_\mu^2\tau_h}{(1+\delta_h)m_\tau^3 \tau_\tau}},
\end{eqnarray}
where $h=\pi, K$.
 An alternative method to test $\tau-\mu$ universality, similar to the $\mu-e$ case, compares the electronic and muonic decay rates and can be expressed as
\begin{eqnarray}\label{gtaumu2}
\left( \frac{A_\tau}{A_\mu}\right)_\tau =\sqrt{R^{\tau}_{\tau/\mu}\frac{\tau_\mu}{\tau_\tau}   \frac{m_\mu^2}{m_\tau^3}(1+\delta_W)(1+\delta_\gamma)}\,.
\end{eqnarray}
In the above equations $m_{e,\mu,\tau}$ are the masses of $e$, $\mu$, and $\tau$, $\tau_{\tau,h}$ are the lifetimes of the particles $\tau$ and $h$, and $\delta_{h,W,\gamma}$ are the weak and electromagnetic radiative corrections (see Ref.~\cite{Pich:2013lsa} and references therein for details). Experimentally, these tests have been carried out at $B$-factories where, at the nominal center-of-mass energy of 10.58 GeV/c$^2$, thanks to a cross section of 0.919 nb, these machines are "$\tau$-Factories" \textit{de facto} that produce large numbers of $\tau$ pairs. 

Both the BaBar and the CLEO Collaborations performed the LFU tests according to Eq.~(\ref{gmue})~\cite{BaBar:2009lyd} and Eq.~(\ref{gtaumu})~\cite{CLEO:1996oro}, while only CLEO performed the measurement according to Eq.~(\ref{gtaumu2}). In the reaction $e^+e^- \to \tau^+\tau^-$ at a $B$-Factory one can use the decay of the $\tau^+$ to tag and study the $\tau^-$(and \textit{vice versa}). Typically one uses either the so-called 3$\times$1 $\tau$ topology, with  the decay $\tau^+ \to \pi^+\pi^+\pi^- \bar{\nu_\tau}$ as a tag and then a study on the other $\tau^-$ is performed, or the so-called 1$\times$1 topology in which both taus decay with one prong (lepton or hadron) and a neutrino. While the latest BaBar measurement only focused on the  3$\times$1 topology, the latest study from the CLEO collaborations also used the 1$\times$1. 

Tests of LFU are precise measurements for which, in addition to sizable amounts of data, one needs to control systematic effects when determining the branching ratios. In the most recent results from BaBar, for example, $R^\tau_{\mu/e}$ and $\left(\frac{A_\mu}{A_e}\right)_\tau$ have been determined with a precision of $ 0.4\% (0.16\%_{\rm stat}\bigoplus  0.36\%_{\rm sys})$ and $ 0.2\%$~\cite{BaBar:2009lyd}, respectively, where the leading systematic uncertainty (0.32$\%$) originated from particle identification; similarly, $R^\pi_{\tau/\mu}$ and $\left(\frac{A_\tau}{A_\mu}\right)_\pi$  were determined with a precision of $ 0.63\% (0.14\%_{\rm stat}\bigoplus0.61\%_{\rm sys})$ and $0.57\%$ where the dominant systematic source originated again from particle identification.

The results from BaBar and CLEO, are also used to obtain the latest HFLAV combination which includes 176 measurements and 89 constraints in $\tau$ processes~\cite{HFLAV:2019otj}; for purely leptonic $\tau$ decays, these are
\begin{eqnarray}\label{eq:g_ratios_HFLAV}
\left(\frac{A_\tau}{A_\mu}\right)_\tau&=&1.0010\pm0.0014 \,,\\ \left(\frac{A_\tau}{A_e}\right)_\tau&=&1.0029\pm0.0014\,,\;{\rm and}\\ \left(\frac{A_\mu}{A_e}\right)_\tau&=&1.0018\pm0.0014\,.
\end{eqnarray}
During the preparation of this manuscript,  new values of ($A_\tau/A_\mu$)$_h$ ($h=\pi,K$),  where the radiative corrections were computed including the lightest multiplets of spin-one heavy states in ChPT, were given in Ref.~\cite{Arroyo-Urena:2021nil}; these new values are
\begin{eqnarray}\label{eq:g_ratios_Ref66}
\left(\frac{A_\tau}{A_\mu}\right)_\pi&=&0.9964\pm0.0038\,,\;{\rm and}\\ \left(\frac{A_\tau}{A_\mu}\right)_K&=&0.9857\pm0.0078\,,
\end{eqnarray}
with the following correlation coefficients~\cite{HFLAV:2019otj}
\begin{equation}
   \begin{matrix} 
    (A_\tau/A_\mu)_\tau & 1  \\
   (A_\tau/A_e)_\tau & 0.51 & 1 \\
   (A_\mu/A_e)_\tau & -0.50 & 0.49 & 1 \\
   (A_\tau/A_\mu)_\pi & 0.23& 0.25 & 0.02 & 1\\
   (A_\tau/A_\mu)_K & 0.11 & 0.10 & -0.01 & 0.06 & 1  \\
    & (A_\tau/A_\mu)_\tau & (A_\tau/A_e)_\tau & (A_\mu/A_e)_\tau & (A_\tau/A_\mu)_\pi  &  (A_\tau/A_\mu)_K 
   \end{matrix} 
\end{equation}
and with 100$\%$ correlation between $(A_\tau/A_\mu)_\tau$, $(A_\tau/A_e)_\tau$, and $(A_\mu/A_e)_\tau$~\cite{HFLAV:2019otj}.

\section{Beyond the SM (BSM) Analysis}
\label{NPsection}

Let us now interpret the experimental bounds for LFUV in the charged current in terms of constraints on NP. For this we will first study effective operators (i.e. modified $W\ell\nu$ couplings and four-fermion operators) and then consider simplified models which give rise to the corresponding Wilson coefficients. In this context we will highlight a possible correlation between the CAA and non-resonant di-lepton searches at the LHC and finally study NP models with focus on the ones motivated by the CAA.

\subsection{Effective Field Theory}

\subsubsection{Modified $W\ell\nu$ couplings}
\label{sec:vertex}

All observables discussed in this review are sensitive to modified $W$ couplings to leptons. In order to investigate their effects, let us therefore use the parameterization\footnote{
In the conventions of Ref.~\cite{Pich:2013lsa} one has $1+\epsilon_{ii}-\epsilon_{jj}=g_i/g_j$, or equivalently $g_i = g_j (1 + \epsilon_{ii}-\epsilon_{jj})$, with $i,j=e,\mu,\tau$. 
}
\begin{equation}
\Lagr \supset
-i\frac{{g_2}}{{\sqrt 2 }}{{\bar \ell }_i}{\gamma ^\mu }{P_L}{\nu _j}{W_\mu^- }\left( {{\delta _{ij}} + {\epsilon_{ij}}} \right) +h.c.\,,
\label{couplings}
\end{equation}
where $i,j = e$,$\mu$,$\tau$ and with the SM $SU(2)_L$ gauge coupling $g_2$ recovered in the limit $\epsilon_{ij}\to 0$. Here we neglected possible effects of the PMNS matrix which drop out in the limit of vanishing neutrino masses. Furthermore, in the following we will disregard flavor-violating couplings ($\eps _{ij}$ with $i\neq j$) since they are tightly bounded by radiative lepton decays $\ell\to \ell^\prime\gamma$ and lead to effects in LFUV observables that do not interfere with the SM and are thus suppressed. Note that in Eq.~\eqref{couplings} we simply parameterize the BSM effect by $\eps_{ij}$ but do not consider the $SU(2)_L$ gauge invariance in SM EFT, which we will discuss later. 

For the phenomenological analysis, note that all LFUV observables (encoded in direct ratios) depend, at leading order, on differences  $\epsilon_{aa}-\epsilon_{bb}$  ($a\neq b$)  while the deficit in first row CKM unitarity, related to the determination of $V_{ud}$, is to a good approximation only sensitive to $\epsilon_{\mu\mu}$~\cite{Crivellin:2020lzu}. In order to see the latter, the first crucial observation is that in order to extract $V_{ud}$ from beta decays, one needs the Fermi constant determined from the muon lifetime~\cite{Tishchenko:2012ie}
\beq
\frac{1}{\tau_{\mu}}=\frac{(G_F^{\Lagr})^2m_{\mu}^5}{192\pi^3}(1+\Delta q)(1+\varepsilon_{ee}+\varepsilon_{\mu\mu})^2\,.
\eeq
Here $G_F^{\Lagr}$ is the Fermi constant appearing in the Lagrangian (excluding BSM contamination) and $\Delta q$ subsumes the phase space, QED, and electroweak radiative corrections. Therefore, the Fermi constant measured in muon decay and extracted under the SM assumption ($G_F$), is related to the one at the Lagrangian level as
\beq
G_F=G_F^{\Lagr}(1+\eps_{ee}+\eps_{\mu\mu})\,.
\label{GFmod}
\eeq
Thus 
\beq
V_{ud}^\beta=V_{ud}^\Lagr\big(1-\eps_{\mu\mu}\big)\,,
\eeq
where $V_{ij}^\Lagr$ denotes CKM matrix elements without any BSM contamination which  by definition fulfills CKM unitarity. Taking into account that the first row and column CKM unitarity relations are very much dominated by $V_{ud}$, being by far the biggest element of the CKM matrix, we find to a good approximation
\begin{equation}
\eps_{\mu\mu}\approx0.00098\pm0.00027\,,
\end{equation}
reflecting the corresponding $3.7\,\sigma$ tension.

\begin{figure}[t!]
	\centering
	\includegraphics[width=1\textwidth]{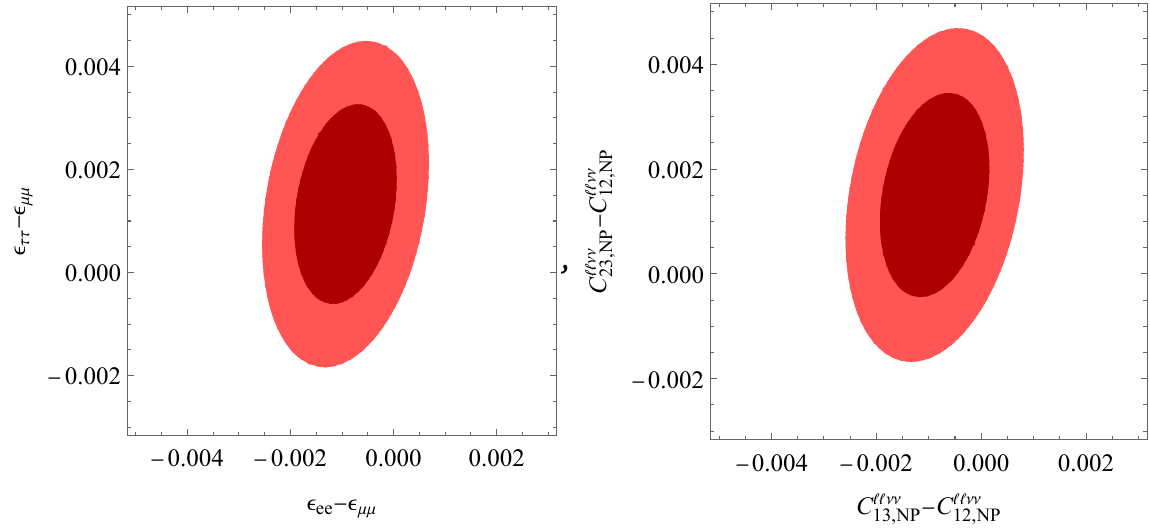}
	\caption{Left: Global fit in the   $\epsilon_{\tau\tau}-\epsilon_{\mu\mu}$  vs. $\epsilon_{ee}-\epsilon_{\mu\mu}$  plane, including kaon, pion, and tau decays, quantifying LFU in the charged current. Right: Global fit in the  $C^{\ell\ell\nu\nu}_{23}-C^{\ell\ell\nu\nu}_{12}$  vs. $C^{\ell\ell\nu\nu}_{13}-C^{\ell\ell\nu\nu}_{12}$ plane from leptonic tau and muon decays.  Uncertainties are shown for one (dark color) and two (lighter color) standard deviations.}  
	\label{eps}	
\end{figure}

We can now reparametrize the NP effects by writing
\begin{equation}
\epsilon_{ee}-\epsilon_{\mu\mu}\,,\qquad \epsilon_{\tau\tau}-\epsilon_{\mu\mu}\;\;{\rm and}\;\;\epsilon_{\mu\mu}\,,
\end{equation}
such that differences are direct measures of LFU and are constrained by the corresponding ratios. This allows us to perform a global fit in the $\epsilon_{\tau\tau}-\epsilon_{\mu\mu}$ ~$vs.$~ $\epsilon_{ee}-\epsilon_{\mu\mu} $ plane which is uncorrelated  with  $\epsilon_{\mu\mu}$  taking into account all LFU ratios discussed previously (including correlations among them). The result in shown in Fig.~\ref{eps} (left). From this depiction, we see that while the hypothesis of LFU in the charged current is compatible with data at the $2\sigma$ level, one observes a slight preference for negative values of $\epsilon_{ee-\mu\mu}$. 

\subsubsection{4-lepton Operators} 
\label{4-fermi}

It is clear that 4-lepton operators only enter purely leptonic decays. Furthermore, as (in the limit of vanishing masses of the final state leptons) only left-handed vector operators with the same flavor structure as the SM lead to interference with the SM in these decays, we can focus on them and write
\begin{equation}
{\cal L}_{4\ell}=-\frac{g_2^2}{2m_W^2}C_{fi}^{\ell\ell\nu\nu}\bar \ell_f\gamma_\mu P_L\ell_i \bar \nu_i\gamma^\mu P_L\nu_f\,,
\end{equation}
 with $C_{fi}^{\ell\ell\nu\nu}=1+C_{fi,{\rm NP}}^{\ell\ell\nu\nu}$. The effects of $C_{fi,{\rm NP}}^{\ell\ell\nu\nu}$ are similar to the ones of modified $W-\ell-\nu$ couplings and we can consider the three parameters $C_{12,{\rm NP}}^{\ell\ell\nu\nu}$, $C_{13,{\rm NP}}^{\ell\ell\nu\nu}-C_{12,{\rm NP}}^{\ell\ell\nu\nu}$ and $C_{23,{\rm NP}}^{\ell\ell\nu\nu}-C_{12,{\rm NP}}^{\ell\ell\nu\nu}$. However, in this case $C_{12}$ is not only determined from the CAA but also has an impact on the global EW fit as it modifies the determination of the Fermi constant from muon decay~\cite{Marciano:1999ih,Crivellin:2021njn}. In fact, they turn out to prefer opposite signs
\begin{equation}
\begin{aligned}
\left. {C_{12,{\rm NP}}^{\ell\ell\nu\nu}} \right|_{\rm CAA} &\approx0.00098\pm0.00027\,,\textrm{and}\\
\left. {C_{12,{\rm NP}}^{\ell\ell\nu\nu}} \right|_{\rm EW} &\approx-0.00067\pm0.00033\,.
\end{aligned}
\end{equation}
$C_{13,{\rm NP}}^{\ell\ell\nu\nu}-C_{12,{\rm NP}}^{\ell\ell\nu\nu}$ as well as $C_{23,{\rm NP}}^{\ell\ell\nu\nu}-C_{12,{\rm NP}}^{\ell\ell\nu\nu}$ are determined from 
the ratios of rates
$\tau\to\mu\nu\nu/\tau\to e\nu\nu$, $\tau\to\mu\nu\nu/\mu\to e\nu\nu$ and $\tau\to e\nu\nu/\mu\to e\nu\nu$, while all ratios involving mesons remain unaffected. Therefore, we find the global fit shown in Fig.~\ref{eps} (right).

\subsubsection{2-quark-2-lepton Operators}

Concerning 2-quark-2-lepton operators, both left-handed vector operators and scalar ones are relevant as they interfere with the SM contribution. In fact, the latter have enhanced effects in $R^P_{e/\mu}$:
\begin{equation}
\left(\dfrac{A_\mu}{A_e}\right)_{R^{P}_{e/\mu}}=
\dfrac{C^{V_L\mu}_{fi}-C^{V_R\mu}_{fi}+\dfrac{m_P^2}{\left(m_{u_f}+m_{d_i}\right) m_\mu} \left(C_{fi}^{S_R\mu}-C_{fi}^{L\mu}\right)}{C^{V_L e}_{fi}-C^{V_R e}_{fi}+\dfrac{m_P^2}{\left(m_{u_f}+m_{d_i}\right) m_e} \left(C_{fi}^{S_Re}-C_{fi}^{S_L e}\right)}\,,\label{4-fermi-R}
\end{equation}
with $u_f=u$ and $d_i=d(s)$ for $P=\pi(K)$, defined via the Lagragian
\begin{equation}
L_{ud\ell\nu}=-\dfrac{g_2^2}{2m_W^2}V_{fi}\sum_{A=L,R}\left(C^{V_A\ell}_{fi}\bar u_f\gamma^\mu P_A d_i\bar\ell\gamma_\mu P_L \nu_\ell +C^{S_A\ell}_{fi}\bar u_f P_{A} d_i \bar \ell P_L \nu_\ell\right)\,,
\end{equation}
where $C^{V_L\ell}_{fi}=1+C^{V_L\ell}_{fi,{\rm NP}}$. All other Wilson coefficients are zero within the SM. 

From the CAA we find
\begin{equation}
 C^{V_L e}_{11,{\rm NP}}=-0.00098\pm0.00027\,,   
\end{equation}
and from LFUV ratios the bounds can be directly read off using Eqs.~(\ref{gpi}) and (\ref{gK}). For the three body decays involved in $R^{K\to \pi}_{e/\mu}$, the last term of Eq.~(\ref{4-fermi-R}) should be omitted and the sign in front of the right-handed vector Wilson coefficient changes. Note that, in principle, constraints from the ratios like ${\rm Br}[K\to\mu\nu]/{\rm Br}[\pi\to\mu\nu]$ (see e.g. Ref.~\cite{FlaviaNetWorkingGrouponKaonDecays:2008hpm} for an overview) have to be taken into account, especially in the case of scalar operators, even though these are not measures of LFUV.

\subsubsection{$SU(2)_L$ Gauge Invariance}

Interestingly, assuming NP above the EW scale which respects $SU(2)_L$ gauge symmetry, any modification of the left-handed charged current also leads to a modification of a neutral current~\cite{Cirigliano:2009wk}. The relevant effective operators in explicit $SU(2)_L$ gauge invariant language (see Refs.~\cite{Buchmuller:1985jz,Grzadkowski:2010es} for the definitions of the conventions) are contained in the Lagraingain
\begin{equation}
\begin{aligned}
L=\dfrac{1}{\Lambda^2}&\left([C_{\ell q}^{(3)}]_{ijkl} (\bar{L}_i\gamma^{\mu}\tau^IL_j)(\bar{Q}_k\gamma_{\mu}\tau^IQ_l)+[C_{\ell\ell}^{(3)}]_{ijkl} (\bar{L}_i\gamma^{\mu}\tau^IL_j)(\bar{L}_k\gamma_{\mu}\tau^IL_l)+\right.
\\  &\;\;\;\;\left. C_{\phi \ell }^{\left( 1 \right)ij} {\phi ^\dag }i\mathord{\buildrel{\lower3pt\hbox{$\scriptscriptstyle\leftrightarrow$}} 
	\over D} _\mu\phi  \, {{\bar L}^i}{\gamma ^\mu }{L^j}  +C_{\phi \ell }^{\left( 3 \right)ij} {\phi ^\dag }i\mathord{\buildrel{\lower3pt\hbox{$\scriptscriptstyle\leftrightarrow$}} 
	\over D} _\mu ^I\phi  \, {{\bar L}^i}{\tau ^I}{\gamma ^\mu }{L^j} \right) \,. 
\end{aligned}
\end{equation}
Making the identification with the operators discussed in the previous subsections we find for modified $W\ell\nu$ couplings
\begin{equation}
  \epsilon_{fi}  =\frac{v^{2}}{\Lambda^{2}} C_{\phi e}^{fi}\,,
\end{equation}
with $v=\sqrt{2}\times174\,{\rm GeV}$ and $\Lambda$ is the mass scale of NP. This means that at the same time   $Z\to\nu\nu$ and/or $Z\to \ell\ell$ effects appear. However, in order to avoid effects in stringently constrained $Z$ couplings to charged leptons~\cite{ALEPH:2005ab}, one can assume $C^{(1)}_{\phi\ell}=-C^{(3)}_{\phi\ell}$. In this case, the constructive effect in $Z\to\nu_\mu\nu_\mu$ can be compensated by a destructive effect in $Z\to\nu_e\nu_e$, as only the sum of the three flavors is measured, resulting in a very good fit to data~\cite{Coutinho:2019aiy}. 

For 4-lepton operators
\begin{equation}
\begin{aligned}
C_{fi,{\rm NP}}^{\ell\ell\nu\nu}&=-\dfrac{4m_W^2}{g_2^2\Lambda^2}[C_{\ell q}^{(3)}]_{ffii}\,.
\end{aligned}
\end{equation}
Here the implications are that a purely leptonic charged current operator with first generation leptons in general also gives rise to a neutral current with electrons which is subject to LEP bounds from non-resonant di-lepton searches~\cite{ALEPH:2013dgf}. 

Finally, for the 4-fermion operators, working in the down-basis, we find
\begin{equation}
\begin{aligned}
C_{fi,{\rm NP}}^{V_L\ell_j}&=-\dfrac{4m_W^2}{g_2^2\Lambda^2}[C_{\ell q}^{(3)}]_{jjfi}  \,.
\end{aligned}
\end{equation}
This means that for left-handed $\bar u d \bar \nu e$ operators affecting beta decays, $\bar u u \bar e e$ and/or $\bar dd \bar e e$ operators are  also generated. This has interesting implications for an explanation of the CAA, as it leads to an additional  neutral $\bar uu \ell\ell$ and $\bar dd \ell\ell$ current after EW symmetry breaking. $[Q_{\ell q}^{(3)}]_{1111}$ therefore also contributes to non-resonant di-electron production at the LHC, which is tailored to search for heavy NP that is above the direct production reach~\cite{Eichten:1984eu,Eichten:1983hw}. The latest di-lepton results from ATLAS and CMS are presented in Ref.~\cite{Aad:2020otl} and Ref.~\cite{Sirunyan:2021khd}, respectively. CMS observed a slight excess in the di-electron cross section at high invariant lepton mass and computed the double ratio 
\begin{equation}
R^{\rm Data}_{\mu^+\mu^-/e^+e^-}/R^{\rm MC}_{\mu^+\mu^-/e^+e^-}\,,
\end{equation}
in order to reduce the uncertainties~\cite{Greljo:2017vvb}. This means that they provide the relative signal strength for muons vs. electrons, $R^{\rm Data}_{\mu^+\mu^-/e^+e^-}$, divided by the SM expectation obtained from Monte Carlo simulations, $R^{\rm MC}_{\mu^+\mu^-/e^+e^-}$. Taking this into account, we find that the best fit value for the Wilson coefficient is 
\begin{equation}
[C_{\ell q}^{(3)}]_{1111}/\Lambda^2\approx1.0 /(10\,{\rm TeV})^2\,,
\end{equation}
with $\Delta \chi^2 \equiv \chi^2 - \chi^2_{SM} \approx-10$ and $0.3 /(10\,{\rm TeV})^2\lessapprox [C_{\ell q}^{(3)}]_{1111}\lessapprox1.8 /(10\,{\rm TeV})^2$
. Note that this value is compatible with the corresponding ATLAS bounds and $\left( \frac{A_\mu}{A_e}  \right)_{ R^{\pi}_{e/ \mu} }$ in Eq. ~\ref{gpi}. Treating the ATLAS exclusion as a hard cut, we therefore find that at $95\%$ CL
\begin{equation}
0.6/(10{\rm TeV})^2\lessapprox [C_{\ell q}^{(3)}]_{1111}/\Lambda^2 \lessapprox 1.4/(10{\rm TeV})^2 \,,
\end{equation}
which predicts
\begin{equation}
1.0004\lessapprox \left( \frac{A_\mu}{A_e}  \right)_{ R^{\pi}_{e/ \mu} }\lessapprox 1.0009\,,
\end{equation}
at the 95\% CL.

\subsection{Simplified NP models} 

The following NP models (see Ref.~\cite{deBlas:2017xtg} for a complete categorization of tree-level extensions of the SM) give contributions to the effective operators discussed in the last subsection, which have a relevant impact on our observables limiting LFUV.

\subsubsection{$W^\prime$ boson}

In order to be relevant for our observables, a $W^\prime$ bosons must be the component of a $SU(2)_L$ triplet $X_a^\mu$ with hypercharge 0 such that it can have coupling to left-handed fermions
\begin{equation}
    L_{X^\mu_a}=-g_{ji}^{\ell} X_{a}^{\mu} \bar{L}_{j} \gamma_{\mu} \frac{\tau^{a}}{2} L_{i}-g_{j i}^{q} X_{a}^{\mu} \bar{Q}_{j} \gamma_{\mu} \frac{\tau^{a}}{2} Q_{i}\,.
\end{equation}
These couplings can contribute to LFUV observables in several ways~\cite{Crivellin:2020ebi}:
\begin{itemize}
    \item Modified $W\ell\nu$ coupling via mixing with the SM $W$. In this case it generates $C^{(3)}_{\phi\ell}$
such that $W\ell\nu$, $Z\nu\nu$ and $Z\nu\nu$ couplings are affected. As discussed in the previous subsection, this leads to limited effects in $W\ell\nu$ due to the stringent constraints from $Z$ decays. Global fits to $C^{(3)}_{\phi\ell}$ can be found in Refs.~\cite{Crivellin:2020ebi,Kirk:2020wdk}.
    \item Tree-level effects in  $\ell\to\ell^\prime\nu\nu$. In this case the Wilson coefficient $C_{\ell \ell}^{(3)}$ is generated which results after EW symmetry breaking in
    \begin{equation}
    C_{fi}^{\ell\ell\nu\nu,{\rm NP}}=\dfrac{g^\ell_{ff}}{g^\ell_{ii} }  \dfrac{m_W^2}{m_{W^\prime}^2}\,,  
    \end{equation}
    and the bounds of Sec.~\ref{4-fermi} can be used.   
    \item Tree-level effects in $d\to u e\nu$. Similarly, if $W'$ possesses couplings to quarks and leptons, it leads to tree-level effects $d\to u e\nu$ via the Wilson coefficient $C_{qq}^{(3)}$ and the bounds from kaon and pion decays apply. 
    \end{itemize}
Fro example, left-handed $W^\prime$ bosons appear as excitations of the SM $W$ in ~\cite{Weinberg:1962hj,Susskind:1978ms} or extra-dimensional models~\cite{Randall:1999ee} as well as in theories with several $SU(2)_L$ gauge groups~\cite{Malkawi:1996fs,Hsieh:2010zr}.

 \subsubsection{Vector-like Leptons}  
 
Vector-like leptons (VLL), such as right-handed neutrinos~\cite{Lee:1977tib}, affect $W-\ell-\nu$ coupling via their mixing with SM leptons and are EW symmetry breaking. There are 5 representations of vector-like leptons that can couple to SM leptons and the Higgs and mix with the former after EW symmetry breaking. These are represented as 

 \begin{equation}
   \begin{array}{l|ccc} 
& S U(3) & S U(2)_{L} & U(1)_{Y} \\
\hline \ell & 1 & 2 & -1 / 2 \\
\mathrm{e} & 1 & 1 & -1 \\
\phi & 1 & 2 & 1 / 2 \\
\hline \mathrm{N} & 1 & 1 & 0 \\
\mathrm{E} & 1 & 1 & -1 \\
\Delta_{1}=\left(\Delta_{1}^{0}, \Delta_{1}^{-}\right) & 1 & 2 & -1 / 2 \\
\Delta_{3}=\left(\Delta_{3}^{-}, \Delta_{3}^{--}\right) & 1 & 2 & -3 / 2 \\
\Sigma_{0}=\left(\Sigma_{0}^{+}, \Sigma_{0}^{0}, \Sigma_{0}^{-}\right) & 1 & 3 & 0 \\
\Sigma_{1}=\left(\Sigma_{1}^{0}, \Sigma_{1}^{-}, \Sigma_{1}^{--}\right) & 1 & 3 & -1
\end{array}   
 \end{equation}
which result in the following Wilson coefficients
  \begin{equation}
 \begin{aligned}
\frac{C_{\phi \ell}^{(1) i j}}{\Lambda^{2}} &=\frac{\lambda_{N}^{i} \lambda_{N}^{j \dagger}}{4 M_{N}^{2}}-\frac{\lambda_{E}^{i} \lambda_{E}^{j \dagger}}{4 M_{E}^{2}}+\frac{3}{16} \frac{\lambda_{\Sigma_{0}}^{i \dagger} \lambda_{\Sigma_{0}}^{j}}{M_{\Sigma_{0}}^{2}}-\frac{3}{16} \frac{\lambda_{\Sigma_{1}}^{i \dagger} \lambda_{\Sigma_{1}}^{j}}{M_{\Sigma_{1}}^{2}} \,,\\
\frac{C_{\phi \ell}^{(3) i j}}{\Lambda^{2}} &=-\frac{\lambda_{N}^{i} \lambda_{N}^{j \dagger}}{4 M_{N}^{2}}-\frac{\lambda_{E}^{i} \lambda_{E}^{j \dagger}}{4 M_{E}^{2}}+\frac{1}{16} \frac{\lambda_{\Sigma_{0}}^{i \dagger} \lambda_{\Sigma_{0}}^{j}}{M_{\Sigma_{0}}^{2}}+\frac{1}{16} \frac{\lambda_{\Sigma_{1}}^{j \dagger} \lambda_{\Sigma_{1}}^{i}}{M_{\Sigma_{1}}^{2}} \,,\\
\frac{C_{\phi \mathrm{e}}^{i j}}{\Lambda^{2}} &=\frac{\lambda_{\Delta_{1}}^{i \dagger} \lambda_{\Delta_{1}}^{j}}{2 M_{\Delta_{1}}^{2}}-\frac{\lambda_{\Delta_{3}}^{i \dagger} \lambda_{\Delta_{3}}^{j}}{2 M_{\Delta_{3}}^{2}}\,,
\end{aligned}
\end{equation}
where $\lambda$ are the couplings of the VLLs to the SM leptons and the Higgs doublet (see Ref.~\cite{Crivellin:2020ebi} for details). Note that only $C_{\phi \ell}^{(3)}$ generates a charged current and affects our LFUV observables while the other coefficients enter $Z\ell\ell$ and $Z\nu\nu$ couplings entering the global EW fit~\cite{delAguila:2008pw}. While for $C_{\phi \ell}^{(3)}$ we can apply the bounds from the previous subsection, in order to take into account all effects, a global fit is necessary for which we refer the reader to Ref.~\cite{Crivellin:2020ebi}.

VLLs are predicted in many SM extensions, such as Grand Unified Theories~\cite{Hewett:1988xc,Langacker:1980js,delAguila:1982fs}, composite models or models with extra dimensions~\cite{Antoniadis:1990ew,ArkaniHamed:1998kx,Csaki:2004ay,ArkaniHamed:2001nc,ArkaniHamed:2002qy,Perelstein:2005ka,delAguila:2010vg,Carmona:2013cq} and 
are involved in the type I~\cite{Minkowski:1977sc,Lee:1977tib} and type III~\cite{Foot:1988aq} seesaw mechanisms.

\subsubsection{Singly charged $SU(2)_L$ singlet scalar}

As it is a $SU(2)_L \times SU(3)_C$ singlet, $\phi^+$ has hypercharge +1 and allows only for Yukawa-type interactions with leptons. The  model is particularly interesting in the context of LFUV; due to  hermicity of the Lagrangian it has anti-symmetric (i.e. off-diagonal) couplings:
\begin{align}
\mathcal{L} = - \frac{\lambda_{ij}}{2}\, \bar{L}^c_{a,i}\, \varepsilon_{ab}\, L_{b,j} \, \Phi^+ + {\rm h.c.}\,,
\end{align}
but not with quarks. Here $L$ is the left-handed $SU(2)_L$ lepton doublet, $c$ stands for charge conjugation, $a$ and $b$ are $SU(2)_L$ indices, $i$ and $j$ are flavor indices and $\varepsilon_{ab}$ is the two-dimensional antisymmetric tensor. Note that without loss of generality, $\lambda_{ij}$ can be chosen to be antisymmetric in flavor space, $\lambda_{ji}=-\lambda_{ij}$, such that $\lambda_{ii}=0$ and our free parameters are $\lambda_{12}$, $\lambda_{13}$, and $\lambda_{23}$ giving
\begin{equation}
 C_{fi,{\rm NP}}^{\ell\ell\nu\nu}=\frac{\left\vert \lambda_{fi}^2\right\vert}{g_2^2}\frac{m_W^2}{m_\phi^2}\,.  
\end{equation}
This means that the effect of $\ell\to\ell^\prime\nu\overline\nu$ is necessarily constructive, such that the CAA can be solved~\cite{Crivellin:2020klg,Felkl:2021qdn,Marzocca:2021azj} and also $C_{23,{\rm NP}}^{\ell\ell\nu\nu}$ has the right sign preferred by the fit. However, in order to not violate the bounds from $\mu\to e\gamma$, $\lambda_{13}$ must be very close to zero~\cite{Crivellin:2020klg}. 

Singly charged scalars have been proposed within the Babu-Zee model~\cite{Zee:1985id,Babu:1988ki} and studied in  Refs.~\cite{Krauss:2002px,Nebot:2007bc,Cai:2014kra,Cheung:2004xm,Ahriche:2014xra,Chen:2014ska,Ahriche:2015loa,Herrero-Garcia:2014hfa,Herrero-Garcia:2017xdu,CentellesChulia:2018gwr,Babu:2019mfe} as part of a larger NP spectrum, mostly with the aim of generating neutrino masses at loop level.

\subsubsection{Scalar $SU(2)_L$ Triplet}
This fermion couples to a lepton doublet and a charge conjugated one as
\begin{align}
\mathcal{L_T} = - \frac{\kappa_{ij}}{2}\, \bar{L}^c_{a,i}\, \varepsilon_{ab} \tau^I_{cd}\, L_{d,j} \, \Phi^T + {\rm h.c.}\,,
\label{Lag}
\end{align}
resulting in
\begin{equation}
 C_{fi,{\rm NP}}^{\ell\ell\nu\nu}=-\frac{\left\vert \kappa_{fi}^2\right\vert}{g_2^2}\frac{2m_W^2}{m_\phi^2}\,,  
\end{equation}
such that the effect is always destructive in $\ell\to\ell^\prime\nu\overline\nu$.

This scalar can generate neutrino masses within the type-II seesaw model~\cite{Konetschny:1977bn,Magg:1980ut,Schechter:1980gr,Cheng:1980qt,Mohapatra:1980yp}. While in general the tiny neutrino masses require very small couplings $\kappa$ (for TeV scale masses), the contribution to the Weinberg operator\cite{Weinberg:1979sa} can be suppressed, e.g. by an approximate Baryon number symmetry, such that phenomenologically relevant effects in LFUV are possible.

\subsubsection{$SU(2)_L$ Neutral Vector Boson ($Z^\prime$):}
A $Z^\prime$ boson which is an $SU(2)_L$ singlet only interferes with the SM amplitudes for $\ell\to\ell^\prime\nu\nu$ if it has couplings to lepton doublets
\begin{equation}
    L_{Z^\prime}=-i g^\ell_{fi}\bar L_f \gamma^\mu L_i Z^\prime_\mu\,.
\end{equation}
Furthermore, these couplings must be flavor violating such that
\begin{equation}
\begin{aligned}
C_{fi,{\rm NP}}^{\ell\ell\nu\nu}&=\dfrac{2m_W^2}{g_2^2 M_{Z^\prime}^2} |g^\ell_{fi}|^2\,.
\end{aligned}
\end{equation}
Note that  the effect in $\ell\to\ell^\prime\nu\overline\nu$ is necessarily constructive~\cite{Buras:2021btx}.

There is a huge literature on $Z^\prime$ bosons (see Ref.~\cite{Langacker:2000ju} for an overview), which could again be excitations of the SM $Z$ and $\gamma$, but in this case also originate from a gauged $U(1)$ flavor symmetry like $L_\mu-L_\tau$ or $B-L$.

\subsubsection{Leptoquark}	

There are 10 representations of leptoquarks (five scalar and five vector)  with gauge invariant couplings to quarks and leptons~\cite{Buchmuller:1986zs}. Out of these, six representations generate a charged current, two a vector current, two a scalar current and two simultaneously a vector and a scalar current. While those with a scalar current are stringently constrained by $R^\pi_{e/\mu}$, for vector currents other observables, like di-lepton pairs, low energy parity violation or kaon decays, are, in general, more constraining. For a recent comprehensive analysis we refer  to Ref.~\cite{Crivellin:2021egp,Crivellin:2021bkd}.

Leptoquarks arise  in the Pati-Salam model~\cite{Pati:1974yy} $SU(5)$ Grand Unified theories~\cite{Georgi:1974sy,Dimopoulos:1980hn} and in the R-parity violating Mimimal Supersymmetric Model (MSSM, see e.g. Ref.~\cite{Barbier:2004ez} for a review) and were studied in the context of LFUV in kaon, tau and pion decays in Ref.~\cite{Davidson:1993qk,Campbell:2003ir,RamseyMusolf:2007yb,Dorsner:2016wpm,Mandal:2019gff}.

\subsubsection{Charged Higgs}

Charged Higgs bosons  have been considered in the context of leptonic meson decays 
~\cite{Crivellin:2013wna}, in particular, $R^{K}_{\mu/e}$ in the context of the MSSM~\cite{Masiero:2005wr,Girrbach:2012km}. Furthermore, the type-X two-Higgs-doublet model (2HDM) is constrained by loop effects in $\tau\to\mu\nu_\tau\nu_\mu$~\cite{Krawczyk:2004na,Broggio:2014mna,Chun:2016hzs} relevant at large values of $\tan\beta$.

\section{Future Prospects}
\label{sec:future}

\subsection{Pion and Kaon Experiments}
\label{PiKExperiments}
 The PIENU \cite{PiENu:2015seu, AguilarArevalo:2010xi} and  PEN \cite{Pocanic1, Glaser:2018aat, Pocanic:2014jka} experiments aim to further improve the precision of the $\pi\to e \nu$ branching ratio $R^{\pi}_{e/\mu}$. However, even when these goals are realized, an experimental improvement by more than an order of magnitude in uncertainty  is warranted to confront the SM prediction and to search for non-SM effects.   A developing proposal \nexp~\cite{pioneer1} aims for an improvement in precision for $R^{\pi}_{e/\mu}$ by an order of magnitude making the experimental uncertainty comparable to the theoretical uncertainty in Eq.~\ref{Remu_SM}. To reach  very high precision requires high statistics as well as extensive evaluation of systematic uncertainties, backgrounds, biases, and distortions in the data selection criteria. Like  PIENU~\cite{Aguilar-Arevalo:2015cdf} and  PEN~\cite{Pocanic1}, \nexp~will be done using stopped pions that decay at rest. Principal features of the experiment include a fully active Si tracking stopping target~\cite{Sadrozinski:2017qpv} and a high resolution calorimeter, both of which contribute to suppression of systematic effects. 

PIONEER also aims to improve the precision for the branching ratio of pion beta decay $\pi^+ \rightarrow \pi^0 e^+ \nu (\gamma)$~\cite{Pocanic:2003pf}.
Pion beta decay,  while providing  the theoretically cleanest determination of the CKM matrix element $V_{ud}$, is currently not competitive: $V_{ud} = 0.9739(28)_{\textrm{exp}}(1)_{\textrm{th}}$, where the experimental uncertainty comes almost entirely from the  $\pi^+ \rightarrow \pi^0 e^+ \nu (\gamma)$ branching ratio~\cite{Pocanic:2003pf} (the pion lifetime contributes ${\delta}V_{ud} = 0.0001$), and the theory uncertainty has been reduced from $({\delta}V_{ud})_{\textrm{th}} = 0.0005$~\cite{Sirlin:1977sv,Cirigliano:2002ng,Passera:2011ae} to $({\delta}V_{ud})_{\textrm{th}} = 0.0001$ via a lattice QCD calculation of the radiative corrections~\cite{Feng:2020zdc}. As pointed out in Ref.~\cite{Czarnecki:2019iwz} even a three-fold improvement in  precision of the pion beta decay branching ratio compared to Ref.~\cite{Pocanic:2003pf} would allow for a 0.2\% accuracy in the determination of the ratio $V_{us}/V_{ud}$ from $R_V \equiv \Gamma (K \to \pi \ell \nu (\gamma))/\Gamma (\pi^+ \to \pi^0 e^+ \nu (\gamma))$,  competitive with the existing determination based on $R_A \equiv \Gamma (K \to \mu \nu) / \Gamma (\pi \to \mu \nu)$.  

Regarding other kaon decays, the TREK~\cite{Kohl:2016afc} experiment may provide additional information on  $R^{K}_{ e/\mu}$ and NA62 may make further measurements relevant to LFUV~\cite{NA62:2020mcv,NA62:2020fhy}.

\subsection{Tau Experiments}
The Belle II experiment is currently collecting data at and near the center-of-mass energy of 10.58 GeV/c$^2$ and is expected to obtain a total integrated luminosity of 
$50$ ab$^{-1}$, equivalent to $45$ billion $\tau-$pairs~\cite{Belle-II:2010dht, Belle-II:2018jsg}. Although precision measurements are ultimately dominated by the capabilities of experiments to limit systematic factors, an improvement in the determination of all LFU parameters in $\tau$ decays is to be expected.

The proposed future circular collider (FCC) plans to collide  electrons and positrons  (FCC-ee) at different center-of-mass energies, including a 4-year high-statistics run at the $Z$ pole~\cite{FCC:2018evy}. Assuming an instantaneous luminosity of 2.3$\times 10^{36}$cm$^{-2}$s$^{-1}$ and four interaction regions, this would translate into 1.7$\times 10^{11}$ $\tau-$pairs produced in $Z\to \tau^+\tau^-$ reactions available for precision studies of $\tau$ properties and polarization at the FCC-ee~\cite{Dam:2021ibi}.
A rich $\tau$ physics program would also be expected by the proposed Circular Electron Positron Collider (CEPC), where 30 billion $\tau$ pairs could be produced at the Z pole~\cite{CEPCStudyGroup:2018ghi}. 

For both the FCC-ee and CEPC, the same considerations regarding the control of systematic effects hold as for Belle II. Although collecting high-statistics samples of $\tau$ is a \textit{conditio sine qua non} to perform precision tests of the standard model via suppressed or forbidden processes in $\tau$ decays, systematic effects must be understood and kept under control. All the above-mentioned experiments will possibly reach a statistical precision at the level of $10^{-4}-10^{-5}$ on a number of LFU parameters, given the large amount of $\tau$ decays that will be produced and analyzed, but systematic effects will dominate the final precision. Experiments planned in the future have the advantage of designing the detectors to minimize  potential systematic effects.

\subsection{Theory}

While the ratios measuring LFU are, in general, theoretically very clean, the uncertainty in the extraction of $V_{ud}$ from  super-allowed beta decays is still limited by the theoretical error. Moreover, expected experimental improvements  in the neutron lifetime and decay asymmetry will enhance the relative impact of theoretical uncertainties in the extraction of $V_{ud}$ from neutron decay. The theoretical uncertainty in both cases arises from electromagnetic radiative corrections.  Progress in the next decade can be expected on a number of fronts. First, at the single nucleon level, one expects results on radiative corrections to neutron decay from lattice QCD. The technology to perform these calculations in meson systems ($K^+ \to \mu^+ \nu_\mu$ and $\pi^+ \to \pi^0 e^+ \nu_e$)
has been demonstrated~\cite{DiCarlo:2019thl,Feng:2020zdc}, but not yet applied  to nucleons. 
Second, one can expect progress in the analysis of few- and many-body effects in nuclear transitions, both with dispersive techniques (see Ref.~\cite{Gorchtein:2018fxl}) and through the development of chiral effective field theory few-body transitions operators to $O(G_F \alpha)$ coupled to first-principle many-body methods. 

\section{Conclusions}
\label{sec:conclusions}

In this article we reviewed the status of the searches for LFUV in the charged current involving pions, kaons, taus, and beta decays. Averaging the values found in Sec.~\ref{piondecays}, Sec.~\ref{kaondecays}, and Sec.~\ref{taudecays} we find that the ratio of the $W$ couplings to leptons are
\begin{equation}
\begin{aligned}
   \frac{g_\mu}{g_e}= & \, 1+\epsilon_{\mu\mu}-\epsilon_{ee}=1.0009\pm0.0006 \,,\\
      \frac{g_\tau}{g_\mu}= & \,  1+\epsilon_{\tau\tau}-\epsilon_{\mu\mu}=1.0013\pm0.0013 \,,\rm{and}\\
         \frac{g_\tau}{g_e}= & \, 1+\epsilon_{\tau\tau}-\epsilon_{ee}=1.0022\pm0.0013 \,.
           \end{aligned}
\end{equation} 
Note that these individual  results for $g_i/g_j$ use  values of the others that minimize the $\chi^2$ of 1d fits and should thus be understood separately i.e. no correlations between the three ratios are taken into account. These  high precision tests  of LVU which agree well with the SM are particularly interesting in light of the experimental hints for LFUV in semi-leptonic $B$ decays, the anomalous magnetic moment of the muon, and the Cabbibo angle anomaly.

Taking into account the current experimental and theoretical results, we performed a combined fit to modified $W-\ell-\nu$ couplings as shown the left plot of Fig.~\ref{eps}. However, the results can also be interpreted in terms of 4-lepton and 2-quark-2-lepton operators. We also reviewed the experimental and theoretical prospects for the many observables. The proposed PIONEER experiment aims to improve the test of $e-\mu$
universality in pion decay and the  pion beta decay determination of $V_{ud}$.  Furthermore, BELLE II measurements will  lead to significant improvements in LFU tests with tau leptons. Finally, the proposed  FCC-ee and CEPC accelerators offer intriguing  possibilities for flavor physics and, in particular, for tau decays given the expected large samples of $B$ mesons and $\tau$ leptons that could be produced by these facilities.

\begin{acknowledgements}

D.B. is supported by NSERC  grant no.  SAPPJ-2018-0017 (Canada). A.C. gratefully acknowledges the support by the Swiss National Science Foundation under Project No.\ PP00P21\_76884. G.I. was supported by the European Research Council under the grant agreement no. 947006 - ``InterLeptons".  V.C. is supported by the US Department of Energy through the Office of Nuclear Physics and the LDRD program at Los Alamos National Laboratory. Los Alamos National Laboratory is operated by Triad National Security, LLC, for the National Nuclear Security Administration of U.S. Department of Energy \end{acknowledgements}

\bibliography{LFUV2}

\begin{thebibliography}{192}%
\makeatletter
\providecommand \@ifxundefined [1]{%
 \@ifx{#1\undefined}
}%
\providecommand \@ifnum [1]{%
 \ifnum #1\expandafter \@firstoftwo
 \else \expandafter \@secondoftwo
 \fi
}%
\providecommand \@ifx [1]{%
 \ifx #1\expandafter \@firstoftwo
 \else \expandafter \@secondoftwo
 \fi
}%
\providecommand \natexlab [1]{#1}%
\providecommand \enquote  [1]{``#1''}%
\providecommand \bibnamefont  [1]{#1}%
\providecommand \bibfnamefont [1]{#1}%
\providecommand \citenamefont [1]{#1}%
\providecommand \href@noop [0]{\@secondoftwo}%
\providecommand \href [0]{\begingroup \@sanitize@url \@href}%
\providecommand \@href[1]{\@@startlink{#1}\@@href}%
\providecommand \@@href[1]{\endgroup#1\@@endlink}%
\providecommand \@sanitize@url [0]{\catcode `\\12\catcode `\$12\catcode
  `\&12\catcode `\#12\catcode `\^12\catcode `\_12\catcode `\%12\relax}%
\providecommand \@@startlink[1]{}%
\providecommand \@@endlink[0]{}%
\providecommand \url  [0]{\begingroup\@sanitize@url \@url }%
\providecommand \@url [1]{\endgroup\@href {#1}{\urlprefix }}%
\providecommand \urlprefix  [0]{URL }%
\providecommand \Eprint [0]{\href }%
\providecommand \doibase [0]{http://dx.doi.org/}%
\providecommand \selectlanguage [0]{\@gobble}%
\providecommand \bibinfo  [0]{\@secondoftwo}%
\providecommand \bibfield  [0]{\@secondoftwo}%
\providecommand \translation [1]{[#1]}%
\providecommand \BibitemOpen [0]{}%
\providecommand \bibitemStop [0]{}%
\providecommand \bibitemNoStop [0]{.\EOS\space}%
\providecommand \EOS [0]{\spacefactor3000\relax}%
\providecommand \BibitemShut  [1]{\csname bibitem#1\endcsname}%
\let\auto@bib@innerbib\@empty
\bibitem [{\citenamefont {Aad}\ \emph {et~al.}(2012)\citenamefont {Aad} \emph
  {et~al.}}]{ATLAS:2012yve}%
  \BibitemOpen
  \bibfield  {author} {\bibinfo {author} {\bibfnamefont {G.}~\bibnamefont
  {Aad}} \emph {et~al.} (\bibinfo {collaboration} {ATLAS}),\ }\href {\doibase
  10.1016/j.physletb.2012.08.020} {\bibfield  {journal} {\bibinfo  {journal}
  {Phys. Lett. B}\ }\textbf {\bibinfo {volume} {716}},\ \bibinfo {pages} {1}
  (\bibinfo {year} {2012})},\ \Eprint {http://arxiv.org/abs/1207.7214}
  {arXiv:1207.7214 [hep-ex]} \BibitemShut {NoStop}%
\bibitem [{\citenamefont {Chatrchyan}\ \emph {et~al.}(2012)\citenamefont
  {Chatrchyan} \emph {et~al.}}]{CMS:2012qbp}%
  \BibitemOpen
  \bibfield  {author} {\bibinfo {author} {\bibfnamefont {S.}~\bibnamefont
  {Chatrchyan}} \emph {et~al.} (\bibinfo {collaboration} {CMS}),\ }\href
  {\doibase 10.1016/j.physletb.2012.08.021} {\bibfield  {journal} {\bibinfo
  {journal} {Phys. Lett. B}\ }\textbf {\bibinfo {volume} {716}},\ \bibinfo
  {pages} {30} (\bibinfo {year} {2012})},\ \Eprint
  {http://arxiv.org/abs/1207.7235} {arXiv:1207.7235 [hep-ex]} \BibitemShut
  {NoStop}%
\bibitem [{\citenamefont {Zyla}\ \emph
  {et~al.}(2020{\natexlab{a}})\citenamefont {Zyla} \emph
  {et~al.}}]{ParticleDataGroup:2020ssz}%
  \BibitemOpen
  \bibfield  {author} {\bibinfo {author} {\bibfnamefont {P.~A.}\ \bibnamefont
  {Zyla}} \emph {et~al.} (\bibinfo {collaboration} {Particle Data Group}),\
  }\href {\doibase 10.1093/ptep/ptaa104} {\bibfield  {journal} {\bibinfo
  {journal} {PTEP}\ }\textbf {\bibinfo {volume} {2020}},\ \bibinfo {pages}
  {083C01} (\bibinfo {year} {2020}{\natexlab{a}})}\BibitemShut {NoStop}%
\bibitem [{\citenamefont {Crivellin}\ and\ \citenamefont
  {Hoferichter}(2021)}]{Crivellin:2021sff}%
  \BibitemOpen
  \bibfield  {author} {\bibinfo {author} {\bibfnamefont {A.}~\bibnamefont
  {Crivellin}}\ and\ \bibinfo {author} {\bibfnamefont {M.}~\bibnamefont
  {Hoferichter}},\ }\href {\doibase 10.1126/science.abk2450} {\bibfield
  {journal} {\bibinfo  {journal} {Science}\ }\textbf {\bibinfo {volume}
  {374}},\ \bibinfo {pages} {1051} (\bibinfo {year} {2021})},\ \Eprint
  {http://arxiv.org/abs/2111.12739} {arXiv:2111.12739 [hep-ph]} \BibitemShut
  {NoStop}%
\bibitem [{\citenamefont {Buddenbrock}\ \emph {et~al.}(2019)\citenamefont
  {Buddenbrock}, \citenamefont {Cornell}, \citenamefont {Fang}, \citenamefont
  {Fadol~Mohammed}, \citenamefont {Kumar}, \citenamefont {Mellado},\ and\
  \citenamefont {Tomiwa}}]{Buddenbrock:2019tua}%
  \BibitemOpen
  \bibfield  {author} {\bibinfo {author} {\bibfnamefont {S.}~\bibnamefont
  {Buddenbrock}}, \bibinfo {author} {\bibfnamefont {A.~S.}\ \bibnamefont
  {Cornell}}, \bibinfo {author} {\bibfnamefont {Y.}~\bibnamefont {Fang}},
  \bibinfo {author} {\bibfnamefont {A.}~\bibnamefont {Fadol~Mohammed}},
  \bibinfo {author} {\bibfnamefont {M.}~\bibnamefont {Kumar}}, \bibinfo
  {author} {\bibfnamefont {B.}~\bibnamefont {Mellado}}, \ and\ \bibinfo
  {author} {\bibfnamefont {K.~G.}\ \bibnamefont {Tomiwa}},\ }\href {\doibase
  10.1007/JHEP10(2019)157} {\bibfield  {journal} {\bibinfo  {journal} {JHEP}\
  }\textbf {\bibinfo {volume} {10}},\ \bibinfo {pages} {157} (\bibinfo {year}
  {2019})},\ \Eprint {http://arxiv.org/abs/1901.05300} {arXiv:1901.05300
  [hep-ph]} \BibitemShut {NoStop}%
\bibitem [{\citenamefont {Crivellin}\ \emph
  {et~al.}(2021{\natexlab{a}})\citenamefont {Crivellin}, \citenamefont {Fang},
  \citenamefont {Fischer}, \citenamefont {Kumar}, \citenamefont {Kumar},
  \citenamefont {Malwa}, \citenamefont {Mellado}, \citenamefont {Rapheeha},
  \citenamefont {Ruan},\ and\ \citenamefont {Sha}}]{Crivellin:2021ubm}%
  \BibitemOpen
  \bibfield  {author} {\bibinfo {author} {\bibfnamefont {A.}~\bibnamefont
  {Crivellin}}, \bibinfo {author} {\bibfnamefont {Y.}~\bibnamefont {Fang}},
  \bibinfo {author} {\bibfnamefont {O.}~\bibnamefont {Fischer}}, \bibinfo
  {author} {\bibfnamefont {A.}~\bibnamefont {Kumar}}, \bibinfo {author}
  {\bibfnamefont {M.}~\bibnamefont {Kumar}}, \bibinfo {author} {\bibfnamefont
  {E.}~\bibnamefont {Malwa}}, \bibinfo {author} {\bibfnamefont
  {B.}~\bibnamefont {Mellado}}, \bibinfo {author} {\bibfnamefont
  {N.}~\bibnamefont {Rapheeha}}, \bibinfo {author} {\bibfnamefont
  {X.}~\bibnamefont {Ruan}}, \ and\ \bibinfo {author} {\bibfnamefont
  {Q.}~\bibnamefont {Sha}},\ }\href@noop {} {\  (\bibinfo {year}
  {2021}{\natexlab{a}})},\ \Eprint {http://arxiv.org/abs/2109.02650}
  {arXiv:2109.02650 [hep-ph]} \BibitemShut {NoStop}%
\bibitem [{\citenamefont {Fischer}\ \emph {et~al.}(2021)\citenamefont {Fischer}
  \emph {et~al.}}]{Fischer:2021sqw}%
  \BibitemOpen
  \bibfield  {author} {\bibinfo {author} {\bibfnamefont {O.}~\bibnamefont
  {Fischer}} \emph {et~al.},\ }\href@noop {} {\  (\bibinfo {year} {2021})},\
  \Eprint {http://arxiv.org/abs/2109.06065} {arXiv:2109.06065 [hep-ph]}
  \BibitemShut {NoStop}%
\bibitem [{\citenamefont {Lees}\ \emph {et~al.}(2012)\citenamefont {Lees} \emph
  {et~al.}}]{Lees:2012xj}%
  \BibitemOpen
  \bibfield  {author} {\bibinfo {author} {\bibfnamefont {J.~P.}\ \bibnamefont
  {Lees}} \emph {et~al.} (\bibinfo {collaboration} {BaBar}),\ }\href {\doibase
  10.1103/PhysRevLett.109.101802} {\bibfield  {journal} {\bibinfo  {journal}
  {Phys. Rev. Lett.}\ }\textbf {\bibinfo {volume} {109}},\ \bibinfo {pages}
  {101802} (\bibinfo {year} {2012})},\ \Eprint {http://arxiv.org/abs/1205.5442}
  {arXiv:1205.5442 [hep-ex]} \BibitemShut {NoStop}%
\bibitem [{\citenamefont {Aaij}\ \emph {et~al.}(2018)\citenamefont {Aaij} \emph
  {et~al.}}]{Aaij:2017deq}%
  \BibitemOpen
  \bibfield  {author} {\bibinfo {author} {\bibfnamefont {R.}~\bibnamefont
  {Aaij}} \emph {et~al.} (\bibinfo {collaboration} {LHCb}),\ }\href {\doibase
  10.1103/PhysRevD.97.072013} {\bibfield  {journal} {\bibinfo  {journal} {Phys.
  Rev. D}\ }\textbf {\bibinfo {volume} {97}},\ \bibinfo {pages} {072013}
  (\bibinfo {year} {2018})},\ \Eprint {http://arxiv.org/abs/1711.02505}
  {arXiv:1711.02505 [hep-ex]} \BibitemShut {NoStop}%
\bibitem [{\citenamefont {Abdesselam}\ \emph {et~al.}(2019)\citenamefont
  {Abdesselam} \emph {et~al.}}]{Abdesselam:2019dgh}%
  \BibitemOpen
  \bibfield  {author} {\bibinfo {author} {\bibfnamefont {A.}~\bibnamefont
  {Abdesselam}} \emph {et~al.} (\bibinfo {collaboration} {Belle}),\ }\href@noop
  {} {\  (\bibinfo {year} {2019})},\ \Eprint {http://arxiv.org/abs/1904.08794}
  {arXiv:1904.08794 [hep-ex]} \BibitemShut {NoStop}%
\bibitem [{\citenamefont {Aaij}\ \emph {et~al.}(2017)\citenamefont {Aaij} \emph
  {et~al.}}]{Aaij:2017vbb}%
  \BibitemOpen
  \bibfield  {author} {\bibinfo {author} {\bibfnamefont {R.}~\bibnamefont
  {Aaij}} \emph {et~al.} (\bibinfo {collaboration} {LHCb}),\ }\href {\doibase
  10.1007/JHEP08(2017)055} {\bibfield  {journal} {\bibinfo  {journal} {JHEP}\
  }\textbf {\bibinfo {volume} {08}},\ \bibinfo {pages} {055} (\bibinfo {year}
  {2017})},\ \Eprint {http://arxiv.org/abs/1705.05802} {arXiv:1705.05802
  [hep-ex]} \BibitemShut {NoStop}%
\bibitem [{\citenamefont {Aaij}\ \emph {et~al.}(2019)\citenamefont {Aaij} \emph
  {et~al.}}]{Aaij:2019wad}%
  \BibitemOpen
  \bibfield  {author} {\bibinfo {author} {\bibfnamefont {R.}~\bibnamefont
  {Aaij}} \emph {et~al.} (\bibinfo {collaboration} {LHCb}),\ }\href {\doibase
  10.1103/PhysRevLett.122.191801} {\bibfield  {journal} {\bibinfo  {journal}
  {Phys. Rev. Lett.}\ }\textbf {\bibinfo {volume} {122}},\ \bibinfo {pages}
  {191801} (\bibinfo {year} {2019})},\ \Eprint
  {http://arxiv.org/abs/1903.09252} {arXiv:1903.09252 [hep-ex]} \BibitemShut
  {NoStop}%
\bibitem [{\citenamefont {Aaij}\ \emph {et~al.}(2021)\citenamefont {Aaij} \emph
  {et~al.}}]{LHCb:2021trn}%
  \BibitemOpen
  \bibfield  {author} {\bibinfo {author} {\bibfnamefont {R.}~\bibnamefont
  {Aaij}} \emph {et~al.} (\bibinfo {collaboration} {LHCb}),\ }\href@noop {} {\
  (\bibinfo {year} {2021})},\ \Eprint {http://arxiv.org/abs/2103.11769}
  {arXiv:2103.11769 [hep-ex]} \BibitemShut {NoStop}%
\bibitem [{\citenamefont {Amhis}\ \emph
  {et~al.}(2021{\natexlab{a}})\citenamefont {Amhis} \emph
  {et~al.}}]{Amhis:2019ckw}%
  \BibitemOpen
  \bibfield  {author} {\bibinfo {author} {\bibfnamefont {Y.~S.}\ \bibnamefont
  {Amhis}} \emph {et~al.} (\bibinfo {collaboration} {HFLAV}),\ }\href {\doibase
  10.1140/epjc/s10052-020-8156-7} {\bibfield  {journal} {\bibinfo  {journal}
  {Eur. Phys. J. C}\ }\textbf {\bibinfo {volume} {81}},\ \bibinfo {pages} {226}
  (\bibinfo {year} {2021}{\natexlab{a}})},\ \Eprint
  {http://arxiv.org/abs/1909.12524} {arXiv:1909.12524 [hep-ex]} \BibitemShut
  {NoStop}%
\bibitem [{\citenamefont {Murgui}\ \emph {et~al.}(2019)\citenamefont {Murgui},
  \citenamefont {Pe\~nuelas}, \citenamefont {Jung},\ and\ \citenamefont
  {Pich}}]{Murgui:2019czp}%
  \BibitemOpen
  \bibfield  {author} {\bibinfo {author} {\bibfnamefont {C.}~\bibnamefont
  {Murgui}}, \bibinfo {author} {\bibfnamefont {A.}~\bibnamefont {Pe\~nuelas}},
  \bibinfo {author} {\bibfnamefont {M.}~\bibnamefont {Jung}}, \ and\ \bibinfo
  {author} {\bibfnamefont {A.}~\bibnamefont {Pich}},\ }\href {\doibase
  10.1007/JHEP09(2019)103} {\bibfield  {journal} {\bibinfo  {journal} {JHEP}\
  }\textbf {\bibinfo {volume} {09}},\ \bibinfo {pages} {103} (\bibinfo {year}
  {2019})},\ \Eprint {http://arxiv.org/abs/1904.09311} {arXiv:1904.09311
  [hep-ph]} \BibitemShut {NoStop}%
\bibitem [{\citenamefont {Shi}\ \emph {et~al.}(2019)\citenamefont {Shi},
  \citenamefont {Geng}, \citenamefont {Grinstein}, \citenamefont {J\"ager},\
  and\ \citenamefont {Martin~Camalich}}]{Shi:2019gxi}%
  \BibitemOpen
  \bibfield  {author} {\bibinfo {author} {\bibfnamefont {R.-X.}\ \bibnamefont
  {Shi}}, \bibinfo {author} {\bibfnamefont {L.-S.}\ \bibnamefont {Geng}},
  \bibinfo {author} {\bibfnamefont {B.}~\bibnamefont {Grinstein}}, \bibinfo
  {author} {\bibfnamefont {S.}~\bibnamefont {J\"ager}}, \ and\ \bibinfo
  {author} {\bibfnamefont {J.}~\bibnamefont {Martin~Camalich}},\ }\href
  {\doibase 10.1007/JHEP12(2019)065} {\bibfield  {journal} {\bibinfo  {journal}
  {JHEP}\ }\textbf {\bibinfo {volume} {12}},\ \bibinfo {pages} {065} (\bibinfo
  {year} {2019})},\ \Eprint {http://arxiv.org/abs/1905.08498} {arXiv:1905.08498
  [hep-ph]} \BibitemShut {NoStop}%
\bibitem [{\citenamefont {Blanke}\ \emph {et~al.}(2019)\citenamefont {Blanke},
  \citenamefont {Crivellin}, \citenamefont {Kitahara}, \citenamefont {Moscati},
  \citenamefont {Nierste},\ and\ \citenamefont
  {Ni\v{s}and\v{z}i\'c}}]{Blanke:2019qrx}%
  \BibitemOpen
  \bibfield  {author} {\bibinfo {author} {\bibfnamefont {M.}~\bibnamefont
  {Blanke}}, \bibinfo {author} {\bibfnamefont {A.}~\bibnamefont {Crivellin}},
  \bibinfo {author} {\bibfnamefont {T.}~\bibnamefont {Kitahara}}, \bibinfo
  {author} {\bibfnamefont {M.}~\bibnamefont {Moscati}}, \bibinfo {author}
  {\bibfnamefont {U.}~\bibnamefont {Nierste}}, \ and\ \bibinfo {author}
  {\bibfnamefont {I.}~\bibnamefont {Ni\v{s}and\v{z}i\'c}},\ }\href {\doibase
  10.1103/PhysRevD.100.035035} {\  (\bibinfo {year} {2019}),\
  10.1103/PhysRevD.100.035035},\ \bibinfo {note} {[Addendum: Phys.Rev.D 100,
  035035 (2019)]},\ \Eprint {http://arxiv.org/abs/1905.08253} {arXiv:1905.08253
  [hep-ph]} \BibitemShut {NoStop}%
\bibitem [{\citenamefont {Kumbhakar}\ \emph {et~al.}(2020)\citenamefont
  {Kumbhakar}, \citenamefont {Alok}, \citenamefont {Kumar},\ and\ \citenamefont
  {Sankar}}]{Kumbhakar:2019avh}%
  \BibitemOpen
  \bibfield  {author} {\bibinfo {author} {\bibfnamefont {S.}~\bibnamefont
  {Kumbhakar}}, \bibinfo {author} {\bibfnamefont {A.~K.}\ \bibnamefont {Alok}},
  \bibinfo {author} {\bibfnamefont {D.}~\bibnamefont {Kumar}}, \ and\ \bibinfo
  {author} {\bibfnamefont {S.~U.}\ \bibnamefont {Sankar}},\ }\href {\doibase
  10.22323/1.364.0272} {\bibfield  {journal} {\bibinfo  {journal} {PoS}\
  }\textbf {\bibinfo {volume} {EPS-HEP2019}},\ \bibinfo {pages} {272} (\bibinfo
  {year} {2020})},\ \Eprint {http://arxiv.org/abs/1909.02840} {arXiv:1909.02840
  [hep-ph]} \BibitemShut {NoStop}%
\bibitem [{\citenamefont {Alguer\'o}\ \emph {et~al.}(2019)\citenamefont
  {Alguer\'o}, \citenamefont {Capdevila}, \citenamefont {Crivellin},
  \citenamefont {Descotes-Genon}, \citenamefont {Masjuan}, \citenamefont
  {Matias}, \citenamefont {Novoa~Brunet},\ and\ \citenamefont
  {Virto}}]{Alguero:2019ptt}%
  \BibitemOpen
  \bibfield  {author} {\bibinfo {author} {\bibfnamefont {M.}~\bibnamefont
  {Alguer\'o}}, \bibinfo {author} {\bibfnamefont {B.}~\bibnamefont
  {Capdevila}}, \bibinfo {author} {\bibfnamefont {A.}~\bibnamefont
  {Crivellin}}, \bibinfo {author} {\bibfnamefont {S.}~\bibnamefont
  {Descotes-Genon}}, \bibinfo {author} {\bibfnamefont {P.}~\bibnamefont
  {Masjuan}}, \bibinfo {author} {\bibfnamefont {J.}~\bibnamefont {Matias}},
  \bibinfo {author} {\bibfnamefont {M.}~\bibnamefont {Novoa~Brunet}}, \ and\
  \bibinfo {author} {\bibfnamefont {J.}~\bibnamefont {Virto}},\ }\href
  {\doibase 10.1140/epjc/s10052-019-7216-3} {\bibfield  {journal} {\bibinfo
  {journal} {Eur. Phys. J. C}\ }\textbf {\bibinfo {volume} {79}},\ \bibinfo
  {pages} {714} (\bibinfo {year} {2019})},\ \bibinfo {note} {[Addendum:
  Eur.Phys.J.C 80, 511 (2020)]},\ \Eprint {http://arxiv.org/abs/1903.09578}
  {arXiv:1903.09578 [hep-ph]} \BibitemShut {NoStop}%
\bibitem [{\citenamefont {Aebischer}\ \emph {et~al.}(2020)\citenamefont
  {Aebischer}, \citenamefont {Altmannshofer}, \citenamefont {Guadagnoli},
  \citenamefont {Reboud}, \citenamefont {Stangl},\ and\ \citenamefont
  {Straub}}]{Aebischer:2019mlg}%
  \BibitemOpen
  \bibfield  {author} {\bibinfo {author} {\bibfnamefont {J.}~\bibnamefont
  {Aebischer}}, \bibinfo {author} {\bibfnamefont {W.}~\bibnamefont
  {Altmannshofer}}, \bibinfo {author} {\bibfnamefont {D.}~\bibnamefont
  {Guadagnoli}}, \bibinfo {author} {\bibfnamefont {M.}~\bibnamefont {Reboud}},
  \bibinfo {author} {\bibfnamefont {P.}~\bibnamefont {Stangl}}, \ and\ \bibinfo
  {author} {\bibfnamefont {D.~M.}\ \bibnamefont {Straub}},\ }\href {\doibase
  10.1140/epjc/s10052-020-7817-x} {\bibfield  {journal} {\bibinfo  {journal}
  {Eur. Phys. J. C}\ }\textbf {\bibinfo {volume} {80}},\ \bibinfo {pages} {252}
  (\bibinfo {year} {2020})},\ \Eprint {http://arxiv.org/abs/1903.10434}
  {arXiv:1903.10434 [hep-ph]} \BibitemShut {NoStop}%
\bibitem [{\citenamefont {Ciuchini}\ \emph {et~al.}(2019)\citenamefont
  {Ciuchini}, \citenamefont {Coutinho}, \citenamefont {Fedele}, \citenamefont
  {Franco}, \citenamefont {Paul}, \citenamefont {Silvestrini},\ and\
  \citenamefont {Valli}}]{Ciuchini:2019usw}%
  \BibitemOpen
  \bibfield  {author} {\bibinfo {author} {\bibfnamefont {M.}~\bibnamefont
  {Ciuchini}}, \bibinfo {author} {\bibfnamefont {A.~M.}\ \bibnamefont
  {Coutinho}}, \bibinfo {author} {\bibfnamefont {M.}~\bibnamefont {Fedele}},
  \bibinfo {author} {\bibfnamefont {E.}~\bibnamefont {Franco}}, \bibinfo
  {author} {\bibfnamefont {A.}~\bibnamefont {Paul}}, \bibinfo {author}
  {\bibfnamefont {L.}~\bibnamefont {Silvestrini}}, \ and\ \bibinfo {author}
  {\bibfnamefont {M.}~\bibnamefont {Valli}},\ }\href {\doibase
  10.1140/epjc/s10052-019-7210-9} {\bibfield  {journal} {\bibinfo  {journal}
  {Eur. Phys. J. C}\ }\textbf {\bibinfo {volume} {79}},\ \bibinfo {pages} {719}
  (\bibinfo {year} {2019})},\ \Eprint {http://arxiv.org/abs/1903.09632}
  {arXiv:1903.09632 [hep-ph]} \BibitemShut {NoStop}%
\bibitem [{\citenamefont {Arbey}\ \emph {et~al.}(2019)\citenamefont {Arbey},
  \citenamefont {Hurth}, \citenamefont {Mahmoudi}, \citenamefont {Santos},\
  and\ \citenamefont {Neshatpour}}]{Arbey:2019duh}%
  \BibitemOpen
  \bibfield  {author} {\bibinfo {author} {\bibfnamefont {A.}~\bibnamefont
  {Arbey}}, \bibinfo {author} {\bibfnamefont {T.}~\bibnamefont {Hurth}},
  \bibinfo {author} {\bibfnamefont {F.}~\bibnamefont {Mahmoudi}}, \bibinfo
  {author} {\bibfnamefont {D.~M.}\ \bibnamefont {Santos}}, \ and\ \bibinfo
  {author} {\bibfnamefont {S.}~\bibnamefont {Neshatpour}},\ }\href {\doibase
  10.1103/PhysRevD.100.015045} {\bibfield  {journal} {\bibinfo  {journal}
  {Phys. Rev. D}\ }\textbf {\bibinfo {volume} {100}},\ \bibinfo {pages}
  {015045} (\bibinfo {year} {2019})},\ \Eprint
  {http://arxiv.org/abs/1904.08399} {arXiv:1904.08399 [hep-ph]} \BibitemShut
  {NoStop}%
\bibitem [{\citenamefont {Feruglio}\ \emph {et~al.}(2017)\citenamefont
  {Feruglio}, \citenamefont {Paradisi},\ and\ \citenamefont
  {Pattori}}]{Feruglio:2016gvd}%
  \BibitemOpen
  \bibfield  {author} {\bibinfo {author} {\bibfnamefont {F.}~\bibnamefont
  {Feruglio}}, \bibinfo {author} {\bibfnamefont {P.}~\bibnamefont {Paradisi}},
  \ and\ \bibinfo {author} {\bibfnamefont {A.}~\bibnamefont {Pattori}},\ }\href
  {\doibase 10.1103/PhysRevLett.118.011801} {\bibfield  {journal} {\bibinfo
  {journal} {Phys. Rev. Lett.}\ }\textbf {\bibinfo {volume} {118}},\ \bibinfo
  {pages} {011801} (\bibinfo {year} {2017})},\ \Eprint
  {http://arxiv.org/abs/1606.00524} {arXiv:1606.00524 [hep-ph]} \BibitemShut
  {NoStop}%
\bibitem [{\citenamefont {Bennett}\ \emph {et~al.}(2006)\citenamefont {Bennett}
  \emph {et~al.}}]{Bennett:2006fi}%
  \BibitemOpen
  \bibfield  {author} {\bibinfo {author} {\bibfnamefont {G.~W.}\ \bibnamefont
  {Bennett}} \emph {et~al.} (\bibinfo {collaboration} {Muon g-2}),\ }\href
  {\doibase 10.1103/PhysRevD.73.072003} {\bibfield  {journal} {\bibinfo
  {journal} {Phys. Rev. D}\ }\textbf {\bibinfo {volume} {73}},\ \bibinfo
  {pages} {072003} (\bibinfo {year} {2006})},\ \Eprint
  {http://arxiv.org/abs/hep-ex/0602035} {arXiv:hep-ex/0602035} \BibitemShut
  {NoStop}%
\bibitem [{\citenamefont {Abi}\ \emph {et~al.}(2021)\citenamefont {Abi} \emph
  {et~al.}}]{Muong-2:2021ojo}%
  \BibitemOpen
  \bibfield  {author} {\bibinfo {author} {\bibfnamefont {B.}~\bibnamefont
  {Abi}} \emph {et~al.} (\bibinfo {collaboration} {Muon g-2}),\ }\href
  {\doibase 10.1103/PhysRevLett.126.141801} {\bibfield  {journal} {\bibinfo
  {journal} {Phys. Rev. Lett.}\ }\textbf {\bibinfo {volume} {126}},\ \bibinfo
  {pages} {141801} (\bibinfo {year} {2021})},\ \Eprint
  {http://arxiv.org/abs/2104.03281} {arXiv:2104.03281 [hep-ex]} \BibitemShut
  {NoStop}%
\bibitem [{\citenamefont {Aoyama}\ \emph {et~al.}(2020)\citenamefont {Aoyama}
  \emph {et~al.}}]{Aoyama:2020ynm}%
  \BibitemOpen
  \bibfield  {author} {\bibinfo {author} {\bibfnamefont {T.}~\bibnamefont
  {Aoyama}} \emph {et~al.},\ }\href {\doibase 10.1016/j.physrep.2020.07.006}
  {\bibfield  {journal} {\bibinfo  {journal} {Phys. Rept.}\ }\textbf {\bibinfo
  {volume} {887}},\ \bibinfo {pages} {1} (\bibinfo {year} {2020})},\ \Eprint
  {http://arxiv.org/abs/2006.04822} {arXiv:2006.04822 [hep-ph]} \BibitemShut
  {NoStop}%
\bibitem [{\citenamefont {Bobeth}\ \emph {et~al.}(2021)\citenamefont {Bobeth},
  \citenamefont {van Dyk}, \citenamefont {Bordone}, \citenamefont {Jung},\ and\
  \citenamefont {Gubernari}}]{Bobeth:2021lya}%
  \BibitemOpen
  \bibfield  {author} {\bibinfo {author} {\bibfnamefont {C.}~\bibnamefont
  {Bobeth}}, \bibinfo {author} {\bibfnamefont {D.}~\bibnamefont {van Dyk}},
  \bibinfo {author} {\bibfnamefont {M.}~\bibnamefont {Bordone}}, \bibinfo
  {author} {\bibfnamefont {M.}~\bibnamefont {Jung}}, \ and\ \bibinfo {author}
  {\bibfnamefont {N.}~\bibnamefont {Gubernari}},\ }\href@noop {} {\  (\bibinfo
  {year} {2021})},\ \Eprint {http://arxiv.org/abs/2104.02094} {arXiv:2104.02094
  [hep-ph]} \BibitemShut {NoStop}%
\bibitem [{\citenamefont {Carvunis}\ \emph {et~al.}(2021)\citenamefont
  {Carvunis}, \citenamefont {Crivellin}, \citenamefont {Guadagnoli},\ and\
  \citenamefont {Gangal}}]{Carvunis:2021dss}%
  \BibitemOpen
  \bibfield  {author} {\bibinfo {author} {\bibfnamefont {A.}~\bibnamefont
  {Carvunis}}, \bibinfo {author} {\bibfnamefont {A.}~\bibnamefont {Crivellin}},
  \bibinfo {author} {\bibfnamefont {D.}~\bibnamefont {Guadagnoli}}, \ and\
  \bibinfo {author} {\bibfnamefont {S.}~\bibnamefont {Gangal}},\ }\href@noop {}
  {\  (\bibinfo {year} {2021})},\ \Eprint {http://arxiv.org/abs/2106.09610}
  {arXiv:2106.09610 [hep-ph]} \BibitemShut {NoStop}%
\bibitem [{\citenamefont {Sirunyan}\ \emph {et~al.}(2021)\citenamefont
  {Sirunyan} \emph {et~al.}}]{Sirunyan:2021khd}%
  \BibitemOpen
  \bibfield  {author} {\bibinfo {author} {\bibfnamefont {A.~M.}\ \bibnamefont
  {Sirunyan}} \emph {et~al.} (\bibinfo {collaboration} {CMS}),\ }\href
  {\doibase 10.1007/JHEP07(2021)208} {\bibfield  {journal} {\bibinfo  {journal}
  {JHEP}\ }\textbf {\bibinfo {volume} {07}},\ \bibinfo {pages} {208} (\bibinfo
  {year} {2021})},\ \Eprint {http://arxiv.org/abs/2103.02708} {arXiv:2103.02708
  [hep-ex]} \BibitemShut {NoStop}%
\bibitem [{\citenamefont {Coutinho}\ \emph {et~al.}(2020)\citenamefont
  {Coutinho}, \citenamefont {Crivellin},\ and\ \citenamefont
  {Manzari}}]{Coutinho:2019aiy}%
  \BibitemOpen
  \bibfield  {author} {\bibinfo {author} {\bibfnamefont {A.~M.}\ \bibnamefont
  {Coutinho}}, \bibinfo {author} {\bibfnamefont {A.}~\bibnamefont {Crivellin}},
  \ and\ \bibinfo {author} {\bibfnamefont {C.~A.}\ \bibnamefont {Manzari}},\
  }\href {\doibase 10.1103/PhysRevLett.125.071802} {\bibfield  {journal}
  {\bibinfo  {journal} {Phys. Rev. Lett.}\ }\textbf {\bibinfo {volume} {125}},\
  \bibinfo {pages} {071802} (\bibinfo {year} {2020})},\ \Eprint
  {http://arxiv.org/abs/1912.08823} {arXiv:1912.08823 [hep-ph]} \BibitemShut
  {NoStop}%
\bibitem [{\citenamefont {Crivellin}\ and\ \citenamefont
  {Hoferichter}(2020)}]{Crivellin:2020lzu}%
  \BibitemOpen
  \bibfield  {author} {\bibinfo {author} {\bibfnamefont {A.}~\bibnamefont
  {Crivellin}}\ and\ \bibinfo {author} {\bibfnamefont {M.}~\bibnamefont
  {Hoferichter}},\ }\href {\doibase 10.1103/PhysRevLett.125.111801} {\bibfield
  {journal} {\bibinfo  {journal} {Phys. Rev. Lett.}\ }\textbf {\bibinfo
  {volume} {125}},\ \bibinfo {pages} {111801} (\bibinfo {year} {2020})},\
  \Eprint {http://arxiv.org/abs/2002.07184} {arXiv:2002.07184 [hep-ph]}
  \BibitemShut {NoStop}%
\bibitem [{\citenamefont {Crivellin}\ \emph {et~al.}(2020)\citenamefont
  {Crivellin}, \citenamefont {Kirk}, \citenamefont {Manzari},\ and\
  \citenamefont {Montull}}]{Crivellin:2020ebi}%
  \BibitemOpen
  \bibfield  {author} {\bibinfo {author} {\bibfnamefont {A.}~\bibnamefont
  {Crivellin}}, \bibinfo {author} {\bibfnamefont {F.}~\bibnamefont {Kirk}},
  \bibinfo {author} {\bibfnamefont {C.~A.}\ \bibnamefont {Manzari}}, \ and\
  \bibinfo {author} {\bibfnamefont {M.}~\bibnamefont {Montull}},\ }\href
  {\doibase 10.1007/JHEP12(2020)166} {\bibfield  {journal} {\bibinfo  {journal}
  {JHEP}\ }\textbf {\bibinfo {volume} {12}},\ \bibinfo {pages} {166} (\bibinfo
  {year} {2020})},\ \Eprint {http://arxiv.org/abs/2008.01113} {arXiv:2008.01113
  [hep-ph]} \BibitemShut {NoStop}%
\bibitem [{\citenamefont {Capdevila}\ \emph {et~al.}(2021)\citenamefont
  {Capdevila}, \citenamefont {Crivellin}, \citenamefont {Manzari},\ and\
  \citenamefont {Montull}}]{Capdevila:2020rrl}%
  \BibitemOpen
  \bibfield  {author} {\bibinfo {author} {\bibfnamefont {B.}~\bibnamefont
  {Capdevila}}, \bibinfo {author} {\bibfnamefont {A.}~\bibnamefont
  {Crivellin}}, \bibinfo {author} {\bibfnamefont {C.~A.}\ \bibnamefont
  {Manzari}}, \ and\ \bibinfo {author} {\bibfnamefont {M.}~\bibnamefont
  {Montull}},\ }\href {\doibase 10.1103/PhysRevD.103.015032} {\bibfield
  {journal} {\bibinfo  {journal} {Phys. Rev. D}\ }\textbf {\bibinfo {volume}
  {103}},\ \bibinfo {pages} {015032} (\bibinfo {year} {2021})},\ \Eprint
  {http://arxiv.org/abs/2005.13542} {arXiv:2005.13542 [hep-ph]} \BibitemShut
  {NoStop}%
\bibitem [{\citenamefont {Crivellin}\ \emph
  {et~al.}(2021{\natexlab{b}})\citenamefont {Crivellin}, \citenamefont
  {Manzari},\ and\ \citenamefont {Montull}}]{Crivellin:2021rbf}%
  \BibitemOpen
  \bibfield  {author} {\bibinfo {author} {\bibfnamefont {A.}~\bibnamefont
  {Crivellin}}, \bibinfo {author} {\bibfnamefont {C.~A.}\ \bibnamefont
  {Manzari}}, \ and\ \bibinfo {author} {\bibfnamefont {M.}~\bibnamefont
  {Montull}},\ }\href@noop {} {\  (\bibinfo {year} {2021}{\natexlab{b}})},\
  \Eprint {http://arxiv.org/abs/2103.12003} {arXiv:2103.12003 [hep-ph]}
  \BibitemShut {NoStop}%
\bibitem [{\citenamefont {Bryman}\ and\ \citenamefont
  {Shrock}(2019)}]{Bryman:2019bjg}%
  \BibitemOpen
  \bibfield  {author} {\bibinfo {author} {\bibfnamefont {D.~A.}\ \bibnamefont
  {Bryman}}\ and\ \bibinfo {author} {\bibfnamefont {R.}~\bibnamefont
  {Shrock}},\ }\href {\doibase 10.1103/PhysRevD.100.073011} {\bibfield
  {journal} {\bibinfo  {journal} {Phys. Rev. D}\ }\textbf {\bibinfo {volume}
  {100}},\ \bibinfo {pages} {073011} (\bibinfo {year} {2019})},\ \Eprint
  {http://arxiv.org/abs/1909.11198} {arXiv:1909.11198 [hep-ph]} \BibitemShut
  {NoStop}%
\bibitem [{\citenamefont {de~Gouv\^ea}\ and\ \citenamefont
  {Kobach}(2016)}]{deGouvea:2015euy}%
  \BibitemOpen
  \bibfield  {author} {\bibinfo {author} {\bibfnamefont {A.}~\bibnamefont
  {de~Gouv\^ea}}\ and\ \bibinfo {author} {\bibfnamefont {A.}~\bibnamefont
  {Kobach}},\ }\href {\doibase 10.1103/PhysRevD.93.033005} {\bibfield
  {journal} {\bibinfo  {journal} {Phys. Rev. D}\ }\textbf {\bibinfo {volume}
  {93}},\ \bibinfo {pages} {033005} (\bibinfo {year} {2016})},\ \Eprint
  {http://arxiv.org/abs/1511.00683} {arXiv:1511.00683 [hep-ph]} \BibitemShut
  {NoStop}%
\bibitem [{\citenamefont {Cortina~Gil}\ \emph
  {et~al.}(2020{\natexlab{a}})\citenamefont {Cortina~Gil} \emph
  {et~al.}}]{NA62:2020mcv}%
  \BibitemOpen
  \bibfield  {author} {\bibinfo {author} {\bibfnamefont {E.}~\bibnamefont
  {Cortina~Gil}} \emph {et~al.} (\bibinfo {collaboration} {NA62}),\ }\href
  {\doibase 10.1016/j.physletb.2020.135599} {\bibfield  {journal} {\bibinfo
  {journal} {Phys. Lett. B}\ }\textbf {\bibinfo {volume} {807}},\ \bibinfo
  {pages} {135599} (\bibinfo {year} {2020}{\natexlab{a}})},\ \Eprint
  {http://arxiv.org/abs/2005.09575} {arXiv:2005.09575 [hep-ex]} \BibitemShut
  {NoStop}%
\bibitem [{\citenamefont {Cortina~Gil}\ \emph {et~al.}(2021)\citenamefont
  {Cortina~Gil} \emph {et~al.}}]{NA62:2021bji}%
  \BibitemOpen
  \bibfield  {author} {\bibinfo {author} {\bibfnamefont {E.}~\bibnamefont
  {Cortina~Gil}} \emph {et~al.} (\bibinfo {collaboration} {NA62}),\ }\href
  {\doibase 10.1016/j.physletb.2021.136259} {\bibfield  {journal} {\bibinfo
  {journal} {Phys. Lett. B}\ }\textbf {\bibinfo {volume} {816}},\ \bibinfo
  {pages} {136259} (\bibinfo {year} {2021})},\ \Eprint
  {http://arxiv.org/abs/2101.12304} {arXiv:2101.12304 [hep-ex]} \BibitemShut
  {NoStop}%
\bibitem [{\citenamefont {Lessa}\ and\ \citenamefont
  {Peres}(2007)}]{Lessa:2007up}%
  \BibitemOpen
  \bibfield  {author} {\bibinfo {author} {\bibfnamefont {A.~P.}\ \bibnamefont
  {Lessa}}\ and\ \bibinfo {author} {\bibfnamefont {O.~L.~G.}\ \bibnamefont
  {Peres}},\ }\href {\doibase 10.1103/PhysRevD.75.094001} {\bibfield  {journal}
  {\bibinfo  {journal} {Phys. Rev. D}\ }\textbf {\bibinfo {volume} {75}},\
  \bibinfo {pages} {094001} (\bibinfo {year} {2007})},\ \Eprint
  {http://arxiv.org/abs/hep-ph/0701068} {arXiv:hep-ph/0701068} \BibitemShut
  {NoStop}%
\bibitem [{\citenamefont {Altmannshofer}\ \emph {et~al.}(2020)\citenamefont
  {Altmannshofer}, \citenamefont {Gori},\ and\ \citenamefont
  {Robinson}}]{Altmannshofer:2019yji}%
  \BibitemOpen
  \bibfield  {author} {\bibinfo {author} {\bibfnamefont {W.}~\bibnamefont
  {Altmannshofer}}, \bibinfo {author} {\bibfnamefont {S.}~\bibnamefont {Gori}},
  \ and\ \bibinfo {author} {\bibfnamefont {D.~J.}\ \bibnamefont {Robinson}},\
  }\href {\doibase 10.1103/PhysRevD.101.075002} {\bibfield  {journal} {\bibinfo
   {journal} {Phys. Rev. D}\ }\textbf {\bibinfo {volume} {101}},\ \bibinfo
  {pages} {075002} (\bibinfo {year} {2020})},\ \Eprint
  {http://arxiv.org/abs/1909.00005} {arXiv:1909.00005 [hep-ph]} \BibitemShut
  {NoStop}%
\bibitem [{\citenamefont {Elahi}\ \emph {et~al.}(2021)\citenamefont {Elahi},
  \citenamefont {Elor},\ and\ \citenamefont {McGehee}}]{Elahi:2021jia}%
  \BibitemOpen
  \bibfield  {author} {\bibinfo {author} {\bibfnamefont {F.}~\bibnamefont
  {Elahi}}, \bibinfo {author} {\bibfnamefont {G.}~\bibnamefont {Elor}}, \ and\
  \bibinfo {author} {\bibfnamefont {R.}~\bibnamefont {McGehee}},\ }\href@noop
  {} {\  (\bibinfo {year} {2021})},\ \Eprint {http://arxiv.org/abs/2109.09751}
  {arXiv:2109.09751 [hep-ph]} \BibitemShut {NoStop}%
\bibitem [{\citenamefont {Aguilar-Arevalo}\ \emph {et~al.}(2018)\citenamefont
  {Aguilar-Arevalo} \emph {et~al.}}]{Aguilar-Arevalo:2017vlf}%
  \BibitemOpen
  \bibfield  {author} {\bibinfo {author} {\bibfnamefont {A.}~\bibnamefont
  {Aguilar-Arevalo}} \emph {et~al.} (\bibinfo {collaboration} {PIENU}),\ }\href
  {\doibase 10.1103/PhysRevD.97.072012} {\bibfield  {journal} {\bibinfo
  {journal} {Phys. Rev. D}\ }\textbf {\bibinfo {volume} {97}},\ \bibinfo
  {pages} {072012} (\bibinfo {year} {2018})},\ \Eprint
  {http://arxiv.org/abs/1712.03275} {arXiv:1712.03275 [hep-ex]} \BibitemShut
  {NoStop}%
\bibitem [{\citenamefont {Aguilar-Arevalo}\ \emph {et~al.}(2019)\citenamefont
  {Aguilar-Arevalo} \emph {et~al.}}]{Aguilar-Arevalo:2019owf}%
  \BibitemOpen
  \bibfield  {author} {\bibinfo {author} {\bibfnamefont {A.}~\bibnamefont
  {Aguilar-Arevalo}} \emph {et~al.} (\bibinfo {collaboration} {PIENU}),\ }\href
  {\doibase 10.1016/j.physletb.2019.134980} {\bibfield  {journal} {\bibinfo
  {journal} {Phys. Lett. B}\ }\textbf {\bibinfo {volume} {798}},\ \bibinfo
  {pages} {134980} (\bibinfo {year} {2019})},\ \Eprint
  {http://arxiv.org/abs/1904.03269} {arXiv:1904.03269 [hep-ex]} \BibitemShut
  {NoStop}%
\bibitem [{\citenamefont {Bryman}\ \emph {et~al.}(2021)\citenamefont {Bryman},
  \citenamefont {Ito},\ and\ \citenamefont {Shrock}}]{Bryman:2021ilc}%
  \BibitemOpen
  \bibfield  {author} {\bibinfo {author} {\bibfnamefont {D.~A.}\ \bibnamefont
  {Bryman}}, \bibinfo {author} {\bibfnamefont {S.}~\bibnamefont {Ito}}, \ and\
  \bibinfo {author} {\bibfnamefont {R.}~\bibnamefont {Shrock}},\ }\href
  {\doibase 10.1103/PhysRevD.104.075032} {\bibfield  {journal} {\bibinfo
  {journal} {Phys. Rev. D}\ }\textbf {\bibinfo {volume} {104}},\ \bibinfo
  {pages} {075032} (\bibinfo {year} {2021})},\ \Eprint
  {http://arxiv.org/abs/2106.02451} {arXiv:2106.02451 [hep-ph]} \BibitemShut
  {NoStop}%
\bibitem [{\citenamefont {Marciano}\ and\ \citenamefont
  {Sirlin}(1993)}]{Marciano:1993sh}%
  \BibitemOpen
  \bibfield  {author} {\bibinfo {author} {\bibfnamefont {W.~J.}\ \bibnamefont
  {Marciano}}\ and\ \bibinfo {author} {\bibfnamefont {A.}~\bibnamefont
  {Sirlin}},\ }\href {\doibase 10.1103/PhysRevLett.71.3629} {\bibfield
  {journal} {\bibinfo  {journal} {Phys. Rev. Lett.}\ }\textbf {\bibinfo
  {volume} {71}},\ \bibinfo {pages} {3629} (\bibinfo {year}
  {1993})}\BibitemShut {NoStop}%
\bibitem [{\citenamefont {Finkemeier}(1996)}]{Finkemeier:1995gi}%
  \BibitemOpen
  \bibfield  {author} {\bibinfo {author} {\bibfnamefont {M.}~\bibnamefont
  {Finkemeier}},\ }\href {\doibase 10.1016/0370-2693(96)01030-1} {\bibfield
  {journal} {\bibinfo  {journal} {Phys. Lett. B}\ }\textbf {\bibinfo {volume}
  {387}},\ \bibinfo {pages} {391} (\bibinfo {year} {1996})},\ \Eprint
  {http://arxiv.org/abs/hep-ph/9505434} {arXiv:hep-ph/9505434} \BibitemShut
  {NoStop}%
\bibitem [{\citenamefont {Cirigliano}\ and\ \citenamefont
  {Rosell}(2007{\natexlab{a}})}]{Cirigliano:2007xi}%
  \BibitemOpen
  \bibfield  {author} {\bibinfo {author} {\bibfnamefont {V.}~\bibnamefont
  {Cirigliano}}\ and\ \bibinfo {author} {\bibfnamefont {I.}~\bibnamefont
  {Rosell}},\ }\href {\doibase 10.1103/PhysRevLett.99.231801} {\bibfield
  {journal} {\bibinfo  {journal} {Phys. Rev. Lett.}\ }\textbf {\bibinfo
  {volume} {99}},\ \bibinfo {pages} {231801} (\bibinfo {year}
  {2007}{\natexlab{a}})},\ \Eprint {http://arxiv.org/abs/0707.3439}
  {arXiv:0707.3439 [hep-ph]} \BibitemShut {NoStop}%
\bibitem [{\citenamefont {Cirigliano}\ and\ \citenamefont
  {Rosell}(2007{\natexlab{b}})}]{Cirigliano:2007ga}%
  \BibitemOpen
  \bibfield  {author} {\bibinfo {author} {\bibfnamefont {V.}~\bibnamefont
  {Cirigliano}}\ and\ \bibinfo {author} {\bibfnamefont {I.}~\bibnamefont
  {Rosell}},\ }\href {\doibase 10.1088/1126-6708/2007/10/005} {\bibfield
  {journal} {\bibinfo  {journal} {JHEP}\ }\textbf {\bibinfo {volume} {10}},\
  \bibinfo {pages} {005} (\bibinfo {year} {2007}{\natexlab{b}})},\ \Eprint
  {http://arxiv.org/abs/0707.4464} {arXiv:0707.4464 [hep-ph]} \BibitemShut
  {NoStop}%
\bibitem [{\citenamefont {Campbell}\ and\ \citenamefont
  {Maybury}(2005)}]{Campbell:2003ir}%
  \BibitemOpen
  \bibfield  {author} {\bibinfo {author} {\bibfnamefont {B.~A.}\ \bibnamefont
  {Campbell}}\ and\ \bibinfo {author} {\bibfnamefont {D.~W.}\ \bibnamefont
  {Maybury}},\ }\href {\doibase 10.1016/j.nuclphysb.2004.12.015} {\bibfield
  {journal} {\bibinfo  {journal} {Nucl. Phys. B}\ }\textbf {\bibinfo {volume}
  {709}},\ \bibinfo {pages} {419} (\bibinfo {year} {2005})},\ \Eprint
  {http://arxiv.org/abs/hep-ph/0303046} {arXiv:hep-ph/0303046} \BibitemShut
  {NoStop}%
\bibitem [{\citenamefont {Weinberg}(1979{\natexlab{a}})}]{Weinberg:1978kz}%
  \BibitemOpen
  \bibfield  {author} {\bibinfo {author} {\bibfnamefont {S.}~\bibnamefont
  {Weinberg}},\ }\href {\doibase 10.1016/0378-4371(79)90223-1} {\bibfield
  {journal} {\bibinfo  {journal} {Physica A}\ }\textbf {\bibinfo {volume}
  {96}},\ \bibinfo {pages} {327} (\bibinfo {year}
  {1979}{\natexlab{a}})}\BibitemShut {NoStop}%
\bibitem [{\citenamefont {Gasser}\ and\ \citenamefont
  {Leutwyler}(1984)}]{Gasser:1983yg}%
  \BibitemOpen
  \bibfield  {author} {\bibinfo {author} {\bibfnamefont {J.}~\bibnamefont
  {Gasser}}\ and\ \bibinfo {author} {\bibfnamefont {H.}~\bibnamefont
  {Leutwyler}},\ }\href {\doibase 10.1016/0003-4916(84)90242-2} {\bibfield
  {journal} {\bibinfo  {journal} {Annals Phys.}\ }\textbf {\bibinfo {volume}
  {158}},\ \bibinfo {pages} {142} (\bibinfo {year} {1984})}\BibitemShut
  {NoStop}%
\bibitem [{\citenamefont {Gasser}\ and\ \citenamefont
  {Leutwyler}(1985)}]{Gasser:1984gg}%
  \BibitemOpen
  \bibfield  {author} {\bibinfo {author} {\bibfnamefont {J.}~\bibnamefont
  {Gasser}}\ and\ \bibinfo {author} {\bibfnamefont {H.}~\bibnamefont
  {Leutwyler}},\ }\href {\doibase 10.1016/0550-3213(85)90492-4} {\bibfield
  {journal} {\bibinfo  {journal} {Nucl. Phys. B}\ }\textbf {\bibinfo {volume}
  {250}},\ \bibinfo {pages} {465} (\bibinfo {year} {1985})}\BibitemShut
  {NoStop}%
\bibitem [{\citenamefont {Knecht}\ \emph {et~al.}(2000)\citenamefont {Knecht},
  \citenamefont {Neufeld}, \citenamefont {Rupertsberger},\ and\ \citenamefont
  {Talavera}}]{Knecht:1999ag}%
  \BibitemOpen
  \bibfield  {author} {\bibinfo {author} {\bibfnamefont {M.}~\bibnamefont
  {Knecht}}, \bibinfo {author} {\bibfnamefont {H.}~\bibnamefont {Neufeld}},
  \bibinfo {author} {\bibfnamefont {H.}~\bibnamefont {Rupertsberger}}, \ and\
  \bibinfo {author} {\bibfnamefont {P.}~\bibnamefont {Talavera}},\ }\href
  {\doibase 10.1007/s100529900265} {\bibfield  {journal} {\bibinfo  {journal}
  {Eur. Phys. J. C}\ }\textbf {\bibinfo {volume} {12}},\ \bibinfo {pages} {469}
  (\bibinfo {year} {2000})},\ \Eprint {http://arxiv.org/abs/hep-ph/9909284}
  {arXiv:hep-ph/9909284} \BibitemShut {NoStop}%
\bibitem [{\citenamefont {Kinoshita}(1959)}]{Kinoshita:1959ha}%
  \BibitemOpen
  \bibfield  {author} {\bibinfo {author} {\bibfnamefont {T.}~\bibnamefont
  {Kinoshita}},\ }\href {\doibase 10.1103/PhysRevLett.2.477} {\bibfield
  {journal} {\bibinfo  {journal} {Phys. Rev. Lett.}\ }\textbf {\bibinfo
  {volume} {2}},\ \bibinfo {pages} {477} (\bibinfo {year} {1959})}\BibitemShut
  {NoStop}%
\bibitem [{\citenamefont {Bijnens}\ \emph {et~al.}(1993)\citenamefont
  {Bijnens}, \citenamefont {Ecker},\ and\ \citenamefont
  {Gasser}}]{Bijnens:1992en}%
  \BibitemOpen
  \bibfield  {author} {\bibinfo {author} {\bibfnamefont {J.}~\bibnamefont
  {Bijnens}}, \bibinfo {author} {\bibfnamefont {G.}~\bibnamefont {Ecker}}, \
  and\ \bibinfo {author} {\bibfnamefont {J.}~\bibnamefont {Gasser}},\ }\href
  {\doibase 10.1016/0550-3213(93)90259-R} {\bibfield  {journal} {\bibinfo
  {journal} {Nucl. Phys. B}\ }\textbf {\bibinfo {volume} {396}},\ \bibinfo
  {pages} {81} (\bibinfo {year} {1993})},\ \Eprint
  {http://arxiv.org/abs/hep-ph/9209261} {arXiv:hep-ph/9209261} \BibitemShut
  {NoStop}%
\bibitem [{\citenamefont {Bryman}\ \emph {et~al.}(2011)\citenamefont {Bryman},
  \citenamefont {Marciano}, \citenamefont {Tschirhart},\ and\ \citenamefont
  {Yamanaka}}]{Bryman:2011zz}%
  \BibitemOpen
  \bibfield  {author} {\bibinfo {author} {\bibfnamefont {D.}~\bibnamefont
  {Bryman}}, \bibinfo {author} {\bibfnamefont {W.~J.}\ \bibnamefont
  {Marciano}}, \bibinfo {author} {\bibfnamefont {R.}~\bibnamefont
  {Tschirhart}}, \ and\ \bibinfo {author} {\bibfnamefont {T.}~\bibnamefont
  {Yamanaka}},\ }\href {\doibase 10.1146/annurev-nucl-102010-130431} {\bibfield
   {journal} {\bibinfo  {journal} {Ann. Rev. Nucl. Part. Sci.}\ }\textbf
  {\bibinfo {volume} {61}},\ \bibinfo {pages} {331} (\bibinfo {year}
  {2011})}\BibitemShut {NoStop}%
\bibitem [{\citenamefont {Aguilar-Arevalo}\ \emph
  {et~al.}(2015{\natexlab{a}})\citenamefont {Aguilar-Arevalo} \emph
  {et~al.}}]{Aguilar-Arevalo:2015cdf}%
  \BibitemOpen
  \bibfield  {author} {\bibinfo {author} {\bibfnamefont {A.}~\bibnamefont
  {Aguilar-Arevalo}} \emph {et~al.} (\bibinfo {collaboration} {PiENu}),\ }\href
  {\doibase 10.1103/PhysRevLett.115.071801} {\bibfield  {journal} {\bibinfo
  {journal} {Phys. Rev. Lett.}\ }\textbf {\bibinfo {volume} {115}},\ \bibinfo
  {pages} {071801} (\bibinfo {year} {2015}{\natexlab{a}})},\ \Eprint
  {http://arxiv.org/abs/1506.05845} {arXiv:1506.05845 [hep-ex]} \BibitemShut
  {NoStop}%
\bibitem [{\citenamefont {Zyla}\ \emph
  {et~al.}(2020{\natexlab{b}})\citenamefont {Zyla} \emph
  {et~al.}}]{Zyla:2020zbs}%
  \BibitemOpen
  \bibfield  {author} {\bibinfo {author} {\bibfnamefont {P.~A.}\ \bibnamefont
  {Zyla}} \emph {et~al.} (\bibinfo {collaboration} {Particle Data Group}),\
  }\href {\doibase 10.1093/ptep/ptaa104} {\bibfield  {journal} {\bibinfo
  {journal} {PTEP}\ }\textbf {\bibinfo {volume} {2020}},\ \bibinfo {pages}
  {083C01} (\bibinfo {year} {2020}{\natexlab{b}})}\BibitemShut {NoStop}%
\bibitem [{\citenamefont {Bryman}\ \emph {et~al.}(1986)\citenamefont {Bryman},
  \citenamefont {Dixit}, \citenamefont {Dubois}, \citenamefont {Macdonald},
  \citenamefont {Numao}, \citenamefont {Olaniyi}, \citenamefont {Olin},\ and\
  \citenamefont {Poutissou}}]{Bryman:1985bv}%
  \BibitemOpen
  \bibfield  {author} {\bibinfo {author} {\bibfnamefont {D.~A.}\ \bibnamefont
  {Bryman}}, \bibinfo {author} {\bibfnamefont {M.~S.}\ \bibnamefont {Dixit}},
  \bibinfo {author} {\bibfnamefont {R.}~\bibnamefont {Dubois}}, \bibinfo
  {author} {\bibfnamefont {J.~A.}\ \bibnamefont {Macdonald}}, \bibinfo {author}
  {\bibfnamefont {T.}~\bibnamefont {Numao}}, \bibinfo {author} {\bibfnamefont
  {B.}~\bibnamefont {Olaniyi}}, \bibinfo {author} {\bibfnamefont
  {A.}~\bibnamefont {Olin}}, \ and\ \bibinfo {author} {\bibfnamefont {J.~M.}\
  \bibnamefont {Poutissou}},\ }\href {\doibase 10.1103/PhysRevD.33.1211}
  {\bibfield  {journal} {\bibinfo  {journal} {Phys. Rev. D}\ }\textbf {\bibinfo
  {volume} {33}},\ \bibinfo {pages} {1211} (\bibinfo {year}
  {1986})}\BibitemShut {NoStop}%
\bibitem [{\citenamefont {Britton}\ \emph {et~al.}(1994)\citenamefont {Britton}
  \emph {et~al.}}]{Britton:1993cj}%
  \BibitemOpen
  \bibfield  {author} {\bibinfo {author} {\bibfnamefont {D.~I.}\ \bibnamefont
  {Britton}} \emph {et~al.},\ }\href {\doibase 10.1103/PhysRevD.49.28}
  {\bibfield  {journal} {\bibinfo  {journal} {Phys. Rev. D}\ }\textbf {\bibinfo
  {volume} {49}},\ \bibinfo {pages} {28} (\bibinfo {year} {1994})}\BibitemShut
  {NoStop}%
\bibitem [{\citenamefont {Czapek}\ \emph {et~al.}(1993)\citenamefont {Czapek}
  \emph {et~al.}}]{Czapek:1993kc}%
  \BibitemOpen
  \bibfield  {author} {\bibinfo {author} {\bibfnamefont {G.}~\bibnamefont
  {Czapek}} \emph {et~al.},\ }\href {\doibase 10.1103/PhysRevLett.70.17}
  {\bibfield  {journal} {\bibinfo  {journal} {Phys. Rev. Lett.}\ }\textbf
  {\bibinfo {volume} {70}},\ \bibinfo {pages} {17} (\bibinfo {year}
  {1993})}\BibitemShut {NoStop}%
\bibitem [{\citenamefont {Lazzeroni}\ \emph {et~al.}(2013)\citenamefont
  {Lazzeroni} \emph {et~al.}}]{Lazzeroni:2012cx}%
  \BibitemOpen
  \bibfield  {author} {\bibinfo {author} {\bibfnamefont {C.}~\bibnamefont
  {Lazzeroni}} \emph {et~al.} (\bibinfo {collaboration} {NA62}),\ }\href
  {\doibase 10.1016/j.physletb.2013.01.037} {\bibfield  {journal} {\bibinfo
  {journal} {Phys. Lett. B}\ }\textbf {\bibinfo {volume} {719}},\ \bibinfo
  {pages} {326} (\bibinfo {year} {2013})},\ \Eprint
  {http://arxiv.org/abs/1212.4012} {arXiv:1212.4012 [hep-ex]} \BibitemShut
  {NoStop}%
\bibitem [{\citenamefont {Ambrosino}\ \emph {et~al.}(2009)\citenamefont
  {Ambrosino} \emph {et~al.}}]{Ambrosino:2009aa}%
  \BibitemOpen
  \bibfield  {author} {\bibinfo {author} {\bibfnamefont {F.}~\bibnamefont
  {Ambrosino}} \emph {et~al.} (\bibinfo {collaboration} {KLOE}),\ }\href
  {\doibase 10.1140/epjc/s10052-009-1217-6} {\bibfield  {journal} {\bibinfo
  {journal} {Eur. Phys. J. C}\ }\textbf {\bibinfo {volume} {64}},\ \bibinfo
  {pages} {627} (\bibinfo {year} {2009})},\ \bibinfo {note} {[Erratum:
  Eur.Phys.J. 65, 703 (2010)]},\ \Eprint {http://arxiv.org/abs/0907.3594}
  {arXiv:0907.3594 [hep-ex]} \BibitemShut {NoStop}%
\bibitem [{\citenamefont {Antonelli}\ \emph
  {et~al.}(2010{\natexlab{a}})\citenamefont {Antonelli} \emph
  {et~al.}}]{Antonelli:2009ws}%
  \BibitemOpen
  \bibfield  {author} {\bibinfo {author} {\bibfnamefont {M.}~\bibnamefont
  {Antonelli}} \emph {et~al.},\ }\href {\doibase 10.1016/j.physrep.2010.05.003}
  {\bibfield  {journal} {\bibinfo  {journal} {Phys. Rept.}\ }\textbf {\bibinfo
  {volume} {494}},\ \bibinfo {pages} {197} (\bibinfo {year}
  {2010}{\natexlab{a}})},\ \Eprint {http://arxiv.org/abs/0907.5386}
  {arXiv:0907.5386 [hep-ph]} \BibitemShut {NoStop}%
\bibitem [{\citenamefont {Cirigliano}\ \emph {et~al.}(2002)\citenamefont
  {Cirigliano}, \citenamefont {Knecht}, \citenamefont {Neufeld}, \citenamefont
  {Rupertsberger},\ and\ \citenamefont {Talavera}}]{Cirigliano:2001mk}%
  \BibitemOpen
  \bibfield  {author} {\bibinfo {author} {\bibfnamefont {V.}~\bibnamefont
  {Cirigliano}}, \bibinfo {author} {\bibfnamefont {M.}~\bibnamefont {Knecht}},
  \bibinfo {author} {\bibfnamefont {H.}~\bibnamefont {Neufeld}}, \bibinfo
  {author} {\bibfnamefont {H.}~\bibnamefont {Rupertsberger}}, \ and\ \bibinfo
  {author} {\bibfnamefont {P.}~\bibnamefont {Talavera}},\ }\href {\doibase
  10.1007/s100520100825} {\bibfield  {journal} {\bibinfo  {journal} {Eur. Phys.
  J. C}\ }\textbf {\bibinfo {volume} {23}},\ \bibinfo {pages} {121} (\bibinfo
  {year} {2002})},\ \Eprint {http://arxiv.org/abs/hep-ph/0110153}
  {arXiv:hep-ph/0110153} \BibitemShut {NoStop}%
\bibitem [{\citenamefont {Cirigliano}\ \emph {et~al.}(2004)\citenamefont
  {Cirigliano}, \citenamefont {Neufeld},\ and\ \citenamefont
  {Pichl}}]{Cirigliano:2004pv}%
  \BibitemOpen
  \bibfield  {author} {\bibinfo {author} {\bibfnamefont {V.}~\bibnamefont
  {Cirigliano}}, \bibinfo {author} {\bibfnamefont {H.}~\bibnamefont {Neufeld}},
  \ and\ \bibinfo {author} {\bibfnamefont {H.}~\bibnamefont {Pichl}},\ }\href
  {\doibase 10.1140/epjc/s2004-01745-1} {\bibfield  {journal} {\bibinfo
  {journal} {Eur. Phys. J. C}\ }\textbf {\bibinfo {volume} {35}},\ \bibinfo
  {pages} {53} (\bibinfo {year} {2004})},\ \Eprint
  {http://arxiv.org/abs/hep-ph/0401173} {arXiv:hep-ph/0401173} \BibitemShut
  {NoStop}%
\bibitem [{\citenamefont {Cirigliano}\ \emph {et~al.}(2008)\citenamefont
  {Cirigliano}, \citenamefont {Giannotti},\ and\ \citenamefont
  {Neufeld}}]{Cirigliano:2008wn}%
  \BibitemOpen
  \bibfield  {author} {\bibinfo {author} {\bibfnamefont {V.}~\bibnamefont
  {Cirigliano}}, \bibinfo {author} {\bibfnamefont {M.}~\bibnamefont
  {Giannotti}}, \ and\ \bibinfo {author} {\bibfnamefont {H.}~\bibnamefont
  {Neufeld}},\ }\href {\doibase 10.1088/1126-6708/2008/11/006} {\bibfield
  {journal} {\bibinfo  {journal} {JHEP}\ }\textbf {\bibinfo {volume} {11}},\
  \bibinfo {pages} {006} (\bibinfo {year} {2008})},\ \Eprint
  {http://arxiv.org/abs/0807.4507} {arXiv:0807.4507 [hep-ph]} \BibitemShut
  {NoStop}%
\bibitem [{\citenamefont {Seng}\ \emph
  {et~al.}(2021{\natexlab{a}})\citenamefont {Seng}, \citenamefont {Galviz},
  \citenamefont {Gorchtein},\ and\ \citenamefont {Mei\ss{}ner}}]{Seng:2021boy}%
  \BibitemOpen
  \bibfield  {author} {\bibinfo {author} {\bibfnamefont {C.-Y.}\ \bibnamefont
  {Seng}}, \bibinfo {author} {\bibfnamefont {D.}~\bibnamefont {Galviz}},
  \bibinfo {author} {\bibfnamefont {M.}~\bibnamefont {Gorchtein}}, \ and\
  \bibinfo {author} {\bibfnamefont {U.~G.}\ \bibnamefont {Mei\ss{}ner}},\
  }\href {\doibase 10.1016/j.physletb.2021.136522} {\bibfield  {journal}
  {\bibinfo  {journal} {Phys. Lett. B}\ }\textbf {\bibinfo {volume} {820}},\
  \bibinfo {pages} {136522} (\bibinfo {year} {2021}{\natexlab{a}})},\ \Eprint
  {http://arxiv.org/abs/2103.00975} {arXiv:2103.00975 [hep-ph]} \BibitemShut
  {NoStop}%
\bibitem [{\citenamefont {Moulson}(2017)}]{Moulson:2017ive}%
  \BibitemOpen
  \bibfield  {author} {\bibinfo {author} {\bibfnamefont {M.}~\bibnamefont
  {Moulson}},\ }\href {\doibase 10.22323/1.291.0033} {\bibfield  {journal}
  {\bibinfo  {journal} {PoS}\ }\textbf {\bibinfo {volume} {CKM2016}},\ \bibinfo
  {pages} {033} (\bibinfo {year} {2017})},\ \Eprint
  {http://arxiv.org/abs/1704.04104} {arXiv:1704.04104 [hep-ex]} \BibitemShut
  {NoStop}%
\bibitem [{\citenamefont {Seng}\ \emph
  {et~al.}(2021{\natexlab{b}})\citenamefont {Seng}, \citenamefont {Galviz},
  \citenamefont {Marciano},\ and\ \citenamefont {Mei\ss{}ner}}]{Seng:2021nar}%
  \BibitemOpen
  \bibfield  {author} {\bibinfo {author} {\bibfnamefont {C.-Y.}\ \bibnamefont
  {Seng}}, \bibinfo {author} {\bibfnamefont {D.}~\bibnamefont {Galviz}},
  \bibinfo {author} {\bibfnamefont {W.~J.}\ \bibnamefont {Marciano}}, \ and\
  \bibinfo {author} {\bibfnamefont {U.-G.}\ \bibnamefont {Mei\ss{}ner}},\
  }\href@noop {} {\  (\bibinfo {year} {2021}{\natexlab{b}})},\ \Eprint
  {http://arxiv.org/abs/2107.14708} {arXiv:2107.14708 [hep-ph]} \BibitemShut
  {NoStop}%
\bibitem [{\citenamefont {Glattauer}\ \emph {et~al.}(2016)\citenamefont
  {Glattauer} \emph {et~al.}}]{Glattauer:2015teq}%
  \BibitemOpen
  \bibfield  {author} {\bibinfo {author} {\bibfnamefont {R.}~\bibnamefont
  {Glattauer}} \emph {et~al.} (\bibinfo {collaboration} {Belle}),\ }\href
  {\doibase 10.1103/PhysRevD.93.032006} {\bibfield  {journal} {\bibinfo
  {journal} {Phys. Rev. D}\ }\textbf {\bibinfo {volume} {93}},\ \bibinfo
  {pages} {032006} (\bibinfo {year} {2016})},\ \Eprint
  {http://arxiv.org/abs/1510.03657} {arXiv:1510.03657 [hep-ex]} \BibitemShut
  {NoStop}%
\bibitem [{\citenamefont {Abdesselam}\ \emph {et~al.}(2017)\citenamefont
  {Abdesselam} \emph {et~al.}}]{Abdesselam:2017kjf}%
  \BibitemOpen
  \bibfield  {author} {\bibinfo {author} {\bibfnamefont {A.}~\bibnamefont
  {Abdesselam}} \emph {et~al.} (\bibinfo {collaboration} {Belle}),\ }\href@noop
  {} {\  (\bibinfo {year} {2017})},\ \Eprint {http://arxiv.org/abs/1702.01521}
  {arXiv:1702.01521 [hep-ex]} \BibitemShut {NoStop}%
\bibitem [{\citenamefont {Waheed}\ \emph {et~al.}(2019)\citenamefont {Waheed}
  \emph {et~al.}}]{Waheed:2018djm}%
  \BibitemOpen
  \bibfield  {author} {\bibinfo {author} {\bibfnamefont {E.}~\bibnamefont
  {Waheed}} \emph {et~al.} (\bibinfo {collaboration} {Belle}),\ }\href
  {\doibase 10.1103/PhysRevD.100.052007} {\bibfield  {journal} {\bibinfo
  {journal} {Phys. Rev. D}\ }\textbf {\bibinfo {volume} {100}},\ \bibinfo
  {pages} {052007} (\bibinfo {year} {2019})},\ \bibinfo {note} {[Erratum:
  Phys.Rev.D 103, 079901 (2021)]},\ \Eprint {http://arxiv.org/abs/1809.03290}
  {arXiv:1809.03290 [hep-ex]} \BibitemShut {NoStop}%
\bibitem [{\citenamefont {Cabibbo}(1963)}]{Cabibbo:1963yz}%
  \BibitemOpen
  \bibfield  {author} {\bibinfo {author} {\bibfnamefont {N.}~\bibnamefont
  {Cabibbo}},\ }\href {\doibase 10.1103/PhysRevLett.10.531} {\bibfield
  {journal} {\bibinfo  {journal} {Phys. Rev. Lett.}\ }\textbf {\bibinfo
  {volume} {10}},\ \bibinfo {pages} {531} (\bibinfo {year} {1963})}\BibitemShut
  {NoStop}%
\bibitem [{\citenamefont {Kobayashi}\ and\ \citenamefont
  {Maskawa}(1973)}]{Kobayashi:1973fv}%
  \BibitemOpen
  \bibfield  {author} {\bibinfo {author} {\bibfnamefont {M.}~\bibnamefont
  {Kobayashi}}\ and\ \bibinfo {author} {\bibfnamefont {T.}~\bibnamefont
  {Maskawa}},\ }\href {\doibase 10.1143/PTP.49.652} {\bibfield  {journal}
  {\bibinfo  {journal} {Prog. Theor. Phys.}\ }\textbf {\bibinfo {volume}
  {49}},\ \bibinfo {pages} {652} (\bibinfo {year} {1973})}\BibitemShut
  {NoStop}%
\bibitem [{\citenamefont {Buchmuller}\ and\ \citenamefont
  {Wyler}(1986)}]{Buchmuller:1985jz}%
  \BibitemOpen
  \bibfield  {author} {\bibinfo {author} {\bibfnamefont {W.}~\bibnamefont
  {Buchmuller}}\ and\ \bibinfo {author} {\bibfnamefont {D.}~\bibnamefont
  {Wyler}},\ }\href {\doibase 10.1016/0550-3213(86)90262-2} {\bibfield
  {journal} {\bibinfo  {journal} {Nucl. Phys. B}\ }\textbf {\bibinfo {volume}
  {268}},\ \bibinfo {pages} {621} (\bibinfo {year} {1986})}\BibitemShut
  {NoStop}%
\bibitem [{\citenamefont {Cirigliano}\ \emph {et~al.}(2010)\citenamefont
  {Cirigliano}, \citenamefont {Jenkins},\ and\ \citenamefont
  {Gonzalez-Alonso}}]{Cirigliano:2009wk}%
  \BibitemOpen
  \bibfield  {author} {\bibinfo {author} {\bibfnamefont {V.}~\bibnamefont
  {Cirigliano}}, \bibinfo {author} {\bibfnamefont {J.}~\bibnamefont {Jenkins}},
  \ and\ \bibinfo {author} {\bibfnamefont {M.}~\bibnamefont
  {Gonzalez-Alonso}},\ }\href {\doibase 10.1016/j.nuclphysb.2009.12.020}
  {\bibfield  {journal} {\bibinfo  {journal} {Nucl. Phys. B}\ }\textbf
  {\bibinfo {volume} {830}},\ \bibinfo {pages} {95} (\bibinfo {year} {2010})},\
  \Eprint {http://arxiv.org/abs/0908.1754} {arXiv:0908.1754 [hep-ph]}
  \BibitemShut {NoStop}%
\bibitem [{\citenamefont {Bauman}\ \emph {et~al.}(2013)\citenamefont {Bauman},
  \citenamefont {Erler},\ and\ \citenamefont {Ramsey-Musolf}}]{Bauman:2012fx}%
  \BibitemOpen
  \bibfield  {author} {\bibinfo {author} {\bibfnamefont {S.}~\bibnamefont
  {Bauman}}, \bibinfo {author} {\bibfnamefont {J.}~\bibnamefont {Erler}}, \
  and\ \bibinfo {author} {\bibfnamefont {M.}~\bibnamefont {Ramsey-Musolf}},\
  }\href {\doibase 10.1103/PhysRevD.87.035012} {\bibfield  {journal} {\bibinfo
  {journal} {Phys. Rev. D}\ }\textbf {\bibinfo {volume} {87}},\ \bibinfo
  {pages} {035012} (\bibinfo {year} {2013})},\ \Eprint
  {http://arxiv.org/abs/1204.0035} {arXiv:1204.0035 [hep-ph]} \BibitemShut
  {NoStop}%
\bibitem [{\citenamefont {Hardy}\ and\ \citenamefont
  {Towner}(2020)}]{Hardy:2020qwl}%
  \BibitemOpen
  \bibfield  {author} {\bibinfo {author} {\bibfnamefont {J.~C.}\ \bibnamefont
  {Hardy}}\ and\ \bibinfo {author} {\bibfnamefont {I.~S.}\ \bibnamefont
  {Towner}},\ }\href {\doibase 10.1103/PhysRevC.102.045501} {\bibfield
  {journal} {\bibinfo  {journal} {Phys. Rev. C}\ }\textbf {\bibinfo {volume}
  {102}},\ \bibinfo {pages} {045501} (\bibinfo {year} {2020})}\BibitemShut
  {NoStop}%
\bibitem [{\citenamefont {Seng}\ \emph {et~al.}(2018)\citenamefont {Seng},
  \citenamefont {Gorchtein}, \citenamefont {Patel},\ and\ \citenamefont
  {Ramsey-Musolf}}]{Seng:2018yzq}%
  \BibitemOpen
  \bibfield  {author} {\bibinfo {author} {\bibfnamefont {C.-Y.}\ \bibnamefont
  {Seng}}, \bibinfo {author} {\bibfnamefont {M.}~\bibnamefont {Gorchtein}},
  \bibinfo {author} {\bibfnamefont {H.~H.}\ \bibnamefont {Patel}}, \ and\
  \bibinfo {author} {\bibfnamefont {M.~J.}\ \bibnamefont {Ramsey-Musolf}},\
  }\href {\doibase 10.1103/PhysRevLett.121.241804} {\bibfield  {journal}
  {\bibinfo  {journal} {Phys. Rev. Lett.}\ }\textbf {\bibinfo {volume} {121}},\
  \bibinfo {pages} {241804} (\bibinfo {year} {2018})},\ \Eprint
  {http://arxiv.org/abs/1807.10197} {arXiv:1807.10197 [hep-ph]} \BibitemShut
  {NoStop}%
\bibitem [{\citenamefont {Czarnecki}\ \emph {et~al.}(2019)\citenamefont
  {Czarnecki}, \citenamefont {Marciano},\ and\ \citenamefont
  {Sirlin}}]{Czarnecki:2019mwq}%
  \BibitemOpen
  \bibfield  {author} {\bibinfo {author} {\bibfnamefont {A.}~\bibnamefont
  {Czarnecki}}, \bibinfo {author} {\bibfnamefont {W.~J.}\ \bibnamefont
  {Marciano}}, \ and\ \bibinfo {author} {\bibfnamefont {A.}~\bibnamefont
  {Sirlin}},\ }\href {\doibase 10.1103/PhysRevD.100.073008} {\bibfield
  {journal} {\bibinfo  {journal} {Phys. Rev. D}\ }\textbf {\bibinfo {volume}
  {100}},\ \bibinfo {pages} {073008} (\bibinfo {year} {2019})},\ \Eprint
  {http://arxiv.org/abs/1907.06737} {arXiv:1907.06737 [hep-ph]} \BibitemShut
  {NoStop}%
\bibitem [{\citenamefont {Towner}(1992)}]{Towner:1992xm}%
  \BibitemOpen
  \bibfield  {author} {\bibinfo {author} {\bibfnamefont {I.~S.}\ \bibnamefont
  {Towner}},\ }\href {\doibase 10.1016/0375-9474(92)90170-O} {\bibfield
  {journal} {\bibinfo  {journal} {Nucl. Phys. A}\ }\textbf {\bibinfo {volume}
  {540}},\ \bibinfo {pages} {478} (\bibinfo {year} {1992})}\BibitemShut
  {NoStop}%
\bibitem [{\citenamefont {Seng}\ \emph {et~al.}(2019)\citenamefont {Seng},
  \citenamefont {Gorchtein},\ and\ \citenamefont
  {Ramsey-Musolf}}]{Seng:2018qru}%
  \BibitemOpen
  \bibfield  {author} {\bibinfo {author} {\bibfnamefont {C.~Y.}\ \bibnamefont
  {Seng}}, \bibinfo {author} {\bibfnamefont {M.}~\bibnamefont {Gorchtein}}, \
  and\ \bibinfo {author} {\bibfnamefont {M.~J.}\ \bibnamefont
  {Ramsey-Musolf}},\ }\href {\doibase 10.1103/PhysRevD.100.013001} {\bibfield
  {journal} {\bibinfo  {journal} {Phys. Rev. D}\ }\textbf {\bibinfo {volume}
  {100}},\ \bibinfo {pages} {013001} (\bibinfo {year} {2019})},\ \Eprint
  {http://arxiv.org/abs/1812.03352} {arXiv:1812.03352 [nucl-th]} \BibitemShut
  {NoStop}%
\bibitem [{\citenamefont {Gorchtein}(2019)}]{Gorchtein:2018fxl}%
  \BibitemOpen
  \bibfield  {author} {\bibinfo {author} {\bibfnamefont {M.}~\bibnamefont
  {Gorchtein}},\ }\href {\doibase 10.1103/PhysRevLett.123.042503} {\bibfield
  {journal} {\bibinfo  {journal} {Phys. Rev. Lett.}\ }\textbf {\bibinfo
  {volume} {123}},\ \bibinfo {pages} {042503} (\bibinfo {year} {2019})},\
  \Eprint {http://arxiv.org/abs/1812.04229} {arXiv:1812.04229 [nucl-th]}
  \BibitemShut {NoStop}%
\bibitem [{\citenamefont {Gonzalez}\ \emph {et~al.}(2021)\citenamefont
  {Gonzalez} \emph {et~al.}}]{UCNt:2021pcg}%
  \BibitemOpen
  \bibfield  {author} {\bibinfo {author} {\bibfnamefont {F.~M.}\ \bibnamefont
  {Gonzalez}} \emph {et~al.} (\bibinfo {collaboration}
  {UCN\ensuremath{\tau}}),\ }\href {\doibase 10.1103/PhysRevLett.127.162501}
  {\bibfield  {journal} {\bibinfo  {journal} {Phys. Rev. Lett.}\ }\textbf
  {\bibinfo {volume} {127}},\ \bibinfo {pages} {162501} (\bibinfo {year}
  {2021})},\ \Eprint {http://arxiv.org/abs/2106.10375} {arXiv:2106.10375
  [nucl-ex]} \BibitemShut {NoStop}%
\bibitem [{\citenamefont {M\"arkisch}\ \emph {et~al.}(2019)\citenamefont
  {M\"arkisch} \emph {et~al.}}]{Markisch:2018ndu}%
  \BibitemOpen
  \bibfield  {author} {\bibinfo {author} {\bibfnamefont {B.}~\bibnamefont
  {M\"arkisch}} \emph {et~al.},\ }\href {\doibase
  10.1103/PhysRevLett.122.242501} {\bibfield  {journal} {\bibinfo  {journal}
  {Phys. Rev. Lett.}\ }\textbf {\bibinfo {volume} {122}},\ \bibinfo {pages}
  {242501} (\bibinfo {year} {2019})},\ \Eprint
  {http://arxiv.org/abs/1812.04666} {arXiv:1812.04666 [nucl-ex]} \BibitemShut
  {NoStop}%
\bibitem [{\citenamefont {Dubbers}\ and\ \citenamefont
  {M\"arkisch}(2021)}]{Dubbers:2021wqv}%
  \BibitemOpen
  \bibfield  {author} {\bibinfo {author} {\bibfnamefont {D.}~\bibnamefont
  {Dubbers}}\ and\ \bibinfo {author} {\bibfnamefont {B.}~\bibnamefont
  {M\"arkisch}},\ }\href {\doibase 10.1146/annurev-nucl-102419-043156} {\
  (\bibinfo {year} {2021}),\ 10.1146/annurev-nucl-102419-043156},\ \Eprint
  {http://arxiv.org/abs/2106.02345} {arXiv:2106.02345 [nucl-ex]} \BibitemShut
  {NoStop}%
\bibitem [{\citenamefont {Falkowski}\ \emph {et~al.}(2021)\citenamefont
  {Falkowski}, \citenamefont {Gonz\'alez-Alonso},\ and\ \citenamefont
  {Naviliat-Cuncic}}]{Falkowski:2020pma}%
  \BibitemOpen
  \bibfield  {author} {\bibinfo {author} {\bibfnamefont {A.}~\bibnamefont
  {Falkowski}}, \bibinfo {author} {\bibfnamefont {M.}~\bibnamefont
  {Gonz\'alez-Alonso}}, \ and\ \bibinfo {author} {\bibfnamefont
  {O.}~\bibnamefont {Naviliat-Cuncic}},\ }\href {\doibase
  10.1007/JHEP04(2021)126} {\bibfield  {journal} {\bibinfo  {journal} {JHEP}\
  }\textbf {\bibinfo {volume} {04}},\ \bibinfo {pages} {126} (\bibinfo {year}
  {2021})},\ \Eprint {http://arxiv.org/abs/2010.13797} {arXiv:2010.13797
  [hep-ph]} \BibitemShut {NoStop}%
\bibitem [{\citenamefont {Cirigliano}\ \emph {et~al.}(2019)\citenamefont
  {Cirigliano}, \citenamefont {Garcia}, \citenamefont {Gazit}, \citenamefont
  {Naviliat-Cuncic}, \citenamefont {Savard},\ and\ \citenamefont
  {Young}}]{Cirigliano:2019wao}%
  \BibitemOpen
  \bibfield  {author} {\bibinfo {author} {\bibfnamefont {V.}~\bibnamefont
  {Cirigliano}}, \bibinfo {author} {\bibfnamefont {A.}~\bibnamefont {Garcia}},
  \bibinfo {author} {\bibfnamefont {D.}~\bibnamefont {Gazit}}, \bibinfo
  {author} {\bibfnamefont {O.}~\bibnamefont {Naviliat-Cuncic}}, \bibinfo
  {author} {\bibfnamefont {G.}~\bibnamefont {Savard}}, \ and\ \bibinfo {author}
  {\bibfnamefont {A.}~\bibnamefont {Young}},\ }\href@noop {} {\  (\bibinfo
  {year} {2019})},\ \Eprint {http://arxiv.org/abs/1907.02164} {arXiv:1907.02164
  [nucl-ex]} \BibitemShut {NoStop}%
\bibitem [{\citenamefont {Marciano}(2004)}]{Marciano:2004uf}%
  \BibitemOpen
  \bibfield  {author} {\bibinfo {author} {\bibfnamefont {W.~J.}\ \bibnamefont
  {Marciano}},\ }\href {\doibase 10.1103/PhysRevLett.93.231803} {\bibfield
  {journal} {\bibinfo  {journal} {Phys. Rev. Lett.}\ }\textbf {\bibinfo
  {volume} {93}},\ \bibinfo {pages} {231803} (\bibinfo {year} {2004})},\
  \Eprint {http://arxiv.org/abs/hep-ph/0402299} {arXiv:hep-ph/0402299}
  \BibitemShut {NoStop}%
\bibitem [{\citenamefont {Antonelli}\ \emph
  {et~al.}(2010{\natexlab{b}})\citenamefont {Antonelli} \emph
  {et~al.}}]{Antonelli:2010yf}%
  \BibitemOpen
  \bibfield  {author} {\bibinfo {author} {\bibfnamefont {M.}~\bibnamefont
  {Antonelli}} \emph {et~al.} (\bibinfo {collaboration} {FlaviaNet Working
  Group on Kaon Decays}),\ }\href {\doibase 10.1140/epjc/s10052-010-1406-3}
  {\bibfield  {journal} {\bibinfo  {journal} {Eur. Phys. J. C}\ }\textbf
  {\bibinfo {volume} {69}},\ \bibinfo {pages} {399} (\bibinfo {year}
  {2010}{\natexlab{b}})},\ \Eprint {http://arxiv.org/abs/1005.2323}
  {arXiv:1005.2323 [hep-ph]} \BibitemShut {NoStop}%
\bibitem [{\citenamefont {Aoki}\ \emph {et~al.}(2020)\citenamefont {Aoki} \emph
  {et~al.}}]{Aoki:2019cca}%
  \BibitemOpen
  \bibfield  {author} {\bibinfo {author} {\bibfnamefont {S.}~\bibnamefont
  {Aoki}} \emph {et~al.} (\bibinfo {collaboration} {Flavour Lattice Averaging
  Group}),\ }\href {\doibase 10.1140/epjc/s10052-019-7354-7} {\bibfield
  {journal} {\bibinfo  {journal} {Eur. Phys. J. C}\ }\textbf {\bibinfo {volume}
  {80}},\ \bibinfo {pages} {113} (\bibinfo {year} {2020})},\ \Eprint
  {http://arxiv.org/abs/1902.08191} {arXiv:1902.08191 [hep-lat]} \BibitemShut
  {NoStop}%
\bibitem [{\citenamefont {Czarnecki}\ \emph {et~al.}(2020)\citenamefont
  {Czarnecki}, \citenamefont {Marciano},\ and\ \citenamefont
  {Sirlin}}]{Czarnecki:2019iwz}%
  \BibitemOpen
  \bibfield  {author} {\bibinfo {author} {\bibfnamefont {A.}~\bibnamefont
  {Czarnecki}}, \bibinfo {author} {\bibfnamefont {W.~J.}\ \bibnamefont
  {Marciano}}, \ and\ \bibinfo {author} {\bibfnamefont {A.}~\bibnamefont
  {Sirlin}},\ }\href {\doibase 10.1103/PhysRevD.101.091301} {\bibfield
  {journal} {\bibinfo  {journal} {Phys. Rev. D}\ }\textbf {\bibinfo {volume}
  {101}},\ \bibinfo {pages} {091301} (\bibinfo {year} {2020})},\ \Eprint
  {http://arxiv.org/abs/1911.04685} {arXiv:1911.04685 [hep-ph]} \BibitemShut
  {NoStop}%
\bibitem [{\citenamefont {Pich}(2014)}]{Pich:2013lsa}%
  \BibitemOpen
  \bibfield  {author} {\bibinfo {author} {\bibfnamefont {A.}~\bibnamefont
  {Pich}},\ }\href {\doibase 10.1016/j.ppnp.2013.11.002} {\bibfield  {journal}
  {\bibinfo  {journal} {Prog. Part. Nucl. Phys.}\ }\textbf {\bibinfo {volume}
  {75}},\ \bibinfo {pages} {41} (\bibinfo {year} {2014})},\ \Eprint
  {http://arxiv.org/abs/1310.7922} {arXiv:1310.7922 [hep-ph]} \BibitemShut
  {NoStop}%
\bibitem [{\citenamefont {Aubert}\ \emph {et~al.}(2010)\citenamefont {Aubert}
  \emph {et~al.}}]{BaBar:2009lyd}%
  \BibitemOpen
  \bibfield  {author} {\bibinfo {author} {\bibfnamefont {B.}~\bibnamefont
  {Aubert}} \emph {et~al.} (\bibinfo {collaboration} {BaBar}),\ }\href
  {\doibase 10.1103/PhysRevLett.105.051602} {\bibfield  {journal} {\bibinfo
  {journal} {Phys. Rev. Lett.}\ }\textbf {\bibinfo {volume} {105}},\ \bibinfo
  {pages} {051602} (\bibinfo {year} {2010})},\ \Eprint
  {http://arxiv.org/abs/0912.0242} {arXiv:0912.0242 [hep-ex]} \BibitemShut
  {NoStop}%
\bibitem [{\citenamefont {Anastassov}\ \emph {et~al.}(1997)\citenamefont
  {Anastassov} \emph {et~al.}}]{CLEO:1996oro}%
  \BibitemOpen
  \bibfield  {author} {\bibinfo {author} {\bibfnamefont {A.}~\bibnamefont
  {Anastassov}} \emph {et~al.} (\bibinfo {collaboration} {CLEO}),\ }\href
  {\doibase 10.1103/PhysRevD.55.2559} {\bibfield  {journal} {\bibinfo
  {journal} {Phys. Rev. D}\ }\textbf {\bibinfo {volume} {55}},\ \bibinfo
  {pages} {2559} (\bibinfo {year} {1997})},\ \bibinfo {note} {[Erratum:
  Phys.Rev.D 58, 119904 (1998)]}\BibitemShut {NoStop}%
\bibitem [{\citenamefont {Amhis}\ \emph
  {et~al.}(2021{\natexlab{b}})\citenamefont {Amhis} \emph
  {et~al.}}]{HFLAV:2019otj}%
  \BibitemOpen
  \bibfield  {author} {\bibinfo {author} {\bibfnamefont {Y.~S.}\ \bibnamefont
  {Amhis}} \emph {et~al.} (\bibinfo {collaboration} {HFLAV}),\ }\href {\doibase
  10.1140/epjc/s10052-020-8156-7} {\bibfield  {journal} {\bibinfo  {journal}
  {Eur. Phys. J. C}\ }\textbf {\bibinfo {volume} {81}},\ \bibinfo {pages} {226}
  (\bibinfo {year} {2021}{\natexlab{b}})},\ \Eprint
  {http://arxiv.org/abs/1909.12524} {arXiv:1909.12524 [hep-ex]} \BibitemShut
  {NoStop}%
\bibitem [{\citenamefont {Arroyo-Ure\~na}\ \emph {et~al.}(2021)\citenamefont
  {Arroyo-Ure\~na}, \citenamefont {Hern\'andez-Tom\'e}, \citenamefont
  {L\'opez-Castro}, \citenamefont {Roig},\ and\ \citenamefont
  {Rosell}}]{Arroyo-Urena:2021nil}%
  \BibitemOpen
  \bibfield  {author} {\bibinfo {author} {\bibfnamefont {M.~A.}\ \bibnamefont
  {Arroyo-Ure\~na}}, \bibinfo {author} {\bibfnamefont {G.}~\bibnamefont
  {Hern\'andez-Tom\'e}}, \bibinfo {author} {\bibfnamefont {G.}~\bibnamefont
  {L\'opez-Castro}}, \bibinfo {author} {\bibfnamefont {P.}~\bibnamefont
  {Roig}}, \ and\ \bibinfo {author} {\bibfnamefont {I.}~\bibnamefont
  {Rosell}},\ }\href@noop {} {\  (\bibinfo {year} {2021})},\ \Eprint
  {http://arxiv.org/abs/2107.04603} {arXiv:2107.04603 [hep-ph]} \BibitemShut
  {NoStop}%
\bibitem [{\citenamefont {Tishchenko}\ \emph {et~al.}(2013)\citenamefont
  {Tishchenko} \emph {et~al.}}]{Tishchenko:2012ie}%
  \BibitemOpen
  \bibfield  {author} {\bibinfo {author} {\bibfnamefont {V.}~\bibnamefont
  {Tishchenko}} \emph {et~al.} (\bibinfo {collaboration} {MuLan}),\ }\href
  {\doibase 10.1103/PhysRevD.87.052003} {\bibfield  {journal} {\bibinfo
  {journal} {Phys. Rev. D}\ }\textbf {\bibinfo {volume} {87}},\ \bibinfo
  {pages} {052003} (\bibinfo {year} {2013})},\ \Eprint
  {http://arxiv.org/abs/1211.0960} {arXiv:1211.0960 [hep-ex]} \BibitemShut
  {NoStop}%
\bibitem [{\citenamefont {Marciano}(1999)}]{Marciano:1999ih}%
  \BibitemOpen
  \bibfield  {author} {\bibinfo {author} {\bibfnamefont {W.~J.}\ \bibnamefont
  {Marciano}},\ }\href {\doibase 10.1103/PhysRevD.60.093006} {\bibfield
  {journal} {\bibinfo  {journal} {Phys. Rev. D}\ }\textbf {\bibinfo {volume}
  {60}},\ \bibinfo {pages} {093006} (\bibinfo {year} {1999})},\ \Eprint
  {http://arxiv.org/abs/hep-ph/9903451} {arXiv:hep-ph/9903451} \BibitemShut
  {NoStop}%
\bibitem [{\citenamefont {Crivellin}\ \emph
  {et~al.}(2021{\natexlab{c}})\citenamefont {Crivellin}, \citenamefont
  {Hoferichter},\ and\ \citenamefont {Manzari}}]{Crivellin:2021njn}%
  \BibitemOpen
  \bibfield  {author} {\bibinfo {author} {\bibfnamefont {A.}~\bibnamefont
  {Crivellin}}, \bibinfo {author} {\bibfnamefont {M.}~\bibnamefont
  {Hoferichter}}, \ and\ \bibinfo {author} {\bibfnamefont {C.~A.}\ \bibnamefont
  {Manzari}},\ }\href {\doibase 10.1103/PhysRevLett.127.071801} {\bibfield
  {journal} {\bibinfo  {journal} {Phys. Rev. Lett.}\ }\textbf {\bibinfo
  {volume} {127}},\ \bibinfo {pages} {071801} (\bibinfo {year}
  {2021}{\natexlab{c}})},\ \Eprint {http://arxiv.org/abs/2102.02825}
  {arXiv:2102.02825 [hep-ph]} \BibitemShut {NoStop}%
\bibitem [{\citenamefont {Antonelli}\ \emph {et~al.}(2008)\citenamefont
  {Antonelli} \emph {et~al.}}]{FlaviaNetWorkingGrouponKaonDecays:2008hpm}%
  \BibitemOpen
  \bibfield  {author} {\bibinfo {author} {\bibfnamefont {M.}~\bibnamefont
  {Antonelli}} \emph {et~al.} (\bibinfo {collaboration} {FlaviaNet Working
  Group on Kaon Decays}),\ }in\ \href@noop {} {\emph {\bibinfo {booktitle}
  {{5th International Workshop on e+ e- Collisions from Phi to Psi}}}}\
  (\bibinfo {year} {2008})\ \Eprint {http://arxiv.org/abs/0801.1817}
  {arXiv:0801.1817 [hep-ph]} \BibitemShut {NoStop}%
\bibitem [{\citenamefont {Grzadkowski}\ \emph {et~al.}(2010)\citenamefont
  {Grzadkowski}, \citenamefont {Iskrzynski}, \citenamefont {Misiak},\ and\
  \citenamefont {Rosiek}}]{Grzadkowski:2010es}%
  \BibitemOpen
  \bibfield  {author} {\bibinfo {author} {\bibfnamefont {B.}~\bibnamefont
  {Grzadkowski}}, \bibinfo {author} {\bibfnamefont {M.}~\bibnamefont
  {Iskrzynski}}, \bibinfo {author} {\bibfnamefont {M.}~\bibnamefont {Misiak}},
  \ and\ \bibinfo {author} {\bibfnamefont {J.}~\bibnamefont {Rosiek}},\ }\href
  {\doibase 10.1007/JHEP10(2010)085} {\bibfield  {journal} {\bibinfo  {journal}
  {JHEP}\ }\textbf {\bibinfo {volume} {10}},\ \bibinfo {pages} {085} (\bibinfo
  {year} {2010})},\ \Eprint {http://arxiv.org/abs/1008.4884} {arXiv:1008.4884
  [hep-ph]} \BibitemShut {NoStop}%
\bibitem [{\citenamefont {Schael}\ \emph {et~al.}(2006)\citenamefont {Schael}
  \emph {et~al.}}]{ALEPH:2005ab}%
  \BibitemOpen
  \bibfield  {author} {\bibinfo {author} {\bibfnamefont {S.}~\bibnamefont
  {Schael}} \emph {et~al.} (\bibinfo {collaboration} {ALEPH, DELPHI, L3, OPAL,
  SLD, LEP Electroweak Working Group, SLD Electroweak Group, SLD Heavy Flavour
  Group}),\ }\href {\doibase 10.1016/j.physrep.2005.12.006} {\bibfield
  {journal} {\bibinfo  {journal} {Phys. Rept.}\ }\textbf {\bibinfo {volume}
  {427}},\ \bibinfo {pages} {257} (\bibinfo {year} {2006})},\ \Eprint
  {http://arxiv.org/abs/hep-ex/0509008} {arXiv:hep-ex/0509008} \BibitemShut
  {NoStop}%
\bibitem [{\citenamefont {Schael}\ \emph {et~al.}(2013)\citenamefont {Schael}
  \emph {et~al.}}]{ALEPH:2013dgf}%
  \BibitemOpen
  \bibfield  {author} {\bibinfo {author} {\bibfnamefont {S.}~\bibnamefont
  {Schael}} \emph {et~al.} (\bibinfo {collaboration} {ALEPH, DELPHI, L3, OPAL,
  LEP Electroweak}),\ }\href {\doibase 10.1016/j.physrep.2013.07.004}
  {\bibfield  {journal} {\bibinfo  {journal} {Phys. Rept.}\ }\textbf {\bibinfo
  {volume} {532}},\ \bibinfo {pages} {119} (\bibinfo {year} {2013})},\ \Eprint
  {http://arxiv.org/abs/1302.3415} {arXiv:1302.3415 [hep-ex]} \BibitemShut
  {NoStop}%
\bibitem [{\citenamefont {Eichten}\ \emph {et~al.}(1984)\citenamefont
  {Eichten}, \citenamefont {Hinchliffe}, \citenamefont {Lane},\ and\
  \citenamefont {Quigg}}]{Eichten:1984eu}%
  \BibitemOpen
  \bibfield  {author} {\bibinfo {author} {\bibfnamefont {E.}~\bibnamefont
  {Eichten}}, \bibinfo {author} {\bibfnamefont {I.}~\bibnamefont {Hinchliffe}},
  \bibinfo {author} {\bibfnamefont {K.~D.}\ \bibnamefont {Lane}}, \ and\
  \bibinfo {author} {\bibfnamefont {C.}~\bibnamefont {Quigg}},\ }\href
  {\doibase 10.1103/RevModPhys.56.579} {\bibfield  {journal} {\bibinfo
  {journal} {Rev. Mod. Phys.}\ }\textbf {\bibinfo {volume} {56}},\ \bibinfo
  {pages} {579} (\bibinfo {year} {1984})},\ \bibinfo {note} {[Addendum:
  Rev.Mod.Phys. 58, 1065--1073 (1986)]}\BibitemShut {NoStop}%
\bibitem [{\citenamefont {Eichten}\ \emph {et~al.}(1983)\citenamefont
  {Eichten}, \citenamefont {Lane},\ and\ \citenamefont
  {Peskin}}]{Eichten:1983hw}%
  \BibitemOpen
  \bibfield  {author} {\bibinfo {author} {\bibfnamefont {E.}~\bibnamefont
  {Eichten}}, \bibinfo {author} {\bibfnamefont {K.~D.}\ \bibnamefont {Lane}}, \
  and\ \bibinfo {author} {\bibfnamefont {M.~E.}\ \bibnamefont {Peskin}},\
  }\href {\doibase 10.1103/PhysRevLett.50.811} {\bibfield  {journal} {\bibinfo
  {journal} {Phys. Rev. Lett.}\ }\textbf {\bibinfo {volume} {50}},\ \bibinfo
  {pages} {811} (\bibinfo {year} {1983})}\BibitemShut {NoStop}%
\bibitem [{\citenamefont {Aad}\ \emph {et~al.}(2020)\citenamefont {Aad} \emph
  {et~al.}}]{Aad:2020otl}%
  \BibitemOpen
  \bibfield  {author} {\bibinfo {author} {\bibfnamefont {G.}~\bibnamefont
  {Aad}} \emph {et~al.} (\bibinfo {collaboration} {ATLAS}),\ }\href {\doibase
  10.1007/JHEP11(2020)005} {\bibfield  {journal} {\bibinfo  {journal} {JHEP}\
  }\textbf {\bibinfo {volume} {11}},\ \bibinfo {pages} {005} (\bibinfo {year}
  {2020})},\ \bibinfo {note} {[Erratum: JHEP 04, 142 (2021)]},\ \Eprint
  {http://arxiv.org/abs/2006.12946} {arXiv:2006.12946 [hep-ex]} \BibitemShut
  {NoStop}%
\bibitem [{\citenamefont {Greljo}\ and\ \citenamefont
  {Marzocca}(2017)}]{Greljo:2017vvb}%
  \BibitemOpen
  \bibfield  {author} {\bibinfo {author} {\bibfnamefont {A.}~\bibnamefont
  {Greljo}}\ and\ \bibinfo {author} {\bibfnamefont {D.}~\bibnamefont
  {Marzocca}},\ }\href {\doibase 10.1140/epjc/s10052-017-5119-8} {\bibfield
  {journal} {\bibinfo  {journal} {Eur. Phys. J. C}\ }\textbf {\bibinfo {volume}
  {77}},\ \bibinfo {pages} {548} (\bibinfo {year} {2017})},\ \Eprint
  {http://arxiv.org/abs/1704.09015} {arXiv:1704.09015 [hep-ph]} \BibitemShut
  {NoStop}%
\bibitem [{\citenamefont {de~Blas}\ \emph {et~al.}(2018)\citenamefont
  {de~Blas}, \citenamefont {Criado}, \citenamefont {Perez-Victoria},\ and\
  \citenamefont {Santiago}}]{deBlas:2017xtg}%
  \BibitemOpen
  \bibfield  {author} {\bibinfo {author} {\bibfnamefont {J.}~\bibnamefont
  {de~Blas}}, \bibinfo {author} {\bibfnamefont {J.~C.}\ \bibnamefont {Criado}},
  \bibinfo {author} {\bibfnamefont {M.}~\bibnamefont {Perez-Victoria}}, \ and\
  \bibinfo {author} {\bibfnamefont {J.}~\bibnamefont {Santiago}},\ }\href
  {\doibase 10.1007/JHEP03(2018)109} {\bibfield  {journal} {\bibinfo  {journal}
  {JHEP}\ }\textbf {\bibinfo {volume} {03}},\ \bibinfo {pages} {109} (\bibinfo
  {year} {2018})},\ \Eprint {http://arxiv.org/abs/1711.10391} {arXiv:1711.10391
  [hep-ph]} \BibitemShut {NoStop}%
\bibitem [{\citenamefont {Kirk}(2021)}]{Kirk:2020wdk}%
  \BibitemOpen
  \bibfield  {author} {\bibinfo {author} {\bibfnamefont {M.}~\bibnamefont
  {Kirk}},\ }\href {\doibase 10.1103/PhysRevD.103.035004} {\bibfield  {journal}
  {\bibinfo  {journal} {Phys. Rev. D}\ }\textbf {\bibinfo {volume} {103}},\
  \bibinfo {pages} {035004} (\bibinfo {year} {2021})},\ \Eprint
  {http://arxiv.org/abs/2008.03261} {arXiv:2008.03261 [hep-ph]} \BibitemShut
  {NoStop}%
\bibitem [{\citenamefont {Weinberg}(1963)}]{Weinberg:1962hj}%
  \BibitemOpen
  \bibfield  {author} {\bibinfo {author} {\bibfnamefont {S.}~\bibnamefont
  {Weinberg}},\ }\href {\doibase 10.1103/PhysRev.130.776} {\bibfield  {journal}
  {\bibinfo  {journal} {Phys. Rev.}\ }\textbf {\bibinfo {volume} {130}},\
  \bibinfo {pages} {776} (\bibinfo {year} {1963})}\BibitemShut {NoStop}%
\bibitem [{\citenamefont {Susskind}(1979)}]{Susskind:1978ms}%
  \BibitemOpen
  \bibfield  {author} {\bibinfo {author} {\bibfnamefont {L.}~\bibnamefont
  {Susskind}},\ }\href {\doibase 10.1103/PhysRevD.20.2619} {\bibfield
  {journal} {\bibinfo  {journal} {Phys. Rev. D}\ }\textbf {\bibinfo {volume}
  {20}},\ \bibinfo {pages} {2619} (\bibinfo {year} {1979})}\BibitemShut
  {NoStop}%
\bibitem [{\citenamefont {Randall}\ and\ \citenamefont
  {Sundrum}(1999)}]{Randall:1999ee}%
  \BibitemOpen
  \bibfield  {author} {\bibinfo {author} {\bibfnamefont {L.}~\bibnamefont
  {Randall}}\ and\ \bibinfo {author} {\bibfnamefont {R.}~\bibnamefont
  {Sundrum}},\ }\href {\doibase 10.1103/PhysRevLett.83.3370} {\bibfield
  {journal} {\bibinfo  {journal} {Phys. Rev. Lett.}\ }\textbf {\bibinfo
  {volume} {83}},\ \bibinfo {pages} {3370} (\bibinfo {year} {1999})},\ \Eprint
  {http://arxiv.org/abs/hep-ph/9905221} {arXiv:hep-ph/9905221} \BibitemShut
  {NoStop}%
\bibitem [{\citenamefont {Malkawi}\ \emph {et~al.}(1996)\citenamefont
  {Malkawi}, \citenamefont {Tait},\ and\ \citenamefont
  {Yuan}}]{Malkawi:1996fs}%
  \BibitemOpen
  \bibfield  {author} {\bibinfo {author} {\bibfnamefont {E.}~\bibnamefont
  {Malkawi}}, \bibinfo {author} {\bibfnamefont {T.~M.~P.}\ \bibnamefont
  {Tait}}, \ and\ \bibinfo {author} {\bibfnamefont {C.~P.}\ \bibnamefont
  {Yuan}},\ }\href {\doibase 10.1016/0370-2693(96)00859-3} {\bibfield
  {journal} {\bibinfo  {journal} {Phys. Lett. B}\ }\textbf {\bibinfo {volume}
  {385}},\ \bibinfo {pages} {304} (\bibinfo {year} {1996})},\ \Eprint
  {http://arxiv.org/abs/hep-ph/9603349} {arXiv:hep-ph/9603349} \BibitemShut
  {NoStop}%
\bibitem [{\citenamefont {Hsieh}\ \emph {et~al.}(2010)\citenamefont {Hsieh},
  \citenamefont {Schmitz}, \citenamefont {Yu},\ and\ \citenamefont
  {Yuan}}]{Hsieh:2010zr}%
  \BibitemOpen
  \bibfield  {author} {\bibinfo {author} {\bibfnamefont {K.}~\bibnamefont
  {Hsieh}}, \bibinfo {author} {\bibfnamefont {K.}~\bibnamefont {Schmitz}},
  \bibinfo {author} {\bibfnamefont {J.-H.}\ \bibnamefont {Yu}}, \ and\ \bibinfo
  {author} {\bibfnamefont {C.~P.}\ \bibnamefont {Yuan}},\ }\href {\doibase
  10.1103/PhysRevD.82.035011} {\bibfield  {journal} {\bibinfo  {journal} {Phys.
  Rev. D}\ }\textbf {\bibinfo {volume} {82}},\ \bibinfo {pages} {035011}
  (\bibinfo {year} {2010})},\ \Eprint {http://arxiv.org/abs/1003.3482}
  {arXiv:1003.3482 [hep-ph]} \BibitemShut {NoStop}%
\bibitem [{\citenamefont {Lee}\ and\ \citenamefont
  {Shrock}(1977)}]{Lee:1977tib}%
  \BibitemOpen
  \bibfield  {author} {\bibinfo {author} {\bibfnamefont {B.~W.}\ \bibnamefont
  {Lee}}\ and\ \bibinfo {author} {\bibfnamefont {R.~E.}\ \bibnamefont
  {Shrock}},\ }\href {\doibase 10.1103/PhysRevD.16.1444} {\bibfield  {journal}
  {\bibinfo  {journal} {Phys. Rev. D}\ }\textbf {\bibinfo {volume} {16}},\
  \bibinfo {pages} {1444} (\bibinfo {year} {1977})}\BibitemShut {NoStop}%
\bibitem [{\citenamefont {del Aguila}\ \emph {et~al.}(2008)\citenamefont {del
  Aguila}, \citenamefont {de~Blas},\ and\ \citenamefont
  {Perez-Victoria}}]{delAguila:2008pw}%
  \BibitemOpen
  \bibfield  {author} {\bibinfo {author} {\bibfnamefont {F.}~\bibnamefont {del
  Aguila}}, \bibinfo {author} {\bibfnamefont {J.}~\bibnamefont {de~Blas}}, \
  and\ \bibinfo {author} {\bibfnamefont {M.}~\bibnamefont {Perez-Victoria}},\
  }\href {\doibase 10.1103/PhysRevD.78.013010} {\bibfield  {journal} {\bibinfo
  {journal} {Phys. Rev. D}\ }\textbf {\bibinfo {volume} {78}},\ \bibinfo
  {pages} {013010} (\bibinfo {year} {2008})},\ \Eprint
  {http://arxiv.org/abs/0803.4008} {arXiv:0803.4008 [hep-ph]} \BibitemShut
  {NoStop}%
\bibitem [{\citenamefont {Hewett}\ and\ \citenamefont
  {Rizzo}(1989)}]{Hewett:1988xc}%
  \BibitemOpen
  \bibfield  {author} {\bibinfo {author} {\bibfnamefont {J.~L.}\ \bibnamefont
  {Hewett}}\ and\ \bibinfo {author} {\bibfnamefont {T.~G.}\ \bibnamefont
  {Rizzo}},\ }\href {\doibase 10.1016/0370-1573(89)90071-9} {\bibfield
  {journal} {\bibinfo  {journal} {Phys. Rept.}\ }\textbf {\bibinfo {volume}
  {183}},\ \bibinfo {pages} {193} (\bibinfo {year} {1989})}\BibitemShut
  {NoStop}%
\bibitem [{\citenamefont {Langacker}(1981)}]{Langacker:1980js}%
  \BibitemOpen
  \bibfield  {author} {\bibinfo {author} {\bibfnamefont {P.}~\bibnamefont
  {Langacker}},\ }\href {\doibase 10.1016/0370-1573(81)90059-4} {\bibfield
  {journal} {\bibinfo  {journal} {Phys. Rept.}\ }\textbf {\bibinfo {volume}
  {72}},\ \bibinfo {pages} {185} (\bibinfo {year} {1981})}\BibitemShut
  {NoStop}%
\bibitem [{\citenamefont {del Aguila}\ and\ \citenamefont
  {Bowick}(1983)}]{delAguila:1982fs}%
  \BibitemOpen
  \bibfield  {author} {\bibinfo {author} {\bibfnamefont {F.}~\bibnamefont {del
  Aguila}}\ and\ \bibinfo {author} {\bibfnamefont {M.~J.}\ \bibnamefont
  {Bowick}},\ }\href {\doibase 10.1016/0550-3213(83)90316-4} {\bibfield
  {journal} {\bibinfo  {journal} {Nucl. Phys. B}\ }\textbf {\bibinfo {volume}
  {224}},\ \bibinfo {pages} {107} (\bibinfo {year} {1983})}\BibitemShut
  {NoStop}%
\bibitem [{\citenamefont {Antoniadis}(1990)}]{Antoniadis:1990ew}%
  \BibitemOpen
  \bibfield  {author} {\bibinfo {author} {\bibfnamefont {I.}~\bibnamefont
  {Antoniadis}},\ }\href {\doibase 10.1016/0370-2693(90)90617-F} {\bibfield
  {journal} {\bibinfo  {journal} {Phys. Lett. B}\ }\textbf {\bibinfo {volume}
  {246}},\ \bibinfo {pages} {377} (\bibinfo {year} {1990})}\BibitemShut
  {NoStop}%
\bibitem [{\citenamefont {Arkani-Hamed}\ \emph
  {et~al.}(2001{\natexlab{a}})\citenamefont {Arkani-Hamed}, \citenamefont
  {Dimopoulos},\ and\ \citenamefont {March-Russell}}]{ArkaniHamed:1998kx}%
  \BibitemOpen
  \bibfield  {author} {\bibinfo {author} {\bibfnamefont {N.}~\bibnamefont
  {Arkani-Hamed}}, \bibinfo {author} {\bibfnamefont {S.}~\bibnamefont
  {Dimopoulos}}, \ and\ \bibinfo {author} {\bibfnamefont {J.}~\bibnamefont
  {March-Russell}},\ }\href {\doibase 10.1103/PhysRevD.63.064020} {\bibfield
  {journal} {\bibinfo  {journal} {Phys. Rev. D}\ }\textbf {\bibinfo {volume}
  {63}},\ \bibinfo {pages} {064020} (\bibinfo {year} {2001}{\natexlab{a}})},\
  \Eprint {http://arxiv.org/abs/hep-th/9809124} {arXiv:hep-th/9809124}
  \BibitemShut {NoStop}%
\bibitem [{\citenamefont {Csaki}(2004)}]{Csaki:2004ay}%
  \BibitemOpen
  \bibfield  {author} {\bibinfo {author} {\bibfnamefont {C.}~\bibnamefont
  {Csaki}},\ }in\ \href@noop {} {\emph {\bibinfo {booktitle} {{Theoretical
  Advanced Study Institute in Elementary Particle Physics (TASI 2002): Particle
  Physics and Cosmology: The Quest for Physics Beyond the Standard
  Model(s)}}}}\ (\bibinfo {year} {2004})\ pp.\ \bibinfo {pages} {605--698},\
  \Eprint {http://arxiv.org/abs/hep-ph/0404096} {arXiv:hep-ph/0404096}
  \BibitemShut {NoStop}%
\bibitem [{\citenamefont {Arkani-Hamed}\ \emph
  {et~al.}(2001{\natexlab{b}})\citenamefont {Arkani-Hamed}, \citenamefont
  {Cohen},\ and\ \citenamefont {Georgi}}]{ArkaniHamed:2001nc}%
  \BibitemOpen
  \bibfield  {author} {\bibinfo {author} {\bibfnamefont {N.}~\bibnamefont
  {Arkani-Hamed}}, \bibinfo {author} {\bibfnamefont {A.~G.}\ \bibnamefont
  {Cohen}}, \ and\ \bibinfo {author} {\bibfnamefont {H.}~\bibnamefont
  {Georgi}},\ }\href {\doibase 10.1016/S0370-2693(01)00741-9} {\bibfield
  {journal} {\bibinfo  {journal} {Phys. Lett. B}\ }\textbf {\bibinfo {volume}
  {513}},\ \bibinfo {pages} {232} (\bibinfo {year} {2001}{\natexlab{b}})},\
  \Eprint {http://arxiv.org/abs/hep-ph/0105239} {arXiv:hep-ph/0105239}
  \BibitemShut {NoStop}%
\bibitem [{\citenamefont {Arkani-Hamed}\ \emph {et~al.}(2002)\citenamefont
  {Arkani-Hamed}, \citenamefont {Cohen}, \citenamefont {Katz},\ and\
  \citenamefont {Nelson}}]{ArkaniHamed:2002qy}%
  \BibitemOpen
  \bibfield  {author} {\bibinfo {author} {\bibfnamefont {N.}~\bibnamefont
  {Arkani-Hamed}}, \bibinfo {author} {\bibfnamefont {A.~G.}\ \bibnamefont
  {Cohen}}, \bibinfo {author} {\bibfnamefont {E.}~\bibnamefont {Katz}}, \ and\
  \bibinfo {author} {\bibfnamefont {A.~E.}\ \bibnamefont {Nelson}},\ }\href
  {\doibase 10.1088/1126-6708/2002/07/034} {\bibfield  {journal} {\bibinfo
  {journal} {JHEP}\ }\textbf {\bibinfo {volume} {07}},\ \bibinfo {pages} {034}
  (\bibinfo {year} {2002})},\ \Eprint {http://arxiv.org/abs/hep-ph/0206021}
  {arXiv:hep-ph/0206021} \BibitemShut {NoStop}%
\bibitem [{\citenamefont {Perelstein}(2007)}]{Perelstein:2005ka}%
  \BibitemOpen
  \bibfield  {author} {\bibinfo {author} {\bibfnamefont {M.}~\bibnamefont
  {Perelstein}},\ }\href {\doibase 10.1016/j.ppnp.2006.04.001} {\bibfield
  {journal} {\bibinfo  {journal} {Prog. Part. Nucl. Phys.}\ }\textbf {\bibinfo
  {volume} {58}},\ \bibinfo {pages} {247} (\bibinfo {year} {2007})},\ \Eprint
  {http://arxiv.org/abs/hep-ph/0512128} {arXiv:hep-ph/0512128} \BibitemShut
  {NoStop}%
\bibitem [{\citenamefont {del Aguila}\ \emph {et~al.}(2010)\citenamefont {del
  Aguila}, \citenamefont {Carmona},\ and\ \citenamefont
  {Santiago}}]{delAguila:2010vg}%
  \BibitemOpen
  \bibfield  {author} {\bibinfo {author} {\bibfnamefont {F.}~\bibnamefont {del
  Aguila}}, \bibinfo {author} {\bibfnamefont {A.}~\bibnamefont {Carmona}}, \
  and\ \bibinfo {author} {\bibfnamefont {J.}~\bibnamefont {Santiago}},\ }\href
  {\doibase 10.1007/JHEP08(2010)127} {\bibfield  {journal} {\bibinfo  {journal}
  {JHEP}\ }\textbf {\bibinfo {volume} {08}},\ \bibinfo {pages} {127} (\bibinfo
  {year} {2010})},\ \Eprint {http://arxiv.org/abs/1001.5151} {arXiv:1001.5151
  [hep-ph]} \BibitemShut {NoStop}%
\bibitem [{\citenamefont {Carmona}\ and\ \citenamefont
  {Goertz}(2013)}]{Carmona:2013cq}%
  \BibitemOpen
  \bibfield  {author} {\bibinfo {author} {\bibfnamefont {A.}~\bibnamefont
  {Carmona}}\ and\ \bibinfo {author} {\bibfnamefont {F.}~\bibnamefont
  {Goertz}},\ }\href {\doibase 10.1007/JHEP04(2013)163} {\bibfield  {journal}
  {\bibinfo  {journal} {JHEP}\ }\textbf {\bibinfo {volume} {04}},\ \bibinfo
  {pages} {163} (\bibinfo {year} {2013})},\ \Eprint
  {http://arxiv.org/abs/1301.5856} {arXiv:1301.5856 [hep-ph]} \BibitemShut
  {NoStop}%
\bibitem [{\citenamefont {Minkowski}(1977)}]{Minkowski:1977sc}%
  \BibitemOpen
  \bibfield  {author} {\bibinfo {author} {\bibfnamefont {P.}~\bibnamefont
  {Minkowski}},\ }\href {\doibase 10.1016/0370-2693(77)90435-X} {\bibfield
  {journal} {\bibinfo  {journal} {Phys. Lett. B}\ }\textbf {\bibinfo {volume}
  {67}},\ \bibinfo {pages} {421} (\bibinfo {year} {1977})}\BibitemShut
  {NoStop}%
\bibitem [{\citenamefont {Foot}\ \emph {et~al.}(1989)\citenamefont {Foot},
  \citenamefont {Lew}, \citenamefont {He},\ and\ \citenamefont
  {Joshi}}]{Foot:1988aq}%
  \BibitemOpen
  \bibfield  {author} {\bibinfo {author} {\bibfnamefont {R.}~\bibnamefont
  {Foot}}, \bibinfo {author} {\bibfnamefont {H.}~\bibnamefont {Lew}}, \bibinfo
  {author} {\bibfnamefont {X.~G.}\ \bibnamefont {He}}, \ and\ \bibinfo {author}
  {\bibfnamefont {G.~C.}\ \bibnamefont {Joshi}},\ }\href {\doibase
  10.1007/BF01415558} {\bibfield  {journal} {\bibinfo  {journal} {Z. Phys. C}\
  }\textbf {\bibinfo {volume} {44}},\ \bibinfo {pages} {441} (\bibinfo {year}
  {1989})}\BibitemShut {NoStop}%
\bibitem [{\citenamefont {Crivellin}\ \emph
  {et~al.}(2021{\natexlab{d}})\citenamefont {Crivellin}, \citenamefont {Kirk},
  \citenamefont {Manzari},\ and\ \citenamefont {Panizzi}}]{Crivellin:2020klg}%
  \BibitemOpen
  \bibfield  {author} {\bibinfo {author} {\bibfnamefont {A.}~\bibnamefont
  {Crivellin}}, \bibinfo {author} {\bibfnamefont {F.}~\bibnamefont {Kirk}},
  \bibinfo {author} {\bibfnamefont {C.~A.}\ \bibnamefont {Manzari}}, \ and\
  \bibinfo {author} {\bibfnamefont {L.}~\bibnamefont {Panizzi}},\ }\href
  {\doibase 10.1103/PhysRevD.103.073002} {\bibfield  {journal} {\bibinfo
  {journal} {Phys. Rev. D}\ }\textbf {\bibinfo {volume} {103}},\ \bibinfo
  {pages} {073002} (\bibinfo {year} {2021}{\natexlab{d}})},\ \Eprint
  {http://arxiv.org/abs/2012.09845} {arXiv:2012.09845 [hep-ph]} \BibitemShut
  {NoStop}%
\bibitem [{\citenamefont {Felkl}\ \emph {et~al.}(2021)\citenamefont {Felkl},
  \citenamefont {Herrero-Garcia},\ and\ \citenamefont
  {Schmidt}}]{Felkl:2021qdn}%
  \BibitemOpen
  \bibfield  {author} {\bibinfo {author} {\bibfnamefont {T.}~\bibnamefont
  {Felkl}}, \bibinfo {author} {\bibfnamefont {J.}~\bibnamefont
  {Herrero-Garcia}}, \ and\ \bibinfo {author} {\bibfnamefont {M.~A.}\
  \bibnamefont {Schmidt}},\ }\href {\doibase 10.1007/JHEP05(2021)122}
  {\bibfield  {journal} {\bibinfo  {journal} {JHEP}\ }\textbf {\bibinfo
  {volume} {05}},\ \bibinfo {pages} {122} (\bibinfo {year} {2021})},\ \Eprint
  {http://arxiv.org/abs/2102.09898} {arXiv:2102.09898 [hep-ph]} \BibitemShut
  {NoStop}%
\bibitem [{\citenamefont {Marzocca}\ and\ \citenamefont
  {Trifinopoulos}(2021)}]{Marzocca:2021azj}%
  \BibitemOpen
  \bibfield  {author} {\bibinfo {author} {\bibfnamefont {D.}~\bibnamefont
  {Marzocca}}\ and\ \bibinfo {author} {\bibfnamefont {S.}~\bibnamefont
  {Trifinopoulos}},\ }\href {\doibase 10.1103/PhysRevLett.127.061803}
  {\bibfield  {journal} {\bibinfo  {journal} {Phys. Rev. Lett.}\ }\textbf
  {\bibinfo {volume} {127}},\ \bibinfo {pages} {061803} (\bibinfo {year}
  {2021})},\ \Eprint {http://arxiv.org/abs/2104.05730} {arXiv:2104.05730
  [hep-ph]} \BibitemShut {NoStop}%
\bibitem [{\citenamefont {Zee}(1986)}]{Zee:1985id}%
  \BibitemOpen
  \bibfield  {author} {\bibinfo {author} {\bibfnamefont {A.}~\bibnamefont
  {Zee}},\ }\href {\doibase 10.1016/0550-3213(86)90475-X} {\bibfield  {journal}
  {\bibinfo  {journal} {Nucl. Phys. B}\ }\textbf {\bibinfo {volume} {264}},\
  \bibinfo {pages} {99} (\bibinfo {year} {1986})}\BibitemShut {NoStop}%
\bibitem [{\citenamefont {Babu}(1988)}]{Babu:1988ki}%
  \BibitemOpen
  \bibfield  {author} {\bibinfo {author} {\bibfnamefont {K.~S.}\ \bibnamefont
  {Babu}},\ }\href {\doibase 10.1016/0370-2693(88)91584-5} {\bibfield
  {journal} {\bibinfo  {journal} {Phys. Lett. B}\ }\textbf {\bibinfo {volume}
  {203}},\ \bibinfo {pages} {132} (\bibinfo {year} {1988})}\BibitemShut
  {NoStop}%
\bibitem [{\citenamefont {Krauss}\ \emph {et~al.}(2003)\citenamefont {Krauss},
  \citenamefont {Nasri},\ and\ \citenamefont {Trodden}}]{Krauss:2002px}%
  \BibitemOpen
  \bibfield  {author} {\bibinfo {author} {\bibfnamefont {L.~M.}\ \bibnamefont
  {Krauss}}, \bibinfo {author} {\bibfnamefont {S.}~\bibnamefont {Nasri}}, \
  and\ \bibinfo {author} {\bibfnamefont {M.}~\bibnamefont {Trodden}},\ }\href
  {\doibase 10.1103/PhysRevD.67.085002} {\bibfield  {journal} {\bibinfo
  {journal} {Phys. Rev. D}\ }\textbf {\bibinfo {volume} {67}},\ \bibinfo
  {pages} {085002} (\bibinfo {year} {2003})},\ \Eprint
  {http://arxiv.org/abs/hep-ph/0210389} {arXiv:hep-ph/0210389} \BibitemShut
  {NoStop}%
\bibitem [{\citenamefont {Nebot}\ \emph {et~al.}(2008)\citenamefont {Nebot},
  \citenamefont {Oliver}, \citenamefont {Palao},\ and\ \citenamefont
  {Santamaria}}]{Nebot:2007bc}%
  \BibitemOpen
  \bibfield  {author} {\bibinfo {author} {\bibfnamefont {M.}~\bibnamefont
  {Nebot}}, \bibinfo {author} {\bibfnamefont {J.~F.}\ \bibnamefont {Oliver}},
  \bibinfo {author} {\bibfnamefont {D.}~\bibnamefont {Palao}}, \ and\ \bibinfo
  {author} {\bibfnamefont {A.}~\bibnamefont {Santamaria}},\ }\href {\doibase
  10.1103/PhysRevD.77.093013} {\bibfield  {journal} {\bibinfo  {journal} {Phys.
  Rev. D}\ }\textbf {\bibinfo {volume} {77}},\ \bibinfo {pages} {093013}
  (\bibinfo {year} {2008})},\ \Eprint {http://arxiv.org/abs/0711.0483}
  {arXiv:0711.0483 [hep-ph]} \BibitemShut {NoStop}%
\bibitem [{\citenamefont {Cai}\ \emph {et~al.}(2015)\citenamefont {Cai},
  \citenamefont {Clarke}, \citenamefont {Schmidt},\ and\ \citenamefont
  {Volkas}}]{Cai:2014kra}%
  \BibitemOpen
  \bibfield  {author} {\bibinfo {author} {\bibfnamefont {Y.}~\bibnamefont
  {Cai}}, \bibinfo {author} {\bibfnamefont {J.~D.}\ \bibnamefont {Clarke}},
  \bibinfo {author} {\bibfnamefont {M.~A.}\ \bibnamefont {Schmidt}}, \ and\
  \bibinfo {author} {\bibfnamefont {R.~R.}\ \bibnamefont {Volkas}},\ }\href
  {\doibase 10.1007/JHEP02(2015)161} {\bibfield  {journal} {\bibinfo  {journal}
  {JHEP}\ }\textbf {\bibinfo {volume} {02}},\ \bibinfo {pages} {161} (\bibinfo
  {year} {2015})},\ \Eprint {http://arxiv.org/abs/1410.0689} {arXiv:1410.0689
  [hep-ph]} \BibitemShut {NoStop}%
\bibitem [{\citenamefont {Cheung}\ and\ \citenamefont
  {Seto}(2004)}]{Cheung:2004xm}%
  \BibitemOpen
  \bibfield  {author} {\bibinfo {author} {\bibfnamefont {K.}~\bibnamefont
  {Cheung}}\ and\ \bibinfo {author} {\bibfnamefont {O.}~\bibnamefont {Seto}},\
  }\href {\doibase 10.1103/PhysRevD.69.113009} {\bibfield  {journal} {\bibinfo
  {journal} {Phys. Rev. D}\ }\textbf {\bibinfo {volume} {69}},\ \bibinfo
  {pages} {113009} (\bibinfo {year} {2004})},\ \Eprint
  {http://arxiv.org/abs/hep-ph/0403003} {arXiv:hep-ph/0403003} \BibitemShut
  {NoStop}%
\bibitem [{\citenamefont {Ahriche}\ \emph {et~al.}(2014)\citenamefont
  {Ahriche}, \citenamefont {Nasri},\ and\ \citenamefont
  {Soualah}}]{Ahriche:2014xra}%
  \BibitemOpen
  \bibfield  {author} {\bibinfo {author} {\bibfnamefont {A.}~\bibnamefont
  {Ahriche}}, \bibinfo {author} {\bibfnamefont {S.}~\bibnamefont {Nasri}}, \
  and\ \bibinfo {author} {\bibfnamefont {R.}~\bibnamefont {Soualah}},\ }\href
  {\doibase 10.1103/PhysRevD.89.095010} {\bibfield  {journal} {\bibinfo
  {journal} {Phys. Rev. D}\ }\textbf {\bibinfo {volume} {89}},\ \bibinfo
  {pages} {095010} (\bibinfo {year} {2014})},\ \Eprint
  {http://arxiv.org/abs/1403.5694} {arXiv:1403.5694 [hep-ph]} \BibitemShut
  {NoStop}%
\bibitem [{\citenamefont {Chen}\ \emph {et~al.}(2014)\citenamefont {Chen},
  \citenamefont {McDonald},\ and\ \citenamefont {Nasri}}]{Chen:2014ska}%
  \BibitemOpen
  \bibfield  {author} {\bibinfo {author} {\bibfnamefont {C.-S.}\ \bibnamefont
  {Chen}}, \bibinfo {author} {\bibfnamefont {K.~L.}\ \bibnamefont {McDonald}},
  \ and\ \bibinfo {author} {\bibfnamefont {S.}~\bibnamefont {Nasri}},\ }\href
  {\doibase 10.1016/j.physletb.2014.05.082} {\bibfield  {journal} {\bibinfo
  {journal} {Phys. Lett. B}\ }\textbf {\bibinfo {volume} {734}},\ \bibinfo
  {pages} {388} (\bibinfo {year} {2014})},\ \Eprint
  {http://arxiv.org/abs/1404.6033} {arXiv:1404.6033 [hep-ph]} \BibitemShut
  {NoStop}%
\bibitem [{\citenamefont {Ahriche}\ \emph {et~al.}(2016)\citenamefont
  {Ahriche}, \citenamefont {McDonald},\ and\ \citenamefont
  {Nasri}}]{Ahriche:2015loa}%
  \BibitemOpen
  \bibfield  {author} {\bibinfo {author} {\bibfnamefont {A.}~\bibnamefont
  {Ahriche}}, \bibinfo {author} {\bibfnamefont {K.~L.}\ \bibnamefont
  {McDonald}}, \ and\ \bibinfo {author} {\bibfnamefont {S.}~\bibnamefont
  {Nasri}},\ }\href {\doibase 10.1007/JHEP02(2016)038} {\bibfield  {journal}
  {\bibinfo  {journal} {JHEP}\ }\textbf {\bibinfo {volume} {02}},\ \bibinfo
  {pages} {038} (\bibinfo {year} {2016})},\ \Eprint
  {http://arxiv.org/abs/1508.02607} {arXiv:1508.02607 [hep-ph]} \BibitemShut
  {NoStop}%
\bibitem [{\citenamefont {Herrero-Garcia}\ \emph {et~al.}(2014)\citenamefont
  {Herrero-Garcia}, \citenamefont {Nebot}, \citenamefont {Rius},\ and\
  \citenamefont {Santamaria}}]{Herrero-Garcia:2014hfa}%
  \BibitemOpen
  \bibfield  {author} {\bibinfo {author} {\bibfnamefont {J.}~\bibnamefont
  {Herrero-Garcia}}, \bibinfo {author} {\bibfnamefont {M.}~\bibnamefont
  {Nebot}}, \bibinfo {author} {\bibfnamefont {N.}~\bibnamefont {Rius}}, \ and\
  \bibinfo {author} {\bibfnamefont {A.}~\bibnamefont {Santamaria}},\ }\href
  {\doibase 10.1016/j.nuclphysb.2014.06.001} {\bibfield  {journal} {\bibinfo
  {journal} {Nucl. Phys. B}\ }\textbf {\bibinfo {volume} {885}},\ \bibinfo
  {pages} {542} (\bibinfo {year} {2014})},\ \Eprint
  {http://arxiv.org/abs/1402.4491} {arXiv:1402.4491 [hep-ph]} \BibitemShut
  {NoStop}%
\bibitem [{\citenamefont {Herrero-Garc\'\i{}a}\ \emph
  {et~al.}(2017)\citenamefont {Herrero-Garc\'\i{}a}, \citenamefont {Ohlsson},
  \citenamefont {Riad},\ and\ \citenamefont
  {Wir\'en}}]{Herrero-Garcia:2017xdu}%
  \BibitemOpen
  \bibfield  {author} {\bibinfo {author} {\bibfnamefont {J.}~\bibnamefont
  {Herrero-Garc\'\i{}a}}, \bibinfo {author} {\bibfnamefont {T.}~\bibnamefont
  {Ohlsson}}, \bibinfo {author} {\bibfnamefont {S.}~\bibnamefont {Riad}}, \
  and\ \bibinfo {author} {\bibfnamefont {J.}~\bibnamefont {Wir\'en}},\ }\href
  {\doibase 10.1007/JHEP04(2017)130} {\bibfield  {journal} {\bibinfo  {journal}
  {JHEP}\ }\textbf {\bibinfo {volume} {04}},\ \bibinfo {pages} {130} (\bibinfo
  {year} {2017})},\ \Eprint {http://arxiv.org/abs/1701.05345} {arXiv:1701.05345
  [hep-ph]} \BibitemShut {NoStop}%
\bibitem [{\citenamefont {Centelles~Chuli\'a}\ \emph
  {et~al.}(2018)\citenamefont {Centelles~Chuli\'a}, \citenamefont
  {Srivastava},\ and\ \citenamefont {Valle}}]{CentellesChulia:2018gwr}%
  \BibitemOpen
  \bibfield  {author} {\bibinfo {author} {\bibfnamefont {S.}~\bibnamefont
  {Centelles~Chuli\'a}}, \bibinfo {author} {\bibfnamefont {R.}~\bibnamefont
  {Srivastava}}, \ and\ \bibinfo {author} {\bibfnamefont {J.~W.~F.}\
  \bibnamefont {Valle}},\ }\href {\doibase 10.1016/j.physletb.2018.03.046}
  {\bibfield  {journal} {\bibinfo  {journal} {Phys. Lett. B}\ }\textbf
  {\bibinfo {volume} {781}},\ \bibinfo {pages} {122} (\bibinfo {year}
  {2018})},\ \Eprint {http://arxiv.org/abs/1802.05722} {arXiv:1802.05722
  [hep-ph]} \BibitemShut {NoStop}%
\bibitem [{\citenamefont {Babu}\ \emph {et~al.}(2020)\citenamefont {Babu},
  \citenamefont {Dev}, \citenamefont {Jana},\ and\ \citenamefont
  {Thapa}}]{Babu:2019mfe}%
  \BibitemOpen
  \bibfield  {author} {\bibinfo {author} {\bibfnamefont {K.~S.}\ \bibnamefont
  {Babu}}, \bibinfo {author} {\bibfnamefont {P.~S.~B.}\ \bibnamefont {Dev}},
  \bibinfo {author} {\bibfnamefont {S.}~\bibnamefont {Jana}}, \ and\ \bibinfo
  {author} {\bibfnamefont {A.}~\bibnamefont {Thapa}},\ }\href {\doibase
  10.1007/JHEP03(2020)006} {\bibfield  {journal} {\bibinfo  {journal} {JHEP}\
  }\textbf {\bibinfo {volume} {03}},\ \bibinfo {pages} {006} (\bibinfo {year}
  {2020})},\ \Eprint {http://arxiv.org/abs/1907.09498} {arXiv:1907.09498
  [hep-ph]} \BibitemShut {NoStop}%
\bibitem [{\citenamefont {Konetschny}\ and\ \citenamefont
  {Kummer}(1977)}]{Konetschny:1977bn}%
  \BibitemOpen
  \bibfield  {author} {\bibinfo {author} {\bibfnamefont {W.}~\bibnamefont
  {Konetschny}}\ and\ \bibinfo {author} {\bibfnamefont {W.}~\bibnamefont
  {Kummer}},\ }\href {\doibase 10.1016/0370-2693(77)90407-5} {\bibfield
  {journal} {\bibinfo  {journal} {Phys. Lett. B}\ }\textbf {\bibinfo {volume}
  {70}},\ \bibinfo {pages} {433} (\bibinfo {year} {1977})}\BibitemShut
  {NoStop}%
\bibitem [{\citenamefont {Magg}\ and\ \citenamefont
  {Wetterich}(1980)}]{Magg:1980ut}%
  \BibitemOpen
  \bibfield  {author} {\bibinfo {author} {\bibfnamefont {M.}~\bibnamefont
  {Magg}}\ and\ \bibinfo {author} {\bibfnamefont {C.}~\bibnamefont
  {Wetterich}},\ }\href {\doibase 10.1016/0370-2693(80)90825-4} {\bibfield
  {journal} {\bibinfo  {journal} {Phys. Lett. B}\ }\textbf {\bibinfo {volume}
  {94}},\ \bibinfo {pages} {61} (\bibinfo {year} {1980})}\BibitemShut {NoStop}%
\bibitem [{\citenamefont {Schechter}\ and\ \citenamefont
  {Valle}(1980)}]{Schechter:1980gr}%
  \BibitemOpen
  \bibfield  {author} {\bibinfo {author} {\bibfnamefont {J.}~\bibnamefont
  {Schechter}}\ and\ \bibinfo {author} {\bibfnamefont {J.~W.~F.}\ \bibnamefont
  {Valle}},\ }\href {\doibase 10.1103/PhysRevD.22.2227} {\bibfield  {journal}
  {\bibinfo  {journal} {Phys. Rev. D}\ }\textbf {\bibinfo {volume} {22}},\
  \bibinfo {pages} {2227} (\bibinfo {year} {1980})}\BibitemShut {NoStop}%
\bibitem [{\citenamefont {Cheng}\ and\ \citenamefont
  {Li}(1980)}]{Cheng:1980qt}%
  \BibitemOpen
  \bibfield  {author} {\bibinfo {author} {\bibfnamefont {T.~P.}\ \bibnamefont
  {Cheng}}\ and\ \bibinfo {author} {\bibfnamefont {L.-F.}\ \bibnamefont {Li}},\
  }\href {\doibase 10.1103/PhysRevD.22.2860} {\bibfield  {journal} {\bibinfo
  {journal} {Phys. Rev. D}\ }\textbf {\bibinfo {volume} {22}},\ \bibinfo
  {pages} {2860} (\bibinfo {year} {1980})}\BibitemShut {NoStop}%
\bibitem [{\citenamefont {Mohapatra}\ and\ \citenamefont
  {Senjanovic}(1981)}]{Mohapatra:1980yp}%
  \BibitemOpen
  \bibfield  {author} {\bibinfo {author} {\bibfnamefont {R.~N.}\ \bibnamefont
  {Mohapatra}}\ and\ \bibinfo {author} {\bibfnamefont {G.}~\bibnamefont
  {Senjanovic}},\ }\href {\doibase 10.1103/PhysRevD.23.165} {\bibfield
  {journal} {\bibinfo  {journal} {Phys. Rev. D}\ }\textbf {\bibinfo {volume}
  {23}},\ \bibinfo {pages} {165} (\bibinfo {year} {1981})}\BibitemShut
  {NoStop}%
\bibitem [{\citenamefont {Weinberg}(1979{\natexlab{b}})}]{Weinberg:1979sa}%
  \BibitemOpen
  \bibfield  {author} {\bibinfo {author} {\bibfnamefont {S.}~\bibnamefont
  {Weinberg}},\ }\href {\doibase 10.1103/PhysRevLett.43.1566} {\bibfield
  {journal} {\bibinfo  {journal} {Phys. Rev. Lett.}\ }\textbf {\bibinfo
  {volume} {43}},\ \bibinfo {pages} {1566} (\bibinfo {year}
  {1979}{\natexlab{b}})}\BibitemShut {NoStop}%
\bibitem [{\citenamefont {Buras}\ \emph {et~al.}(2021)\citenamefont {Buras},
  \citenamefont {Crivellin}, \citenamefont {Kirk}, \citenamefont {Manzari},\
  and\ \citenamefont {Montull}}]{Buras:2021btx}%
  \BibitemOpen
  \bibfield  {author} {\bibinfo {author} {\bibfnamefont {A.~J.}\ \bibnamefont
  {Buras}}, \bibinfo {author} {\bibfnamefont {A.}~\bibnamefont {Crivellin}},
  \bibinfo {author} {\bibfnamefont {F.}~\bibnamefont {Kirk}}, \bibinfo {author}
  {\bibfnamefont {C.~A.}\ \bibnamefont {Manzari}}, \ and\ \bibinfo {author}
  {\bibfnamefont {M.}~\bibnamefont {Montull}},\ }\href {\doibase
  10.1007/JHEP06(2021)068} {\bibfield  {journal} {\bibinfo  {journal} {JHEP}\
  }\textbf {\bibinfo {volume} {06}},\ \bibinfo {pages} {068} (\bibinfo {year}
  {2021})},\ \Eprint {http://arxiv.org/abs/2104.07680} {arXiv:2104.07680
  [hep-ph]} \BibitemShut {NoStop}%
\bibitem [{\citenamefont {Langacker}\ and\ \citenamefont
  {Plumacher}(2000)}]{Langacker:2000ju}%
  \BibitemOpen
  \bibfield  {author} {\bibinfo {author} {\bibfnamefont {P.}~\bibnamefont
  {Langacker}}\ and\ \bibinfo {author} {\bibfnamefont {M.}~\bibnamefont
  {Plumacher}},\ }\href {\doibase 10.1103/PhysRevD.62.013006} {\bibfield
  {journal} {\bibinfo  {journal} {Phys. Rev. D}\ }\textbf {\bibinfo {volume}
  {62}},\ \bibinfo {pages} {013006} (\bibinfo {year} {2000})},\ \Eprint
  {http://arxiv.org/abs/hep-ph/0001204} {arXiv:hep-ph/0001204} \BibitemShut
  {NoStop}%
\bibitem [{\citenamefont {Buchmuller}\ \emph {et~al.}(1987)\citenamefont
  {Buchmuller}, \citenamefont {Ruckl},\ and\ \citenamefont
  {Wyler}}]{Buchmuller:1986zs}%
  \BibitemOpen
  \bibfield  {author} {\bibinfo {author} {\bibfnamefont {W.}~\bibnamefont
  {Buchmuller}}, \bibinfo {author} {\bibfnamefont {R.}~\bibnamefont {Ruckl}}, \
  and\ \bibinfo {author} {\bibfnamefont {D.}~\bibnamefont {Wyler}},\ }\href
  {\doibase 10.1016/0370-2693(87)90637-X} {\bibfield  {journal} {\bibinfo
  {journal} {Phys. Lett. B}\ }\textbf {\bibinfo {volume} {191}},\ \bibinfo
  {pages} {442} (\bibinfo {year} {1987})},\ \bibinfo {note} {[Erratum:
  Phys.Lett.B 448, 320--320 (1999)]}\BibitemShut {NoStop}%
\bibitem [{\citenamefont {Crivellin}\ \emph
  {et~al.}(2021{\natexlab{e}})\citenamefont {Crivellin}, \citenamefont
  {M\"uller},\ and\ \citenamefont {Schnell}}]{Crivellin:2021egp}%
  \BibitemOpen
  \bibfield  {author} {\bibinfo {author} {\bibfnamefont {A.}~\bibnamefont
  {Crivellin}}, \bibinfo {author} {\bibfnamefont {D.}~\bibnamefont {M\"uller}},
  \ and\ \bibinfo {author} {\bibfnamefont {L.}~\bibnamefont {Schnell}},\ }\href
  {\doibase 10.1103/PhysRevD.103.115023} {\bibfield  {journal} {\bibinfo
  {journal} {Phys. Rev. D}\ }\textbf {\bibinfo {volume} {103}},\ \bibinfo
  {pages} {115023} (\bibinfo {year} {2021}{\natexlab{e}})},\ \Eprint
  {http://arxiv.org/abs/2104.06417} {arXiv:2104.06417 [hep-ph]} \BibitemShut
  {NoStop}%
\bibitem [{\citenamefont {Crivellin}\ \emph
  {et~al.}(2021{\natexlab{f}})\citenamefont {Crivellin}, \citenamefont
  {Hoferichter}, \citenamefont {Kirk}, \citenamefont {Manzari},\ and\
  \citenamefont {Schnell}}]{Crivellin:2021bkd}%
  \BibitemOpen
  \bibfield  {author} {\bibinfo {author} {\bibfnamefont {A.}~\bibnamefont
  {Crivellin}}, \bibinfo {author} {\bibfnamefont {M.}~\bibnamefont
  {Hoferichter}}, \bibinfo {author} {\bibfnamefont {M.}~\bibnamefont {Kirk}},
  \bibinfo {author} {\bibfnamefont {C.~A.}\ \bibnamefont {Manzari}}, \ and\
  \bibinfo {author} {\bibfnamefont {L.}~\bibnamefont {Schnell}},\ }\href@noop
  {} {\  (\bibinfo {year} {2021}{\natexlab{f}})},\ \Eprint
  {http://arxiv.org/abs/2107.13569} {arXiv:2107.13569 [hep-ph]} \BibitemShut
  {NoStop}%
\bibitem [{\citenamefont {Pati}\ and\ \citenamefont
  {Salam}(1974)}]{Pati:1974yy}%
  \BibitemOpen
  \bibfield  {author} {\bibinfo {author} {\bibfnamefont {J.~C.}\ \bibnamefont
  {Pati}}\ and\ \bibinfo {author} {\bibfnamefont {A.}~\bibnamefont {Salam}},\
  }\href {\doibase 10.1103/PhysRevD.10.275} {\bibfield  {journal} {\bibinfo
  {journal} {Phys. Rev. D}\ }\textbf {\bibinfo {volume} {10}},\ \bibinfo
  {pages} {275} (\bibinfo {year} {1974})},\ \bibinfo {note} {[Erratum:
  Phys.Rev.D 11, 703--703 (1975)]}\BibitemShut {NoStop}%
\bibitem [{\citenamefont {Georgi}\ and\ \citenamefont
  {Glashow}(1974)}]{Georgi:1974sy}%
  \BibitemOpen
  \bibfield  {author} {\bibinfo {author} {\bibfnamefont {H.}~\bibnamefont
  {Georgi}}\ and\ \bibinfo {author} {\bibfnamefont {S.~L.}\ \bibnamefont
  {Glashow}},\ }\href {\doibase 10.1103/PhysRevLett.32.438} {\bibfield
  {journal} {\bibinfo  {journal} {Phys. Rev. Lett.}\ }\textbf {\bibinfo
  {volume} {32}},\ \bibinfo {pages} {438} (\bibinfo {year} {1974})}\BibitemShut
  {NoStop}%
\bibitem [{\citenamefont {Dimopoulos}\ \emph {et~al.}(1980)\citenamefont
  {Dimopoulos}, \citenamefont {Raby},\ and\ \citenamefont
  {Susskind}}]{Dimopoulos:1980hn}%
  \BibitemOpen
  \bibfield  {author} {\bibinfo {author} {\bibfnamefont {S.}~\bibnamefont
  {Dimopoulos}}, \bibinfo {author} {\bibfnamefont {S.}~\bibnamefont {Raby}}, \
  and\ \bibinfo {author} {\bibfnamefont {L.}~\bibnamefont {Susskind}},\ }\href
  {\doibase 10.1016/0550-3213(80)90215-1} {\bibfield  {journal} {\bibinfo
  {journal} {Nucl. Phys. B}\ }\textbf {\bibinfo {volume} {173}},\ \bibinfo
  {pages} {208} (\bibinfo {year} {1980})}\BibitemShut {NoStop}%
\bibitem [{\citenamefont {Barbier}\ \emph {et~al.}(2005)\citenamefont {Barbier}
  \emph {et~al.}}]{Barbier:2004ez}%
  \BibitemOpen
  \bibfield  {author} {\bibinfo {author} {\bibfnamefont {R.}~\bibnamefont
  {Barbier}} \emph {et~al.},\ }\href {\doibase 10.1016/j.physrep.2005.08.006}
  {\bibfield  {journal} {\bibinfo  {journal} {Phys. Rept.}\ }\textbf {\bibinfo
  {volume} {420}},\ \bibinfo {pages} {1} (\bibinfo {year} {2005})},\ \Eprint
  {http://arxiv.org/abs/hep-ph/0406039} {arXiv:hep-ph/0406039} \BibitemShut
  {NoStop}%
\bibitem [{\citenamefont {Davidson}\ \emph {et~al.}(1994)\citenamefont
  {Davidson}, \citenamefont {Bailey},\ and\ \citenamefont
  {Campbell}}]{Davidson:1993qk}%
  \BibitemOpen
  \bibfield  {author} {\bibinfo {author} {\bibfnamefont {S.}~\bibnamefont
  {Davidson}}, \bibinfo {author} {\bibfnamefont {D.~C.}\ \bibnamefont
  {Bailey}}, \ and\ \bibinfo {author} {\bibfnamefont {B.~A.}\ \bibnamefont
  {Campbell}},\ }\href {\doibase 10.1007/BF01552629} {\bibfield  {journal}
  {\bibinfo  {journal} {Z. Phys. C}\ }\textbf {\bibinfo {volume} {61}},\
  \bibinfo {pages} {613} (\bibinfo {year} {1994})},\ \Eprint
  {http://arxiv.org/abs/hep-ph/9309310} {arXiv:hep-ph/9309310} \BibitemShut
  {NoStop}%
\bibitem [{\citenamefont {Ramsey-Musolf}\ \emph {et~al.}(2007)\citenamefont
  {Ramsey-Musolf}, \citenamefont {Su},\ and\ \citenamefont
  {Tulin}}]{RamseyMusolf:2007yb}%
  \BibitemOpen
  \bibfield  {author} {\bibinfo {author} {\bibfnamefont {M.~J.}\ \bibnamefont
  {Ramsey-Musolf}}, \bibinfo {author} {\bibfnamefont {S.}~\bibnamefont {Su}}, \
  and\ \bibinfo {author} {\bibfnamefont {S.}~\bibnamefont {Tulin}},\ }\href
  {\doibase 10.1103/PhysRevD.76.095017} {\bibfield  {journal} {\bibinfo
  {journal} {Phys. Rev. D}\ }\textbf {\bibinfo {volume} {76}},\ \bibinfo
  {pages} {095017} (\bibinfo {year} {2007})},\ \Eprint
  {http://arxiv.org/abs/0705.0028} {arXiv:0705.0028 [hep-ph]} \BibitemShut
  {NoStop}%
\bibitem [{\citenamefont {Dor\v{s}ner}\ \emph {et~al.}(2016)\citenamefont
  {Dor\v{s}ner}, \citenamefont {Fajfer}, \citenamefont {Greljo}, \citenamefont
  {Kamenik},\ and\ \citenamefont {Ko\v{s}nik}}]{Dorsner:2016wpm}%
  \BibitemOpen
  \bibfield  {author} {\bibinfo {author} {\bibfnamefont {I.}~\bibnamefont
  {Dor\v{s}ner}}, \bibinfo {author} {\bibfnamefont {S.}~\bibnamefont {Fajfer}},
  \bibinfo {author} {\bibfnamefont {A.}~\bibnamefont {Greljo}}, \bibinfo
  {author} {\bibfnamefont {J.~F.}\ \bibnamefont {Kamenik}}, \ and\ \bibinfo
  {author} {\bibfnamefont {N.}~\bibnamefont {Ko\v{s}nik}},\ }\href {\doibase
  10.1016/j.physrep.2016.06.001} {\bibfield  {journal} {\bibinfo  {journal}
  {Phys. Rept.}\ }\textbf {\bibinfo {volume} {641}},\ \bibinfo {pages} {1}
  (\bibinfo {year} {2016})},\ \Eprint {http://arxiv.org/abs/1603.04993}
  {arXiv:1603.04993 [hep-ph]} \BibitemShut {NoStop}%
\bibitem [{\citenamefont {Mandal}\ and\ \citenamefont
  {Pich}(2019)}]{Mandal:2019gff}%
  \BibitemOpen
  \bibfield  {author} {\bibinfo {author} {\bibfnamefont {R.}~\bibnamefont
  {Mandal}}\ and\ \bibinfo {author} {\bibfnamefont {A.}~\bibnamefont {Pich}},\
  }\href {\doibase 10.1007/JHEP12(2019)089} {\bibfield  {journal} {\bibinfo
  {journal} {JHEP}\ }\textbf {\bibinfo {volume} {12}},\ \bibinfo {pages} {089}
  (\bibinfo {year} {2019})},\ \Eprint {http://arxiv.org/abs/1908.11155}
  {arXiv:1908.11155 [hep-ph]} \BibitemShut {NoStop}%
\bibitem [{\citenamefont {Crivellin}\ \emph {et~al.}(2013)\citenamefont
  {Crivellin}, \citenamefont {Kokulu},\ and\ \citenamefont
  {Greub}}]{Crivellin:2013wna}%
  \BibitemOpen
  \bibfield  {author} {\bibinfo {author} {\bibfnamefont {A.}~\bibnamefont
  {Crivellin}}, \bibinfo {author} {\bibfnamefont {A.}~\bibnamefont {Kokulu}}, \
  and\ \bibinfo {author} {\bibfnamefont {C.}~\bibnamefont {Greub}},\ }\href
  {\doibase 10.1103/PhysRevD.87.094031} {\bibfield  {journal} {\bibinfo
  {journal} {Phys. Rev. D}\ }\textbf {\bibinfo {volume} {87}},\ \bibinfo
  {pages} {094031} (\bibinfo {year} {2013})},\ \Eprint
  {http://arxiv.org/abs/1303.5877} {arXiv:1303.5877 [hep-ph]} \BibitemShut
  {NoStop}%
\bibitem [{\citenamefont {Masiero}\ \emph {et~al.}(2006)\citenamefont
  {Masiero}, \citenamefont {Paradisi},\ and\ \citenamefont
  {Petronzio}}]{Masiero:2005wr}%
  \BibitemOpen
  \bibfield  {author} {\bibinfo {author} {\bibfnamefont {A.}~\bibnamefont
  {Masiero}}, \bibinfo {author} {\bibfnamefont {P.}~\bibnamefont {Paradisi}}, \
  and\ \bibinfo {author} {\bibfnamefont {R.}~\bibnamefont {Petronzio}},\ }\href
  {\doibase 10.1103/PhysRevD.74.011701} {\bibfield  {journal} {\bibinfo
  {journal} {Phys. Rev. D}\ }\textbf {\bibinfo {volume} {74}},\ \bibinfo
  {pages} {011701} (\bibinfo {year} {2006})},\ \Eprint
  {http://arxiv.org/abs/hep-ph/0511289} {arXiv:hep-ph/0511289} \BibitemShut
  {NoStop}%
\bibitem [{\citenamefont {Girrbach}\ and\ \citenamefont
  {Nierste}(2012)}]{Girrbach:2012km}%
  \BibitemOpen
  \bibfield  {author} {\bibinfo {author} {\bibfnamefont {J.}~\bibnamefont
  {Girrbach}}\ and\ \bibinfo {author} {\bibfnamefont {U.}~\bibnamefont
  {Nierste}},\ }\href@noop {} {\  (\bibinfo {year} {2012})},\ \Eprint
  {http://arxiv.org/abs/1202.4906} {arXiv:1202.4906 [hep-ph]} \BibitemShut
  {NoStop}%
\bibitem [{\citenamefont {Krawczyk}\ and\ \citenamefont
  {Temes}(2005)}]{Krawczyk:2004na}%
  \BibitemOpen
  \bibfield  {author} {\bibinfo {author} {\bibfnamefont {M.}~\bibnamefont
  {Krawczyk}}\ and\ \bibinfo {author} {\bibfnamefont {D.}~\bibnamefont
  {Temes}},\ }\href {\doibase 10.1140/epjc/s2005-02370-2} {\bibfield  {journal}
  {\bibinfo  {journal} {Eur. Phys. J. C}\ }\textbf {\bibinfo {volume} {44}},\
  \bibinfo {pages} {435} (\bibinfo {year} {2005})},\ \Eprint
  {http://arxiv.org/abs/hep-ph/0410248} {arXiv:hep-ph/0410248} \BibitemShut
  {NoStop}%
\bibitem [{\citenamefont {Broggio}\ \emph {et~al.}(2014)\citenamefont
  {Broggio}, \citenamefont {Chun}, \citenamefont {Passera}, \citenamefont
  {Patel},\ and\ \citenamefont {Vempati}}]{Broggio:2014mna}%
  \BibitemOpen
  \bibfield  {author} {\bibinfo {author} {\bibfnamefont {A.}~\bibnamefont
  {Broggio}}, \bibinfo {author} {\bibfnamefont {E.~J.}\ \bibnamefont {Chun}},
  \bibinfo {author} {\bibfnamefont {M.}~\bibnamefont {Passera}}, \bibinfo
  {author} {\bibfnamefont {K.~M.}\ \bibnamefont {Patel}}, \ and\ \bibinfo
  {author} {\bibfnamefont {S.~K.}\ \bibnamefont {Vempati}},\ }\href {\doibase
  10.1007/JHEP11(2014)058} {\bibfield  {journal} {\bibinfo  {journal} {JHEP}\
  }\textbf {\bibinfo {volume} {11}},\ \bibinfo {pages} {058} (\bibinfo {year}
  {2014})},\ \Eprint {http://arxiv.org/abs/1409.3199} {arXiv:1409.3199
  [hep-ph]} \BibitemShut {NoStop}%
\bibitem [{\citenamefont {Chun}\ and\ \citenamefont
  {Kim}(2016)}]{Chun:2016hzs}%
  \BibitemOpen
  \bibfield  {author} {\bibinfo {author} {\bibfnamefont {E.~J.}\ \bibnamefont
  {Chun}}\ and\ \bibinfo {author} {\bibfnamefont {J.}~\bibnamefont {Kim}},\
  }\href {\doibase 10.1007/JHEP07(2016)110} {\bibfield  {journal} {\bibinfo
  {journal} {JHEP}\ }\textbf {\bibinfo {volume} {07}},\ \bibinfo {pages} {110}
  (\bibinfo {year} {2016})},\ \Eprint {http://arxiv.org/abs/1605.06298}
  {arXiv:1605.06298 [hep-ph]} \BibitemShut {NoStop}%
\bibitem [{\citenamefont {Aguilar-Arevalo}\ \emph
  {et~al.}(2015{\natexlab{b}})\citenamefont {Aguilar-Arevalo} \emph
  {et~al.}}]{PiENu:2015seu}%
  \BibitemOpen
  \bibfield  {author} {\bibinfo {author} {\bibfnamefont {A.}~\bibnamefont
  {Aguilar-Arevalo}} \emph {et~al.} (\bibinfo {collaboration} {PiENu}),\ }\href
  {\doibase 10.1103/PhysRevLett.115.071801} {\bibfield  {journal} {\bibinfo
  {journal} {Phys. Rev. Lett.}\ }\textbf {\bibinfo {volume} {115}},\ \bibinfo
  {pages} {071801} (\bibinfo {year} {2015}{\natexlab{b}})},\ \Eprint
  {http://arxiv.org/abs/1506.05845} {arXiv:1506.05845 [hep-ex]} \BibitemShut
  {NoStop}%
\bibitem [{\citenamefont {Aguilar-Arevalo}\ \emph {et~al.}(2010)\citenamefont
  {Aguilar-Arevalo} \emph {et~al.}}]{AguilarArevalo:2010xi}%
  \BibitemOpen
  \bibfield  {author} {\bibinfo {author} {\bibfnamefont {A.}~\bibnamefont
  {Aguilar-Arevalo}} \emph {et~al.},\ }\href {\doibase
  10.1016/j.nima.2010.05.037} {\bibfield  {journal} {\bibinfo  {journal} {Nucl.
  Instrum. Meth. A}\ }\textbf {\bibinfo {volume} {621}},\ \bibinfo {pages}
  {188} (\bibinfo {year} {2010})},\ \Eprint {http://arxiv.org/abs/1003.2235}
  {arXiv:1003.2235 [physics.ins-det]} \BibitemShut {NoStop}%
\bibitem [{\citenamefont {{PEN Collaboration}}()}]{Pocanic1}%
  \BibitemOpen
  \bibfield  {author} {\bibinfo {author} {\bibnamefont {{PEN Collaboration}}},\
  }\href@noop {} {}\bibinfo {howpublished}
  {\url{http://pen.phys.virginia.edu/}}\BibitemShut {NoStop}%
\bibitem [{\citenamefont {Glaser}\ \emph {et~al.}(2018)\citenamefont {Glaser}
  \emph {et~al.}}]{Glaser:2018aat}%
  \BibitemOpen
  \bibfield  {author} {\bibinfo {author} {\bibfnamefont {C.~J.}\ \bibnamefont
  {Glaser}} \emph {et~al.} (\bibinfo {collaboration} {PEN}),\ }in\ \href@noop
  {} {\emph {\bibinfo {booktitle} {{13th Conference on the Intersections of
  Particle and Nuclear Physics}}}}\ (\bibinfo {year} {2018})\ \Eprint
  {http://arxiv.org/abs/1812.00782} {arXiv:1812.00782 [hep-ex]} \BibitemShut
  {NoStop}%
\bibitem [{\citenamefont {Pocanic}\ \emph {et~al.}(2014)\citenamefont
  {Pocanic}, \citenamefont {Frlez},\ and\ \citenamefont {van~der
  Schaaf}}]{Pocanic:2014jka}%
  \BibitemOpen
  \bibfield  {author} {\bibinfo {author} {\bibfnamefont {D.}~\bibnamefont
  {Pocanic}}, \bibinfo {author} {\bibfnamefont {E.}~\bibnamefont {Frlez}}, \
  and\ \bibinfo {author} {\bibfnamefont {A.}~\bibnamefont {van~der Schaaf}},\
  }\href {\doibase 10.1088/0954-3899/41/11/114002} {\bibfield  {journal}
  {\bibinfo  {journal} {J. Phys. G}\ }\textbf {\bibinfo {volume} {41}},\
  \bibinfo {pages} {114002} (\bibinfo {year} {2014})},\ \Eprint
  {http://arxiv.org/abs/1407.2865} {arXiv:1407.2865 [hep-ex]} \BibitemShut
  {NoStop}%
\bibitem [{\citenamefont {{PIONEER Collaboration}}(2021)}]{pioneer1}%
  \BibitemOpen
  \bibfield  {author} {\bibinfo {author} {\bibnamefont {{PIONEER
  Collaboration}}},\ }\href@noop {} {}\bibinfo {howpublished}
  {\url{https://www.snowmass21.org/docs/files/summaries/RF/SNOWMASS21-RF2_RF3-048.pdf}}
  (\bibinfo {year} {2021})\BibitemShut {NoStop}%
\bibitem [{\citenamefont {Sadrozinski}\ \emph {et~al.}(2018)\citenamefont
  {Sadrozinski}, \citenamefont {Seiden},\ and\ \citenamefont
  {Cartiglia}}]{Sadrozinski:2017qpv}%
  \BibitemOpen
  \bibfield  {author} {\bibinfo {author} {\bibfnamefont {H.~F.~W.}\
  \bibnamefont {Sadrozinski}}, \bibinfo {author} {\bibfnamefont
  {A.}~\bibnamefont {Seiden}}, \ and\ \bibinfo {author} {\bibfnamefont
  {N.}~\bibnamefont {Cartiglia}},\ }\href {\doibase 10.1088/1361-6633/aa94d3}
  {\bibfield  {journal} {\bibinfo  {journal} {Rept. Prog. Phys.}\ }\textbf
  {\bibinfo {volume} {81}},\ \bibinfo {pages} {026101} (\bibinfo {year}
  {2018})},\ \Eprint {http://arxiv.org/abs/1704.08666} {arXiv:1704.08666
  [physics.ins-det]} \BibitemShut {NoStop}%
\bibitem [{\citenamefont {Pocanic}\ \emph {et~al.}(2004)\citenamefont {Pocanic}
  \emph {et~al.}}]{Pocanic:2003pf}%
  \BibitemOpen
  \bibfield  {author} {\bibinfo {author} {\bibfnamefont {D.}~\bibnamefont
  {Pocanic}} \emph {et~al.},\ }\href {\doibase 10.1103/PhysRevLett.93.181803}
  {\bibfield  {journal} {\bibinfo  {journal} {Phys. Rev. Lett.}\ }\textbf
  {\bibinfo {volume} {93}},\ \bibinfo {pages} {181803} (\bibinfo {year}
  {2004})},\ \Eprint {http://arxiv.org/abs/hep-ex/0312030}
  {arXiv:hep-ex/0312030} \BibitemShut {NoStop}%
\bibitem [{\citenamefont {Sirlin}(1978)}]{Sirlin:1977sv}%
  \BibitemOpen
  \bibfield  {author} {\bibinfo {author} {\bibfnamefont {A.}~\bibnamefont
  {Sirlin}},\ }\href {\doibase 10.1103/RevModPhys.50.573} {\bibfield  {journal}
  {\bibinfo  {journal} {Rev. Mod. Phys.}\ }\textbf {\bibinfo {volume} {50}},\
  \bibinfo {pages} {573} (\bibinfo {year} {1978})},\ \bibinfo {note} {[Erratum:
  Rev.Mod.Phys. 50, 905 (1978)]}\BibitemShut {NoStop}%
\bibitem [{\citenamefont {Cirigliano}\ \emph {et~al.}(2003)\citenamefont
  {Cirigliano}, \citenamefont {Knecht}, \citenamefont {Neufeld},\ and\
  \citenamefont {Pichl}}]{Cirigliano:2002ng}%
  \BibitemOpen
  \bibfield  {author} {\bibinfo {author} {\bibfnamefont {V.}~\bibnamefont
  {Cirigliano}}, \bibinfo {author} {\bibfnamefont {M.}~\bibnamefont {Knecht}},
  \bibinfo {author} {\bibfnamefont {H.}~\bibnamefont {Neufeld}}, \ and\
  \bibinfo {author} {\bibfnamefont {H.}~\bibnamefont {Pichl}},\ }\href
  {\doibase 10.1140/epjc/s2002-01093-2} {\bibfield  {journal} {\bibinfo
  {journal} {Eur. Phys. J. C}\ }\textbf {\bibinfo {volume} {27}},\ \bibinfo
  {pages} {255} (\bibinfo {year} {2003})},\ \Eprint
  {http://arxiv.org/abs/hep-ph/0209226} {arXiv:hep-ph/0209226} \BibitemShut
  {NoStop}%
\bibitem [{\citenamefont {Passera}\ \emph {et~al.}(2011)\citenamefont
  {Passera}, \citenamefont {Philippides},\ and\ \citenamefont
  {Sirlin}}]{Passera:2011ae}%
  \BibitemOpen
  \bibfield  {author} {\bibinfo {author} {\bibfnamefont {M.}~\bibnamefont
  {Passera}}, \bibinfo {author} {\bibfnamefont {K.}~\bibnamefont
  {Philippides}}, \ and\ \bibinfo {author} {\bibfnamefont {A.}~\bibnamefont
  {Sirlin}},\ }\href {\doibase 10.1103/PhysRevD.84.094030} {\bibfield
  {journal} {\bibinfo  {journal} {Phys. Rev. D}\ }\textbf {\bibinfo {volume}
  {84}},\ \bibinfo {pages} {094030} (\bibinfo {year} {2011})},\ \Eprint
  {http://arxiv.org/abs/1109.1069} {arXiv:1109.1069 [hep-ph]} \BibitemShut
  {NoStop}%
\bibitem [{\citenamefont {Feng}\ \emph {et~al.}(2020)\citenamefont {Feng},
  \citenamefont {Gorchtein}, \citenamefont {Jin}, \citenamefont {Ma},\ and\
  \citenamefont {Seng}}]{Feng:2020zdc}%
  \BibitemOpen
  \bibfield  {author} {\bibinfo {author} {\bibfnamefont {X.}~\bibnamefont
  {Feng}}, \bibinfo {author} {\bibfnamefont {M.}~\bibnamefont {Gorchtein}},
  \bibinfo {author} {\bibfnamefont {L.-C.}\ \bibnamefont {Jin}}, \bibinfo
  {author} {\bibfnamefont {P.-X.}\ \bibnamefont {Ma}}, \ and\ \bibinfo {author}
  {\bibfnamefont {C.-Y.}\ \bibnamefont {Seng}},\ }\href {\doibase
  10.1103/PhysRevLett.124.192002} {\bibfield  {journal} {\bibinfo  {journal}
  {Phys. Rev. Lett.}\ }\textbf {\bibinfo {volume} {124}},\ \bibinfo {pages}
  {192002} (\bibinfo {year} {2020})},\ \Eprint
  {http://arxiv.org/abs/2003.09798} {arXiv:2003.09798 [hep-lat]} \BibitemShut
  {NoStop}%
\bibitem [{\citenamefont {Kohl}(2016)}]{Kohl:2016afc}%
  \BibitemOpen
  \bibfield  {author} {\bibinfo {author} {\bibfnamefont {M.}~\bibnamefont
  {Kohl}} (\bibinfo {collaboration} {TREK/E36}),\ }in\ \href@noop {} {\emph
  {\bibinfo {booktitle} {{Physics with Neutral Kaon Beam at JLab Workshop}}}}\
  (\bibinfo {year} {2016})\ pp.\ \bibinfo {pages} {191--197}\BibitemShut
  {NoStop}%
\bibitem [{\citenamefont {Cortina~Gil}\ \emph
  {et~al.}(2020{\natexlab{b}})\citenamefont {Cortina~Gil} \emph
  {et~al.}}]{NA62:2020fhy}%
  \BibitemOpen
  \bibfield  {author} {\bibinfo {author} {\bibfnamefont {E.}~\bibnamefont
  {Cortina~Gil}} \emph {et~al.} (\bibinfo {collaboration} {NA62}),\ }\href
  {\doibase 10.1007/JHEP11(2020)042} {\bibfield  {journal} {\bibinfo  {journal}
  {JHEP}\ }\textbf {\bibinfo {volume} {11}},\ \bibinfo {pages} {042} (\bibinfo
  {year} {2020}{\natexlab{b}})},\ \Eprint {http://arxiv.org/abs/2007.08218}
  {arXiv:2007.08218 [hep-ex]} \BibitemShut {NoStop}%
\bibitem [{\citenamefont {Abe}\ \emph {et~al.}(2010)\citenamefont {Abe} \emph
  {et~al.}}]{Belle-II:2010dht}%
  \BibitemOpen
  \bibfield  {author} {\bibinfo {author} {\bibfnamefont {T.}~\bibnamefont
  {Abe}} \emph {et~al.} (\bibinfo {collaboration} {Belle-II}),\ }\href@noop {}
  {\  (\bibinfo {year} {2010})},\ \Eprint {http://arxiv.org/abs/1011.0352}
  {arXiv:1011.0352 [physics.ins-det]} \BibitemShut {NoStop}%
\bibitem [{\citenamefont {Altmannshofer}\ \emph {et~al.}(2019)\citenamefont
  {Altmannshofer} \emph {et~al.}}]{Belle-II:2018jsg}%
  \BibitemOpen
  \bibfield  {author} {\bibinfo {author} {\bibfnamefont {W.}~\bibnamefont
  {Altmannshofer}} \emph {et~al.} (\bibinfo {collaboration} {Belle-II}),\
  }\href {\doibase 10.1093/ptep/ptz106} {\bibfield  {journal} {\bibinfo
  {journal} {PTEP}\ }\textbf {\bibinfo {volume} {2019}},\ \bibinfo {pages}
  {123C01} (\bibinfo {year} {2019})},\ \bibinfo {note} {[Erratum: PTEP 2020,
  029201 (2020)]},\ \Eprint {http://arxiv.org/abs/1808.10567} {arXiv:1808.10567
  [hep-ex]} \BibitemShut {NoStop}%
\bibitem [{\citenamefont {Abada}\ \emph {et~al.}(2019)\citenamefont {Abada}
  \emph {et~al.}}]{FCC:2018evy}%
  \BibitemOpen
  \bibfield  {author} {\bibinfo {author} {\bibfnamefont {A.}~\bibnamefont
  {Abada}} \emph {et~al.} (\bibinfo {collaboration} {FCC}),\ }\href {\doibase
  10.1140/epjst/e2019-900045-4} {\bibfield  {journal} {\bibinfo  {journal}
  {Eur. Phys. J. ST}\ }\textbf {\bibinfo {volume} {228}},\ \bibinfo {pages}
  {261} (\bibinfo {year} {2019})}\BibitemShut {NoStop}%
\bibitem [{\citenamefont {Dam}(2021)}]{Dam:2021ibi}%
  \BibitemOpen
  \bibfield  {author} {\bibinfo {author} {\bibfnamefont {M.}~\bibnamefont
  {Dam}},\ }\href {\doibase 10.1140/epjp/s13360-021-01894-y} {\bibfield
  {journal} {\bibinfo  {journal} {Eur. Phys. J. Plus}\ }\textbf {\bibinfo
  {volume} {136}},\ \bibinfo {pages} {963} (\bibinfo {year}
  {2021})}\BibitemShut {NoStop}%
\bibitem [{\citenamefont {Dong}\ \emph {et~al.}(2018)\citenamefont {Dong} \emph
  {et~al.}}]{CEPCStudyGroup:2018ghi}%
  \BibitemOpen
  \bibfield  {author} {\bibinfo {author} {\bibfnamefont {M.}~\bibnamefont
  {Dong}} \emph {et~al.} (\bibinfo {collaboration} {CEPC Study Group}),\
  }\href@noop {} {\  (\bibinfo {year} {2018})},\ \Eprint
  {http://arxiv.org/abs/1811.10545} {arXiv:1811.10545 [hep-ex]} \BibitemShut
  {NoStop}%
\bibitem [{\citenamefont {Di~Carlo}\ \emph {et~al.}(2019)\citenamefont
  {Di~Carlo}, \citenamefont {Giusti}, \citenamefont {Lubicz}, \citenamefont
  {Martinelli}, \citenamefont {Sachrajda}, \citenamefont {Sanfilippo},
  \citenamefont {Simula},\ and\ \citenamefont {Tantalo}}]{DiCarlo:2019thl}%
  \BibitemOpen
  \bibfield  {author} {\bibinfo {author} {\bibfnamefont {M.}~\bibnamefont
  {Di~Carlo}}, \bibinfo {author} {\bibfnamefont {D.}~\bibnamefont {Giusti}},
  \bibinfo {author} {\bibfnamefont {V.}~\bibnamefont {Lubicz}}, \bibinfo
  {author} {\bibfnamefont {G.}~\bibnamefont {Martinelli}}, \bibinfo {author}
  {\bibfnamefont {C.~T.}\ \bibnamefont {Sachrajda}}, \bibinfo {author}
  {\bibfnamefont {F.}~\bibnamefont {Sanfilippo}}, \bibinfo {author}
  {\bibfnamefont {S.}~\bibnamefont {Simula}}, \ and\ \bibinfo {author}
  {\bibfnamefont {N.}~\bibnamefont {Tantalo}},\ }\href {\doibase
  10.1103/PhysRevD.100.034514} {\bibfield  {journal} {\bibinfo  {journal}
  {Phys. Rev. D}\ }\textbf {\bibinfo {volume} {100}},\ \bibinfo {pages}
  {034514} (\bibinfo {year} {2019})},\ \Eprint
  {http://arxiv.org/abs/1904.08731} {arXiv:1904.08731 [hep-lat]} \BibitemShut
  {NoStop}%
\end{thebibliography}%

\end{document}